\documentclass[twocolumn,twocolappendix,numberedappendix]{emulateapj}

\usepackage{amsmath,amssymb}
\usepackage{times}
\usepackage{mathrsfs}
\usepackage{bm} 

\newcommand{\beq}{\begin{equation}}
\newcommand{\eeq}{\end{equation}}
\newcommand{\intd}{{\rm d}}
\newcommand{\msun}{\ensuremath{M_\odot}}

\newcommand{\rnu}{\ensuremath{r_{\nu}}}

\newcommand{\texp}{\ensuremath{t_{\rm exp}}}

\newcommand{\qdot}{\ensuremath{\dot{q}}}
\newcommand{\ye}{\ensuremath{Y_{\rm e}}}
\newcommand{\nue}{\ensuremath{\nu_{\rm e}}}
\newcommand{\nuebar}{\ensuremath{\bar{\nu}_{\rm e}}}
\newcommand{\lnue}{\ensuremath{L_{\nue}}}
\newcommand{\lnuebar}{\ensuremath{L_{\nuebar}}}
\newcommand{\lc}{\ensuremath{L^{\rm crit}}}
\newcommand{\lcrit}{\ensuremath{\lnue^{\rm crit}}}
\newcommand{\mdot}{\ensuremath{\dot{M}}}
\newcommand{\rs}{\ensuremath{r_{\rm S}}}
\newcommand{\enue}{\ensuremath{\varepsilon_{\nue}}}
\newcommand{\enuebar}{\ensuremath{\varepsilon_{\nuebar}}}
\newcommand{\gr}{{\tt GR1D}}
\newcommand{\rnue}{\ensuremath{r_{\nue}}}
\newcommand{\rnuebar}{\ensuremath{r_{\nuebar}}}
\newcommand{\pwind}{\ensuremath{P_{\rm wind}}}
\newcommand{\ewind}{\ensuremath{E_{\rm wind}}}
\newcommand{\esn}{\ensuremath{E_{\rm SN}}}
\newcommand{\fsn}{\ensuremath{f_{\rm SN}}}
\newcommand{\vesc}{\ensuremath{v_{\rm esc}}}
\newcommand{\nic}{$^{56}$Ni}
\newcommand{\mni}{\ensuremath{M_{\rm Ni}}}
\newcommand{\ergs}{ergs\,s$^{-1}$}
\newcommand{\fheat}{\ensuremath{f_{\rm heat}}}
\newcommand{\mns}{\ensuremath{\overline{M}_{\rm NS}}}
\newcommand{\sns}{\ensuremath{\sigma_{\rm NS}}}
\newcommand{\mbh}{\ensuremath{\overline{M}_{\rm BH}}}
\newcommand{\sbh}{\ensuremath{\sigma_{\rm BH}}}

\begin{document}

\title{The Landscape of the Neutrino Mechanism of Core-Collapse Supernovae:\\ Neutron Star and Black Hole Mass Functions, Explosion Energies and Nickel Yields}
\shorttitle{The Landscape of the Neutrino Mechanism}
\shortauthors{Pejcha and Thompson} 

\author{ Ond\v{r}ej Pejcha\altaffilmark{1}}
\affil{Department of Astrophysical Sciences, Princeton University, 4 Ivy Lane, Princeton, NJ 08540, USA}
\email{pejcha@astro.princeton.edu}
\and
\author{Todd A. Thompson}
\affil{Department of Astronomy and Center for Cosmology and Astroparticle Physics, The Ohio State University, 140 West 18th Avenue, Columbus, OH 43210, USA}
\email{thompson@astronomy.ohio-state.edu}
\altaffiltext{1}{Hubble and Lyman Spitzer Jr.\ Fellow}

\begin{abstract}
If the neutrino luminosity from the proto-neutron star formed during a massive star core collapse exceeds a critical threshold, a supernova (SN) results. Using spherical quasi-static evolutionary sequences for hundreds of progenitors over a range of metallicities, we study how the explosion threshold maps onto observables, including the fraction of successful explosions, the neutron star (NS) and black hole (BH) mass functions, the explosion energies ($\esn$) and nickel yields ($\mni$), and their mutual correlations. Successful explosions are intertwined with failures in a complex pattern that is not simply related to initial progenitor mass or compactness. We predict that progenitors with initial masses of $15\pm1$, $19\pm1$, and $\sim21-26\,\msun$ are most likely to form BHs, that the BH formation probability is non-zero at solar-metallicity and increases significantly at low metallicity, and that low luminosity, low Ni-yield SNe come from progenitors close to success/failure interfaces. We qualitatively reproduce the observed $\esn-\mni$ correlation, we predict a correlation between the mean and width of the NS mass and $\esn$ distributions, and that the means of the NS and BH mass distributions are correlated.  We show that the observed mean NS mass of $\simeq 1.33\,\msun$ implies that the successful explosion fraction is higher than $0.35$. Overall, we show that the neutrino mechanism can in principle explain the observed properties of SNe and their compact objects. We argue that the rugged landscape of progenitors and outcomes mandates that SN theory should focus on reproducing the wide ranging distributions of observed SN properties.
\end{abstract}

\section{Introduction}


The inevitable collapse of the core of a massive star at the end of its life leaves behind a neutron star or a black hole. A magnificent supernova explosion accompanies many (if potentially not all) core collapses and ejects chemically enriched matter into the interstellar medium. The typical characteristics of massive stars, supernova explosions and their remnants are roughly known. For example, the minimum initial stellar mass to produce a supernova explosion is $8 \pm 1\,\msun$ \citep[e.g.][]{smartt09,smartt_etal09,ibeling13}, the kinetic supernova explosion energies are a few times $10^{51}$\,ergs \citep[e.g.][]{hamuy03,dallora14}, and the neutron star masses at birth are $1.2$ to $1.5\,\msun$ \citep[e.g.][]{ozel12,pejcha_ns,kiziltan13}. Yet, the mapping between the structure of an individual pre-collapse star and the properties of its explosion (if it explodes at all) and its remnant are poorly known.

\subsection{Progenitor structure determines supernova outcome}

The core of a massive star starts to collapse when it reaches the Chandrasekhar mass. Effects of thermal structure, Coulomb corrections, surface pressure, and general relativity \citep{timmes96} result in a range of Chandrasekhar core masses between about $1.2$ and $1.6\,\msun$ at solar metallicity and even higher masses for lower metallicities \citep{woosley02}. Similarly, the densities of the progenitors at the mass coordinate corresponding to a typical neutron star mass can differ by many orders of magnitude \citep[and Fig.~\ref{fig:progenitors} of this paper]{woosley02}.

The collapsing core separates into a homologously contracting inner core and a super-sonically infalling outer core \citep{goldreich80,yahil83}. For a fixed set of physical assumptions, the mass of the inner core $M_{\rm ic} \sim 0.5\,\msun$ is only very weakly dependent on the pre-supernova structure \citep{burrows83,baron85a,baron85b,baron89,myra87,bruenn88,baron90,langanke03}. When the central densities exceed the nuclear density, the adiabatic index of the material increases and the infall rapidly halts. A pressure wave propagates outward from the center and steepens to a shock wave just outside the inner core. As a result, the shock wave is created at approximately the same mass coordinate ($M_{\rm ic}$) for all progenitors, which is thus the common initial condition for all supernovae.

The shock wave does not propagate all the way through the progenitor star: it inevitably turns into a stalled accretion shock by a combination of energy losses due to neutrino emission as the shock wave propagates to the optically-thin surrounding envelope, dissociation of the nuclei across the shock, and the ram pressure of the dense infalling material. The ensuing quasi-static accretion phase typically lasts many dynamical times of the system. Despite decades of modeling effort, the mechanism responsible for shock revival has not been conclusively identified. 

The most studied ``delayed neutrino mechanism'' makes use of energy deposition in the semi-transparent layer below the shock due to the absorption of the neutrinos emanating from the hot proto-neutron star (PNS).
The accretion shock forms an information barrier, where the super-sonically infalling progenitor mantle provides the outer boundary condition of instantaneous mass, momentum, energy, and compositional flux for the sub-sonic region containing the PNS and its overlying semi-transparent settling layer. The PNS evolution proceeds on the neutrino diffusion timescale, which can be substantially longer than the advection or sound-crossing timescale below the shock. Since the PNS is built up by accretion through the shock on the inner core, the PNS radius, luminosity, and temperature encode the time-integrated accretion history and thus reflect the progenitor structure.

Originally developed by \citet{colgate66} and \citet{bethe85}, the neutrino mechanism was formalized by \citet{burrows93}, who calculated the steady-state structure of the region between the PNS surface and the accretion shock during the quasi-static evolutionary phase. For fixed mass accretion rate through the shock they found an upper bound on the neutrino luminosity of the PNS to sustain steady-state accretion. If the actual PNS neutrino luminosity evolves above this {\em critical neutrino luminosity} ($\lc$), a supernova explosion results. The critical luminosity $\lc$ is a function of PNS mass, radius, and mass accretion rate through the shock forming a critical curve or manifold in these parameters. 

In \citet{pt12}, we identified the physics and origin of $\lc$. First, we analyzed isothermal accretion bounded by a shock and found that there exists a critical isothermal sound speed $c_T^{\rm crit}$ above which there is no steady-state accretion solution: it is not possible to simultaneously satisfy the shock jump conditions and the Euler equations \citep{yamasaki05}. We identified the {\em antesonic} condition, which gives the critical condition for explosion $c_T^2/\vesc^2 = 3/16$, where $\vesc$ is the escape velocity. We applied our results to a more complete calculation including degenerate electrons and positrons, electron fraction and radial dependence of neutrino luminosities. We proved a direct equivalence $c_T^{\rm crit}$ and $\lc$ and generalized the antesonic condition as $\max (c_S/\vesc)^2 \simeq 0.19$, where $c_S$ is the adiabatic sound speed and the maximum is searched between the neutrinosphere and the shock. This condition captures $\lc$ to within $5\%$ over a broad range of boundary values and physics.

The formalism of $\lc$ offers a way to study the influence of various physical effects on the supernova mechanism in a controlled setting. For example, the effects of convection, rotation, equation of state on the steady-state calculation of $\lc$, and the associated linear stability were studied by \citet{yamasaki05,yamasaki06,yamasaki07}. The equation-of-state dependence was later studied in time-dependent calculations by \citet{couch13a} and \citet{suwa13}. In \citet{pt12}, we found that including the accretion luminosity of the sub-sonic flow feeds back on the flow structure and non-trivially reduces $\lc$ by up to about $23\%$. In \citet{pejcha_cno}, we showed how collective neutrino oscillations can lower $\lc$ if the oscillations are not suppressed by other effects such a interactions with electrons and positrons. 

Recently, considerable attention within the framework of $\lc$ has been devoted to the differences between simulations with identical physics and boundary conditions, but performed either in 1D, 2D, or 3D, because the most detailed 1D models generically fail to explode, except for the lowest mass progenitors \citep{rampp00,bruenn01,liebendorfer01,mezzacappa01,thompson03,kitaura06,janka08,suwa14}. It has been established that 2D simulations robustly require $\lc$ lower by about $20\%$ compared to 1D \citep{ohnishi06,murphy08}, but it has not been unambiguously determined that the hierarchy continues and $\lc$ is lower in 3D than in 2D. In some cases, $\lc$ in 3D is slightly changed with respect to 2D \citep{iwakami08,nordhaus10,hanke12,takiwaki12,couch13b,couch14,dolence13,dolence14}. The multi-dimensional effects responsible for the reduction of $\lc$ or increase of the core luminosity were argued to be the convection both in the PNS interior \citep{keil96,mezzacappa98,dessart06} and in the heating region between the PNS and the shock \citep{herant92,herant94,janka96,burrows95,fryer00,fryer02,fryer04,buras06,murphy11,burrows12,murphy13,ott13}, the standing accretion shock instability
(SASI) and/or vortical-acoustic instability \citep{foglizzo02,blondin03,blondin06,ohnishi06,iwakami08,marek09,fernandez10,suwa10,hanke12,hanke13}, or their combination \citep{scheck08,muller12b,iwakami14,fernandez14}. Despite the disagreements in the cause for the changes in $\lc$ as a function of dimension of the simulations, {\em the critical curves in 1D, 2D, and 3D are basically parallel to each other for a given set of calculations}. This means that by consistently rescaling the $\lc$ obtained in 1D, it should be possible to obtain a relatively good estimate of $\lc$ in 2D or 3D simulations. 

Fundamentally, the dependence of $\lc$ on the dimension of the calculation is a systematic uncertainty in our understanding of the supernova explosion that affects all calculations in a consistent but not fully known manner. Similarly, there are other new and potentially interesting heating mechanisms such as magnetic dissipation \citep{thompson05}, damping of Alfvenic waves, and magnetic fields, possibly combined with rotation \citep[e.g.][]{iwakami14,nakamura14a,sawai13,sawai14,obergaulinger14}. The differences between the values of $\lc$ in steady-state and time-dependent calculations might as well be important \citep{fernandez12}. These all lead to systematic uncertainties in the normalization and shape of the critical curve that need to be addressed by simulations, but also by understanding their observational consequences. The point of this paper is to investigate the consequences of variations in the normalization and shape of $\lc$ and thus implicitly taking into account the mentioned systematic uncertainties in our understanding of the neutrino mechanism.

Inevitably, the physics of the supernova explosion mechanism ought to be verified by comparing the models with observations. The moment when the neutrino luminosity exceeds $\lc$ determines a number of parameters amenable to observational verification. The mass of the compact remnant ($M_{\rm NS}$ or $M_{\rm BH}$) is set by the duration and rate of the accretion through the shock with corrections due to the mass loss in the neutrino-driven wind from the PNS, simultaneous accretion and explosion, and fallback. The latter two are potentially substantial. The resulting supernova kinetic energy is composed of several contributions \citep[e.g.][]{scheck06,ugliano12,yamamoto13}, including the integrated mechanical power of the neutrino-driven wind $\pwind$, the recombination energy of nucleons and $\alpha$ particles in the layers below the shock that get ejected, $E_{\rm rec}$, and the nuclear burning of material swept-up by the shock. These sources of energy have to overcome the negative binding energy of the stellar material, $E_{\rm bind}$. Nucleosynthesis in the material swept up by the shock during the explosion produces an amount of \nic, which powers a part of the supernova light curve. Other elements are synthesized depending on the explosion conditions as well.

The binding energy of the progenitor gives an immediate constraint on the explosion mechanism. If the major components of the supernova explosion energy are determined by the physics inside of the shock (neutrino-driven wind and recombination) and not by burning of the material swept-up by the shock, the explosion mechanism has no prior ``knowledge'' of how much energy needs to be supplied to make the star explode.
 If an explosion is forced with a pre-defined asymptotic energy that is smaller than the overlying binding energy of the progenitor, significant fallback will occur \citep[e.g.][]{zhang08}.

\subsection{This paper}

It has been demonstrated that the progenitor structure varies drastically even in relatively narrow ranges of progenitor masses and depends on assumptions like nuclear reaction rates, convection and semiconvection, which are systematic and can be highly uncertain \citep{ibeling13,west13,sukhbold14}. This means that collapse calculations of a single progenitor are not statistically meaningful or representative. Instead, the solution to the supernova problem is inherently statistical in nature in the sense that an ensemble of simulations must correctly reproduce the supernova observables. In the ideal case, the observed properties of supernovae and their remnants would be self-consistently predicted by sophisticated multi-dimensional simulations. However, 3D simulations are currently too expensive to be performed in sufficient detail for enough progenitors and 2D simulations of different groups using similar physics and performed on the same four progenitors show qualitative differences ranging from full success to complete failure (\citealp{muller12a}; M\"uller et al., 2013, YITP conference; \citealp{bruenn13}; \citealp{dolence14}). More fundamentally, no simulation can yet capture all of the physics inside supernovae.

Instead of multi-dimensional simulations, we aim to identify the minimum necessary set of assumptions required to properly reflect the variations between different progenitors and make observational predictions, and then construct a semi-analytic quasi-static model of supernovae based on these assumptions. By investigating a wide of range of progenitors, input physics, and parameterizations of the model, we explore the differences between the progenitors and the range of allowed consequences of the delayed neutrino mechanism while comparing the results to observations. More specifically, we introduce artificial explosions at different moments in our supernova evolutionary sequences, which produces a range of outcomes: later explosions result in more massive neutron stars and lower neutrino luminosities at explosion, which leads to smaller explosion energies and lower nickel yields than in earlier explosions. The point of our work is that we do this consistently over whole populations of progenitors. This enables us to study relations between global population properties. For example, we expect that by lowering the fraction of successful explosions we limit the number of stars having a neutron star as a remnant, which can potentially lower the width of the neutron star mass function. There should be accompanying changes in the black hole mass distribution. Our overarching goal is to see whether or not the known physics of the neutrino mechanism, with some scaling, can account for the observations. We also want to highlight any potential inconsistencies and missing pieces.

In the space of sophistication, number of predicted of observational characteristics, and required computational resources our approach is positioned between the simple ``compactness'' estimate of explosion/failure \citep{oconnor11,oconnor13,sukhbold14,kochanek14b} and more sophisticated studies, which either study a range of progenitors but for a single parameterization \citep{ugliano12,suwa14}, or focus on a limited number of stars, but include more sophisticated or multi-dimensional physics or more parameterizations \citep[e.g.][]{scheck06,arcones07,arcones11,yamamoto13}. Our approach is easy and inexpensive to extend to include more progenitors with different physics. The accuracy of the model can be increased through more sophisticated parameterizations and changes to input physics as the understanding of the supernova explosion mechanism improves.

In Section~\ref{sec:setup},  we construct a semi-analytic parameterized model combining the progenitor structure and the physics of the steady-state critical neutrino luminosity. In Section~\ref{sec:method}, we introduce a method for making systematic predictions from the neutrino mechanism. In Section~\ref{sec:results}, we describe our predictions for explosion success or failure, remnant mass, explosion energy, and nickel yields for a several hundred progenitor models for two representative parameterizations of the neutrino mechanism. In Section~\ref{sec:robust}, we study how our predictions change for many different parameterizations of the neutrino mechanism and examine the effects of progenitor metallicity and equation of state. In Section~\ref{sec:disc}, we compare our predictions with previous results and with observed properties. In Section~\ref{sec:constraint}, we illustrate how the neutrino mechanism can be constrained by comparing our results with observations. We conclude in Section~\ref{sec:conclusions}. In Appendix~\ref{app:nu_wind}, we describe how we estimate the time-integrated energy of the neutrino-driven wind. In Appendix~\ref{app:compactness}, we study the ``compactness'' parameter and its internal consistency.

\section{Calculations setup}
\label{sec:setup}


To connect the neutrino mechanism to observations, we need to obtain the time evolution of the critical neutrino luminosity and compare it with the actual neutrino luminosity to infer explosion properties.  In Section~\ref{sec:lcrit_calc}, we describe our calculation of the critical neutrino luminosity. The time evolution of the boundary conditions is given in Section~\ref{sec:gr1d}, and the employed progenitor models are reviewed in Section~\ref{sec:progenitors}.

\subsection{Critical luminosity determination}
\label{sec:lcrit_calc}

Our calculation of the critical neutrino luminosity follows \citet{pt12} with several small changes. We solve the steady-state Euler equations for density $\rho$, velocity $v$, and temperature $T$ between the neutrinosphere $\rnu$ and the shock $\rs$
\begin{eqnarray}
\frac{1}{\rho}\frac{\intd \rho}{\intd r} + \frac{1}{v}\frac{\intd v}{\intd r} + \frac{2}{r} &=& 0,\\
v\frac{\intd v}{\intd r} + \frac{1}{\rho}\frac{\intd P}{\intd r} &=& -\frac{GM}{r^2},\\
\frac{\intd\epsilon}{\intd r} - \frac{P}{\rho^2}\frac{\intd \rho}{\intd r} &=& \frac{\qdot}{v},
\end{eqnarray}
where $P$ is the gas pressure, $M$ is the mass within the neutrinosphere, $\epsilon$ is the internal specific energy of the gas, and $\qdot$ is the net heating rate. These equations couple
through the equation of state (EOS), $P(\rho, T, \ye)$ and $\epsilon(\rho, T, \ye)$, where $T$ is the
gas temperature, to the equation for electron fraction $\ye$. To take into account the self-coupling of the flow through the accretion luminosity \citep{pt12}, we solve for the radial dependence of $\lnue$ and $\lnuebar$
\begin{eqnarray}
\frac{\intd \lnue}{\intd r} &=& -4\pi r^2 \rho \qdot_{\nue},\\
\frac{\intd \lnuebar}{\intd r} &=& -4\pi r^2 \rho \qdot_{\nuebar},
\end{eqnarray}
where $\qdot_{\nue}$ and $\qdot_{\nuebar}$ are the net heating rates due to absorption and emission of electron neutrinos and antineutrinos, respectively, $\qdot = \qdot_{\nue}+\qdot_{\nuebar}$.

In our neutrino physics we use prescriptions of heating, cooling, opacity, and reaction rates for charged-current processes with neutrons and protons from \citet{scheck06}. We assume that the electron neutrino and antineutrino energies $\enue$ and $\enuebar$, which we define as rms energies of the Fermi-Dirac distribution, are independent of radius. We use the Helmholtz EOS \citep{timmes00}, which includes photons, an ideal gas of ions, and an electron-positron gas with an arbitrary degree of degeneracy. We assume that the PNS mass $M$ dominates over the mass between the neutrinosphere and the accretion shock. The Helmholtz EOS conveniently supplies smooth derivatives of thermodynamic variables, which aids convergence of the relaxation code.

In our time-independent calculation, the mass accretion rate $\mdot = -4\pi r^2\rho v$ is constant throughout the region, which is achieved by applying this constraint at the inner boundary. We also prescribe $\lnue(\rnu)$ and $\lnuebar(\rnu)$. At the outer boundary, we apply momentum and energy shock jump conditions assuming that the matter has velocity equal to $0.25$ of the free fall velocity \citep{yahil82,bethe90}, has negligible thermal pressure, and is composed of iron, $\ye(\rs) = 26/56$.

For the given values of $\lnue(\rnu)$ and $\lnuebar(\rnu)$, we find the radial profiles of $\rho$, $v$, $T$, $\ye$, $\lnue$, $\lnuebar$, and $\rs$, which is the eigenvalue of the problem. For each set of $M$, $\mdot$, $\rnu$, $\enue$, and $\enuebar$ we find the critical luminosity $\lcrit$ by increasing $\lnue$, while keeping the ratio $\lnue/\lnuebar$ fixed and using the previous solution as an initial guess for the next one. $\lcrit$ is determined to a specified accuracy using bisection. Additional details on the procedure and discussion of the results are given by \citet{pt12}.

\subsection{Evolution of boundary conditions}
\label{sec:gr1d}

\begin{figure}
\plotone{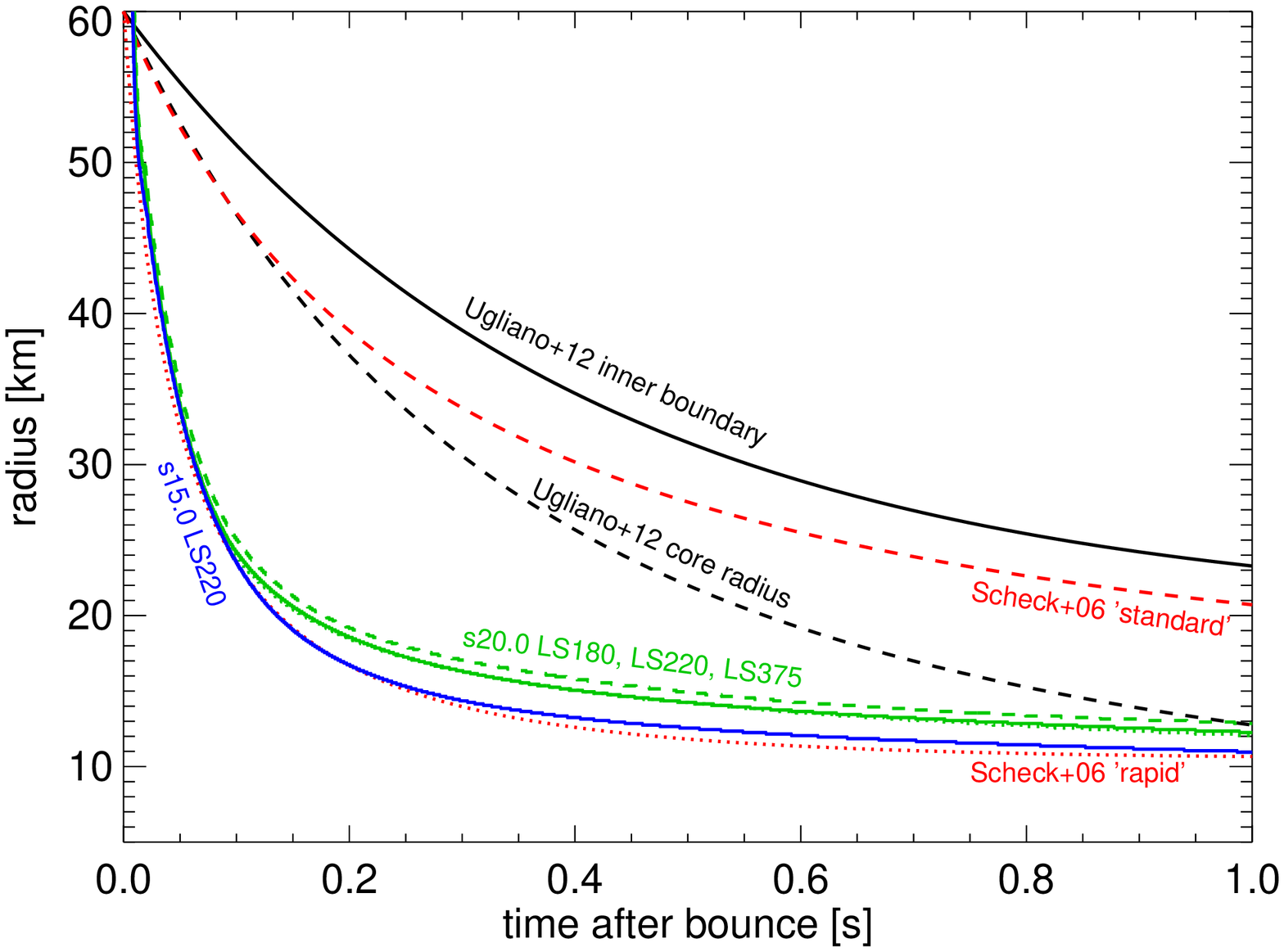}
\caption{Radial evolution of the $M=1.1\,\msun$ coordinate calculated with \gr\ for the s20.0WHW02 progenitor evolved with the three \citet{lattimer91} EOS (green lines) and the s15.0WHW02 progenitor evolved with LS220 EOS (blue line). The dotted and dashed red lines show the ``rapid'' and ``standard'' contraction, respectively, of the excised inner boundary at $1.1\,\msun$ mass coordinate of \citet{scheck06} assuming an initial radius of 60\,km. The inner boundary trajectory of \citet{ugliano12} is shown with solid black line along with the assumed evolution of the PNS core radius (dashed black line).}
\label{fig:m11}
\end{figure}

To calculate $\lcrit(t)$, we need the time evolution of the boundary conditions $M$, $\mdot$, $\rnu$, $\lnue(\rnu)$, $\lnuebar(\rnu)$, $\enue$, and $\enuebar$. This evolution encodes the structure of the progenitor either directly ($\mdot$ at the shock, and its time integrated value $M$), or through the physics of neutrino interactions with matter ($\rnu$, neutrino luminosities and energies). We calculate the time evolution of these quantities using the open-source spherically-symmetric code for stellar collapse \gr\footnote{\url{http://www.stellarcollapse.org}} \citep{oconnor10}. Features of \gr\ relevant to our study include robust calculation of many different progenitors with many EOSs, a neutrino-leakage/heating scheme for the postbounce epoch, and the ability to follow the postbounce evolution for many seconds or until a collapse to a black hole, which can be unambiguously identified in the code. \gr\ has been successfully used to study collapse to a black hole for both non-rotating \citep{oconnor11} and rotating \citep{dessart12} progenitors.

We perform the collapse calculations with most of the fiducial collapse settings in \gr. We do not include nucleon-nucleon bremsstrahlung, because we found that it results in the code crashing much earlier and has only small effect on the results. To prevent problems with the EOS, we limit the minimum temperature in a zone to 1\,MeV if there are significantly warmer zones above this zone. This acts as an artificial heating deep inside the proto-neutron star, but the total amount of artificial heat added this way should be very small. We also partially fixed the inconsistency in the neutrino pressure, as described in Section 2.2 of \citet{oconnor11}. We used the \citet{lattimer91} EOS with incompressibilities of 180, 220, and 375\,MeV (hereafter abbreviated as LS180, LS220, and LS375), and the \citet{shen11} EOS (hereafter abbreviated as HShen). For each progenitor, we ran the calculation for 6 seconds (core bounce typically occurs $0.3$\,s after the start of the calculations) or until black hole formation (determined by the lapse function in the center) or until the code crashes for numerical reasons (typically this happens after more than 3\,s of post-bounce evolution). Since \gr\ is a serial code, the typical runtime of a single progenitor in \gr\ of a couple of days allows us to run many different progenitors with different EOS on a modest cluster.

For each \gr\ run, we extract the post-bounce time evolution of $\rnue$, $\rnuebar$, and $\lnue$, $\lnuebar$, $\enue$, and $\enuebar$ at their corresponding neutrinospheres and at higher optical depths $\tau$. To suppress noise and oscillatory behavior in these quantities, we smooth them with a Gaussian with a width of $5$\,ms. Although the mass accretion rate and accreted mass can be quite well estimated analytically, we extract these quantities from \gr\ to maintain time synchronization with the neutrino luminosities and energies.

In some parametric studies, the inner regions of the simulation are excised to allow for longer timesteps. For example, \citet{scheck06} and \citet{ugliano12} remove the inner $1.1\,\msun$ of the PNS and instead prescribe the contraction of the inner boundary, where they assume hydrostatic equilibrium. In Figure~\ref{fig:m11}, we compare the radial evolution of the $M=1.1\,\msun$ coordinate for two progenitors in \gr, which fully includes the inner regions of the PNS, with the prescriptions used in \citet{scheck06} and \citet{ugliano12} assuming the same initial and final radii. We do not see much difference between the individual EOSs and only a small differences between the two progenitors. \gr\ produces very fast PNS contraction, similar to the ``rapid'' case of \citet{scheck06}, but very different from the ``standard'' case of the same authors, which was also used in \citet{ugliano12}. This rapid contraction has important consequences for the evolution of $\lcrit$ we describe in Section~\ref{sec:method}.

\gr\ uses a gray neutrino-leakage scheme after bounce, which is less sophisticated than energy-dependent radiation transport used in self-consistent studies. One unfavorable outcome of the neutrino treatment in \gr\ is that in some situations $\enue > \enuebar$ even though $\rnue > \rnuebar$. Since the neutrino heating is proportional to the square of the neutrino energy, this can have some effect on the calculations. As a result, for each progenitor we run two sets of steady-state models with one set using the neutrino energies from \gr\ and the other having the neutrino energies calculated from the luminosities and neutrinosphere radii assuming black-body emission. We discuss the implications in Section~\ref{sec:method}. It is also worth pointing out that the neutrino leakage scheme in \gr\ somewhat underestimates neutrinosphere radii compared to a two-moment scheme (Ott 2013, private communication). The associated changes in neutrino energies and luminosities on $\lcrit$ should have a systematic effect on all progenitors. Since we investigate many different parameterizations of the neutrino mechanism, such a systematic effect is in some sense taken into account and it does not likely introduce large differences between individual progenitors. However, systematic changes with different neutrino physics will be the subject of future work.

\begin{figure}
\plotone{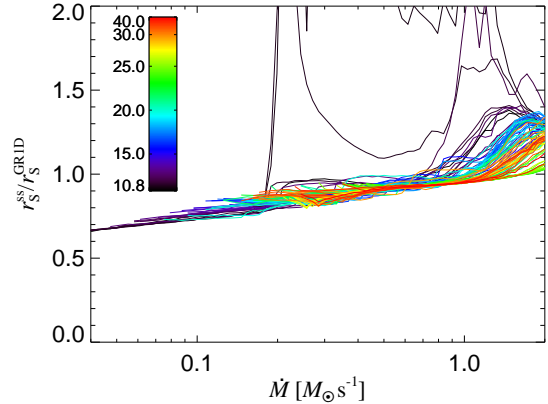}
\caption{Ratio of shock radii from our steady-state code, $\rs^{\rm ss}$, to the corresponding shock radii calculated with \gr\, $\rs^{\rm GR1D}$ as a function of $\mdot$. Each line corresponds to the evolution of a different progenitor from the sWHW02 progenitor series. Each color corresponds to a different initial progenitor mass, as given in the colorbar.}
\label{fig:rshock}
\end{figure}

Our steady-state code and \gr\ use different microphysics and treatment of neutrinos. A sensitive diagnostic of the differences between codes is the shock radius $\rs$ \citep{pt12}. In Figure~\ref{fig:rshock}, we show the ratio of shock radii calculated with our steady-state code and with \gr\ for the solar-metallicity progenitors of \citet{woosley02} (sWHW02) and the HShen EOS. The steady-state code systematically underestimates $\rs$ for small $\mdot$, likely due to differences in the physics and/or the treatment of gravity, but overall the absolute scale is well reproduced. The trend with $\mdot$ is implicitly taken into account by our exploration of many different parameterizations of the critical curve in Section~\ref{sec:correlations}. A part of the relatively small scatter around the trend is due to an overestimate of $\rs$ in the steady-state code when a sudden drop in density reaches the shock. At this  point, $\mdot$ drops while other parameters like PNS mass, radius, and luminosity remain essentially unchanged, which increases $\rs$ and lowers the critical luminosity. The dynamical terms in \gr\ damp the shock expansion to some extent and these effects are not captured by the steady-state code. Finally, for two progenitors (s11.2WHW02 and s11.4WHW02) the steady-state code grossly overestimates the shock radii. In these progenitors, the density drop associated with the edge of the iron core is accreted significantly earlier than in any other progenitor (about 50 to 60\,ms after bounce for s11.2WHW02 and s11.4WHW02 and more than 90\,ms for the remaining sWHW02 progenitors) and the assumption of steady-state is likely not valid for these progenitors. In the subsequent discussion, we assume that these two progenitors explode very soon after bounce. We note that the s11.2WHW02 model has been frequently used for core-collapse supernova simulations \citep[e.g.][]{buras06,burrows07b,muller12a,takiwaki12}, sometimes producing successful explosions, but the $\mdot$ evolution is in fact quite different from most of the other \citet{woosley02} progenitors.


\subsection{Progenitors}
\label{sec:progenitors}

\begin{deluxetable*}{cccccc}
\tablecolumns{6}
\tablewidth{0pt}
\tabletypesize{\scriptsize} 
\tablecaption{Progenitor models}
\tablehead{\colhead{Reference} & \colhead{Model series} & \colhead{Metallicity} & \colhead{Mass range $[M_\sun]$} & \colhead{Number of models} & \colhead{Equations of state}}
\startdata
\citet{woosley02} & zWHW02 & 0 & 11.0 -- 40 & 30 & HShen\\
\citet{woosley02} & uWHW02 & $10^{-4}Z_\sun$ & 11.0 -- 60 & 246 & HShen\\
\citet{woosley02} & sWHW02 & $Z_\sun$ & 10.8 -- 40 & 100 & LS180, LS220, LS375, HShen\\
\citet{limongi06} & LC06A  & $Z_\sun$ & 11 -- 120 & 15 & HShen \\
\citet{limongi06} & LC06B  & $Z_\sun$ & 40 -- 120 & 4 & HShen  
\enddata 
\label{tab:progenitors}
\end{deluxetable*}

\begin{figure*}
\plotone{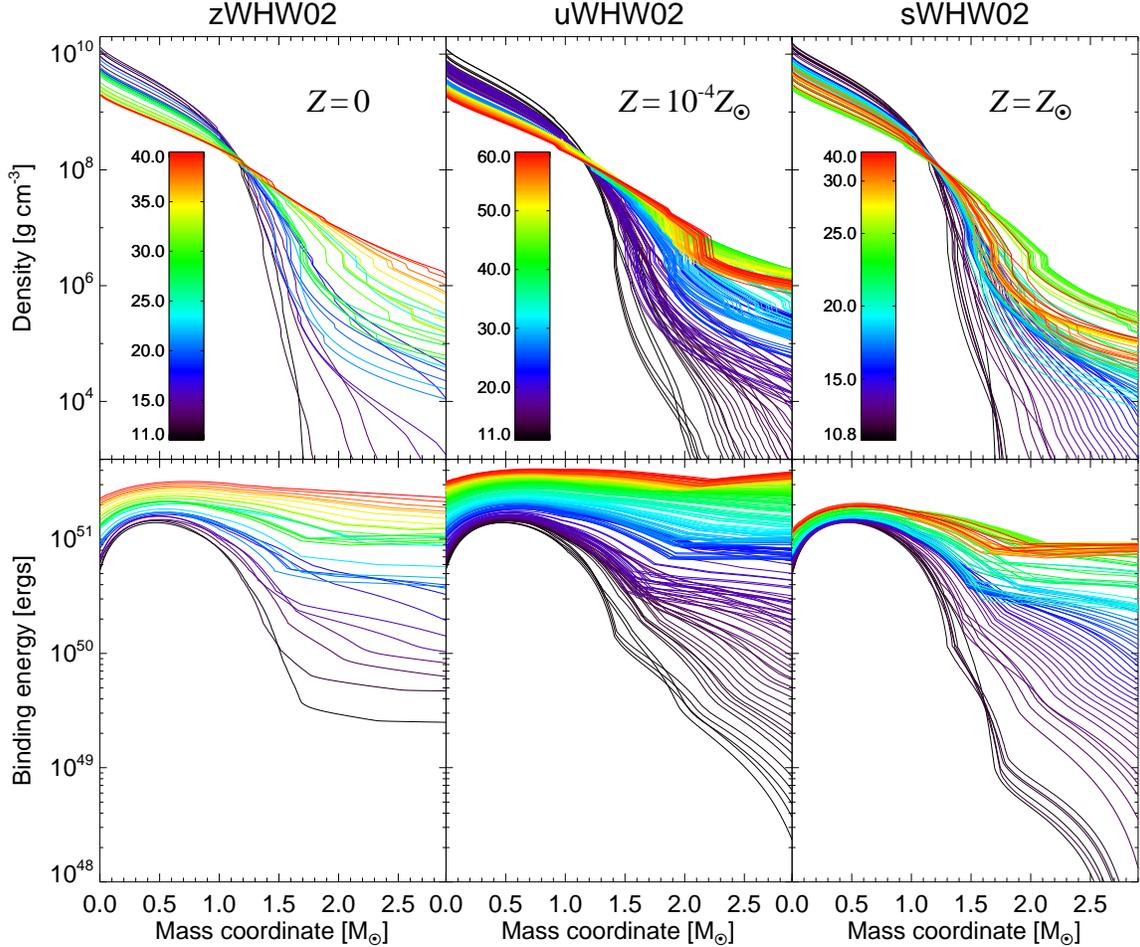}
\caption{Progenitor structure of the \citet{woosley02} pre-supernovae stellar models. The upper panels show the density profiles as a function of enclosed mass, while the lower panels display the absolute value of the binding energy above a particular mass coordinate. The binding energy is calculated by correcting the gravitational binding energy for the thermal energy of the progenitors. Each column displays progenitors with different metallicity (see Table~\ref{tab:progenitors}) and the color encodes the initial progenitor mass. For solar-metallicity progenitors, the color mapping is nonlinear, which reflects the structure of the progenitor grid.}
\label{fig:progenitors}
\end{figure*}

\begin{figure}
\plotone{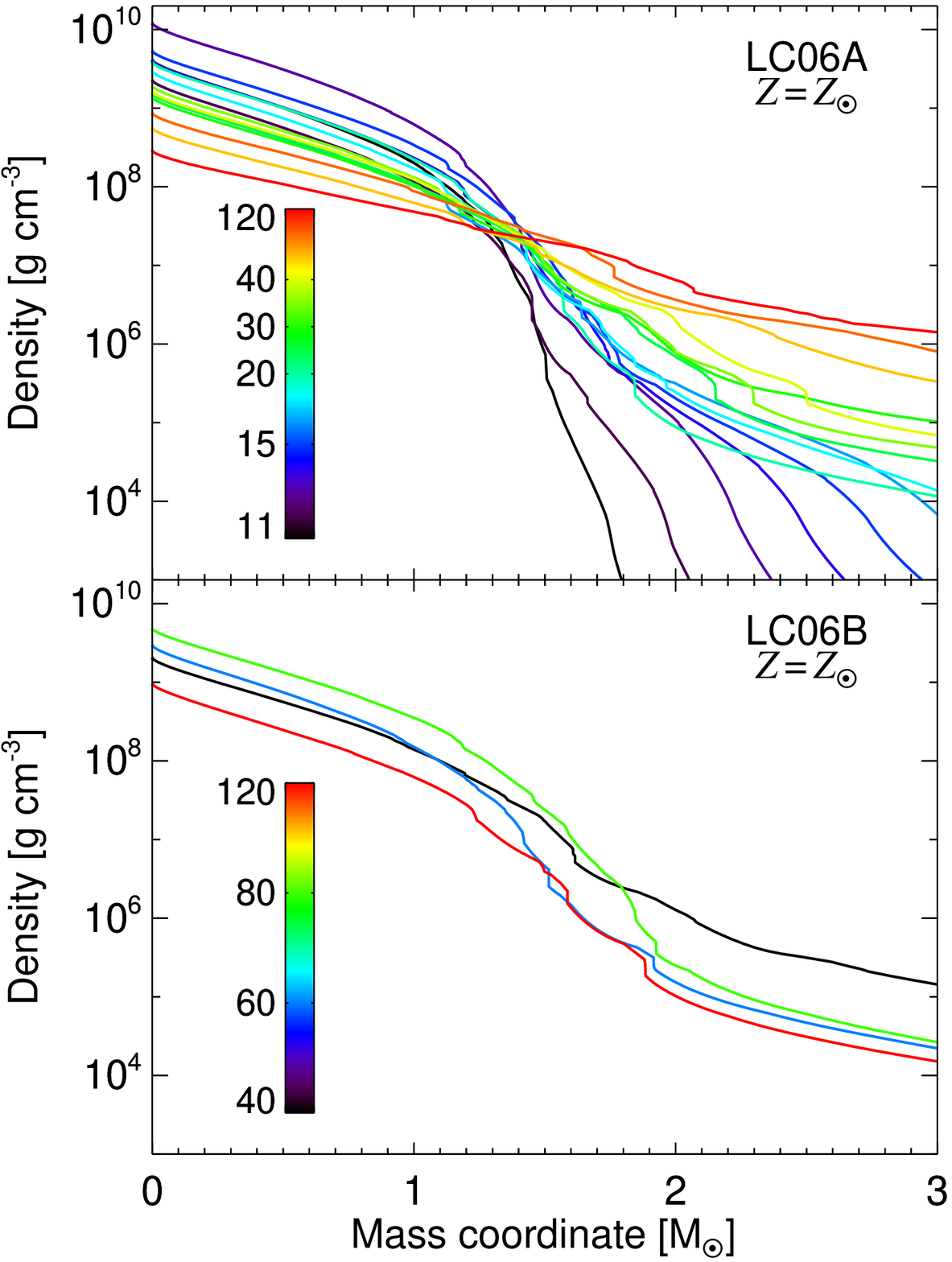}
\caption{Density profiles of the \citet{limongi06} solar-metallicity progenitors. The initial mass of the progenitors is encoded in the color of the lines.}
\label{fig:limongi}
\end{figure}

Our computational setup allows us to study a large number of progenitors with various EOSs. We employ the non-rotating progenitors of \citet{woosley02}\footnote{\url{http://2sn.org/stellarevolution/data.shtml}.}, which include progenitors with solar (sWHW02), $10^{-4}$ solar (uWHW02), and zero metallicity (zWHW02). We also use solar-metallicity \citet{limongi06}\footnote{\url{http://www.iasf-roma.inaf.it/orfeo/public\_html/sotto.html}.} progenitors (LC06A), which include a set of progenitors with the \citet{langer89} prescription for Wolf-Rayet winds (LC06B). Progenitor models and the EOSs used in our codes are summarized in Table~\ref{tab:progenitors}.

The pre-supernova progenitor structures of \citet{woosley02} can be readily collapsed in \gr, but \citet{limongi06} did not publish velocity fields for their pre-collapse progenitors. The assumption of zero initial velocity prevents core bounce in most \citet{limongi06} progenitors in \gr. To mitigate this problem, we calculated a generic velocity profile as a function of enclosed mass from the solar-metallicity \citet{woosley02} progenitors by scaling the velocity fields to the mass of the iron core and then taking the average. After applying this mean velocity field to the \citet{limongi06} progenitors, we observed core bounce in all progenitors except 120.0LC06B, which required additionally lowering $\ye$ by $0.02$ in the iron core.

In Figures~\ref{fig:progenitors} and \ref{fig:limongi}, we show the structure of the density and binding energy of the progenitor models in the range of mass coordinate relevant for supernova explosions. The presupernova density profiles become more shallow with increasing initial mass in the zero metallicity and to some extent also in $10^{-4}$ solar metallicity progenitors. For solar metallicity stars with initial mass $\mathscr{M} \gtrsim 25\,\msun$, stellar winds lower the mass of the star already on the main sequence, which results in steeper density profiles than in stars with the same initial mass but lower metallicity.

This trend is reflected also in the progenitor binding energy above a particular mass coordinate, which is shown in the lower panels of Figure~\ref{fig:progenitors}\footnote{The thermal energy structure is not available online for the \citet{limongi06} progenitors, which does not allow a straightforward calculation of the binding energy.}. The binding energy includes contributions from the gravitational potential energy and the thermal energy of the progenitor. The binding energies above the mass coordinates corresponding to the expected neutron star masses (between about $1.3$ and $2.0\,\msun$) imply the minimum amount of energy necessarily supplied by the explosion mechanism to have a successful explosion producing a neutron star as a remnant. The zero and $10^{-4}$ solar metallicity progenitors generally have higher binding energies than solar-metallicity ones. Low-metallicity progenitors with $\mathscr{M} \gtrsim 25\,\msun$ will likely have explosion energies greater than $10^{51}$\,ergs, if successful explosion results, because the binding energy does not significantly drop with increasing mass coordinate and forcing an explosion with asymptotic energy lower than the binding energy would likely result in significant fallback and a massive remnant black hole \citep[e.g.][]{zhang08}. We evaluate the black hole formation as a function of metallicity in Section~\ref{sec:metallicity}. For low-mass progenitors, the binding energy is negligible if a sufficient fraction of the star is allowed to accrete and hence the supernova explosion mechanism can, in principle, produce successful explosions with energies of the order of $10^{50}$\,ergs or even less.

\section{A method for systematically exploring observable predictions of the neutrino mechanism}
\label{sec:method}

Supernovae do not generally explode in 1D simulations except for progenitors with a tenuous envelope on top of an O-Ne-Mg core. However, 1D simulations are computationally cheap so that they can be used to study the parameter space of the transition from accretion to explosion. The necessary artificial explosions can be achieved in a variety of ways. For example, \citet{ugliano12} parameterize the contraction of the PNS core (and hence neutrino luminosities) as we show in Figure~\ref{fig:m11}. The parameter values required to match the explosion energy of SN1987A in their simulations mean that the PNS core contracts below the inner boundary of the simulations almost immediately, as we illustrate in Figure~\ref{fig:m11}.

We take a different approach, which allows us to explore consequences of many different parameterizations of the neutrino mechanism. As described in Section~\ref{sec:setup}, we calculate the time evolution of the critical neutrino luminosity $\lcrit$. We study the part of the parameter space where the steady-state approximation used to calculate $\lcrit$ is approximately valid, which we estimate to be $\mdot < 2\,\msun\,$s$^{-1}$. 

\begin{figure}
\plotone{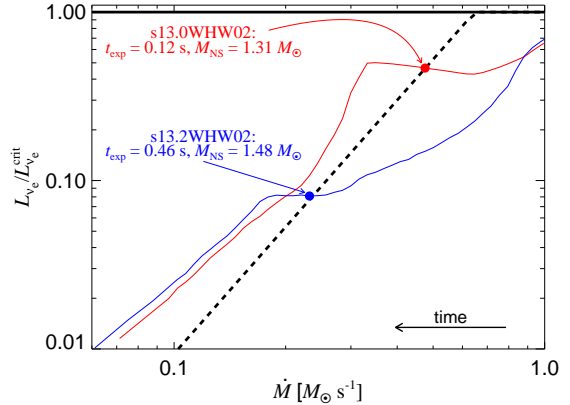}
\caption{Ratio of the electron neutrino luminosity at the neutrinosphere to the critical value, $\lnue/\lcrit$, as a function of the mass accretion rate  $\mdot$ through the shock for the s13.0WHW02 (red) and s13.2WHW02 (blue) models. The time evolution proceeds from high to low $\mdot$, as indicated by the arrow. We assume that explosion occurs when  $\lnue/\lcrit$ gets above an artificial explosion threshold (thick black dashed line), which is a function of $\mdot$ only with $(p,q) = (0.3, 2.5)$ in Equation~(\ref{eq:artificial}). This threshold is lower than $\lnue/\lcrit = 1$ (thick solid line). The structure of different progenitors determines the moment of the explosion and associated parameters such as the remnant mass.}
\label{fig:principle} 
\end{figure}

In Figure~\ref{fig:principle}, we illustrate our logic for determining which stars explode and what are the explosion properties. We show the ratio of the actual luminosity of the core to the critical value, $\lnue/\lcrit$, as a function of the mass accretion rate through the shock for two progenitors with similar initial mass. The behavior of other progenitors is qualitatively similar. We see that the ratio $\lnue/\lcrit$ decreases with decreasing $\mdot$ and remains below unity for all considered $\mdot$. This is consistent with 1D supernova simulations not exploding except for the lowest-mass progenitors with a tenuous envelope on top of a dense core \citep{kitaura06,burrows07a}. 

Since the supernova explosion mechanism is unknown and the importance of many pieces of physics are not properly constrained, we assume that an explosion occurs when $\lnue/\lcrit$ gets above a particular value, which is lower than unity in the 1D scale of our model. We also assume that the explosion threshold is a function of only $\mdot$. There are several reasons for this. First, $\mdot$ is a monotonic function of time and reflects the instantaneous structure of the progenitor layers coming through the shock. Second, the physical effects that are believed to play an important role in the supernova explosion mechanism depend on $\mdot$. In particular, convection below the shock is driven by neutrino heating from hotter layers at smaller radii and a significant fraction of the neutrino emission comes from the accretion luminosity \citep[e.g.][]{pt12}, which is directly proportional to $\mdot$. Similarly, the SASI is an acoustic-advective cycle \citep[e.g.][]{foglizzo02,foglizzo06,foglizzo07} and as such depends on the rate with which matter is advected. Third, the critical curves $\lcrit(\mdot)$ evaluated in 1D, 2D, and 3D calculations are nearly parallel to each other. This means that the multi-dimensional effects can be approximated by changing the normalization of $\lcrit(\mdot)$. Finally, the scatter about the systematic trend between the steady-state code and \gr\ in Figure~\ref{fig:rshock} is smaller when plotted as a function of $\mdot$ instead of post-bounce time, which suggests that some of the systematic effects are better taken into account by this choice. Thus, we assume that the explosion threshold is a power law in $\mdot$ so that explosions occur when
\beq
\frac{\lnue}{\lcrit} > p\left(\frac{\mdot}{0.4\ \msun\,{\rm s}^{-1}}\right)^q,
\label{eq:artificial}
\eeq
where $p>0$ and $q>0$ are free parameters, and $\mdot$ is normalized to a typical value of $0.4\,\msun$\,s$^{-1}$. We set the maximum value to be $\lnue/\lcrit = 1$.  Equation~(\ref{eq:artificial}) implies that we are modifying $\lcrit$ as a function of $\mdot$ only, $\lcrit \rightarrow \lcrit \times p (\mdot/0.4\,\msun\,{\rm s}^{-1})^q$. By changing the values of $p$ and $q$, we can observe how the properties of artificial explosions change and identify common trends of the neutrino mechanism.

This procedure is illustrated in Figure~\ref{fig:principle} for progenitors s13.0WHW02 and s13.2WHW02 for one possible parameterization of the explosion mechanism with $(p,q) = (0.3,2.5)$. Despite their very similar initial mass, the pre-supernova structure of these two stars is very different: unlike s13.2WHW02, s13.0WHW02 exhibits a strong drop in $\mdot$, but the neutrinosphere radii, neutrino energies and luminosities (if measured at the neutrinosphere or deeper) are affected much less, which lowers $\lcrit$. As a result, $\lnue/\lcrit$ increases, which is visible as a bump in the red line in Figure~\ref{fig:principle}, and s13.0WHW02 crosses the explosion threshold of Equation~(\ref{eq:artificial}). For s13.2WHW02, the threshold is crossed at much lower $\mdot$ at later times. This will have implications for the observational properties of the resulting supernovae. The mass of the PNS at the moment of explosion is about $0.17\,\msun$ higher in s13.2WHW02 due to the longer accretion phase. Differences can also be expected in explosion energy, or nucleosynthetic production. Many other parameterization with different $p$ and $q$ can be drawn in Figure~\ref{fig:principle} resulting in explosions at different moments, or one or both progenitors not exploding. We investigate the implications of varying these parameterizations in Section~\ref{sec:correlations}.

\begin{figure}
\plotone{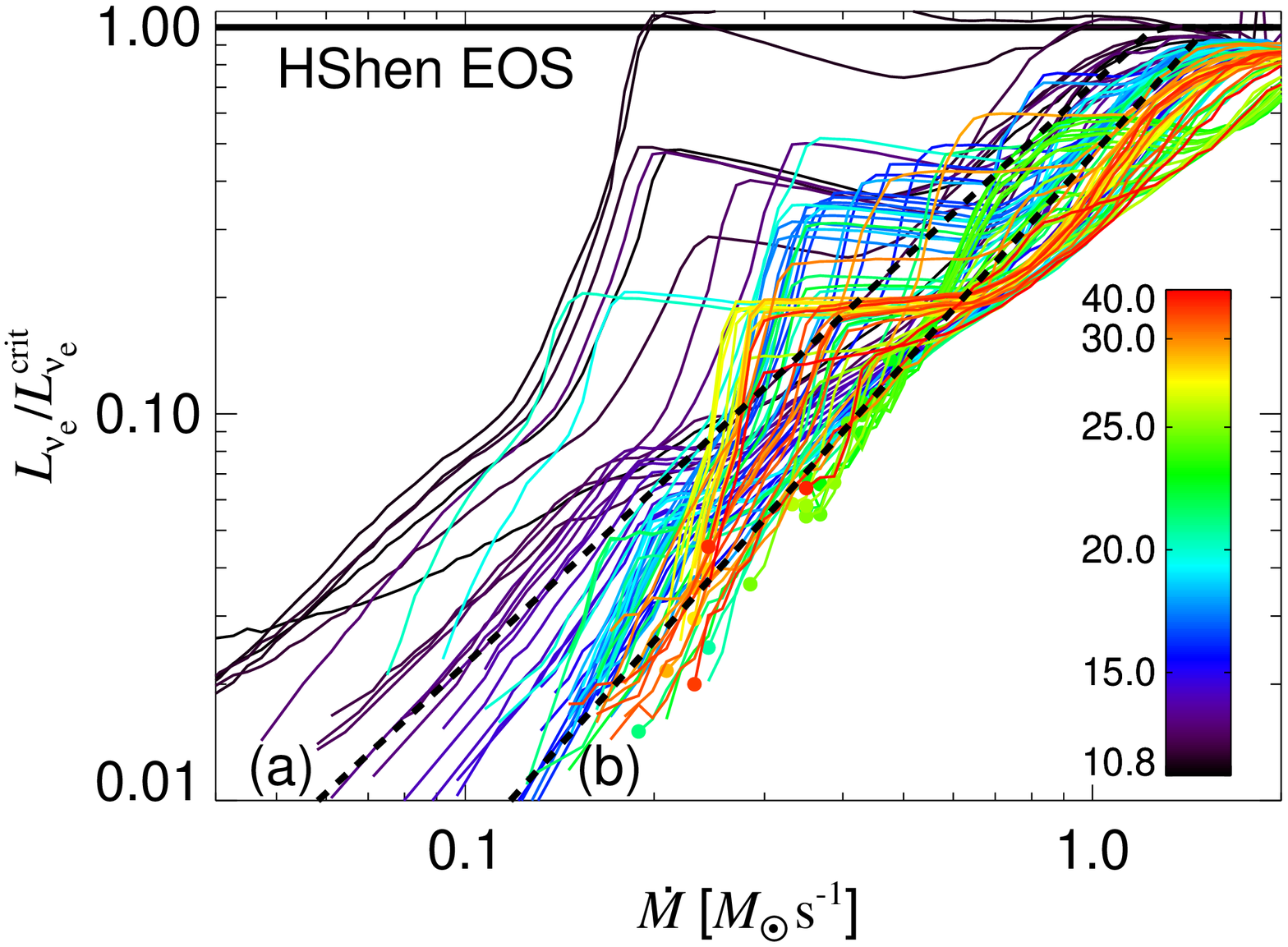}
\plotone{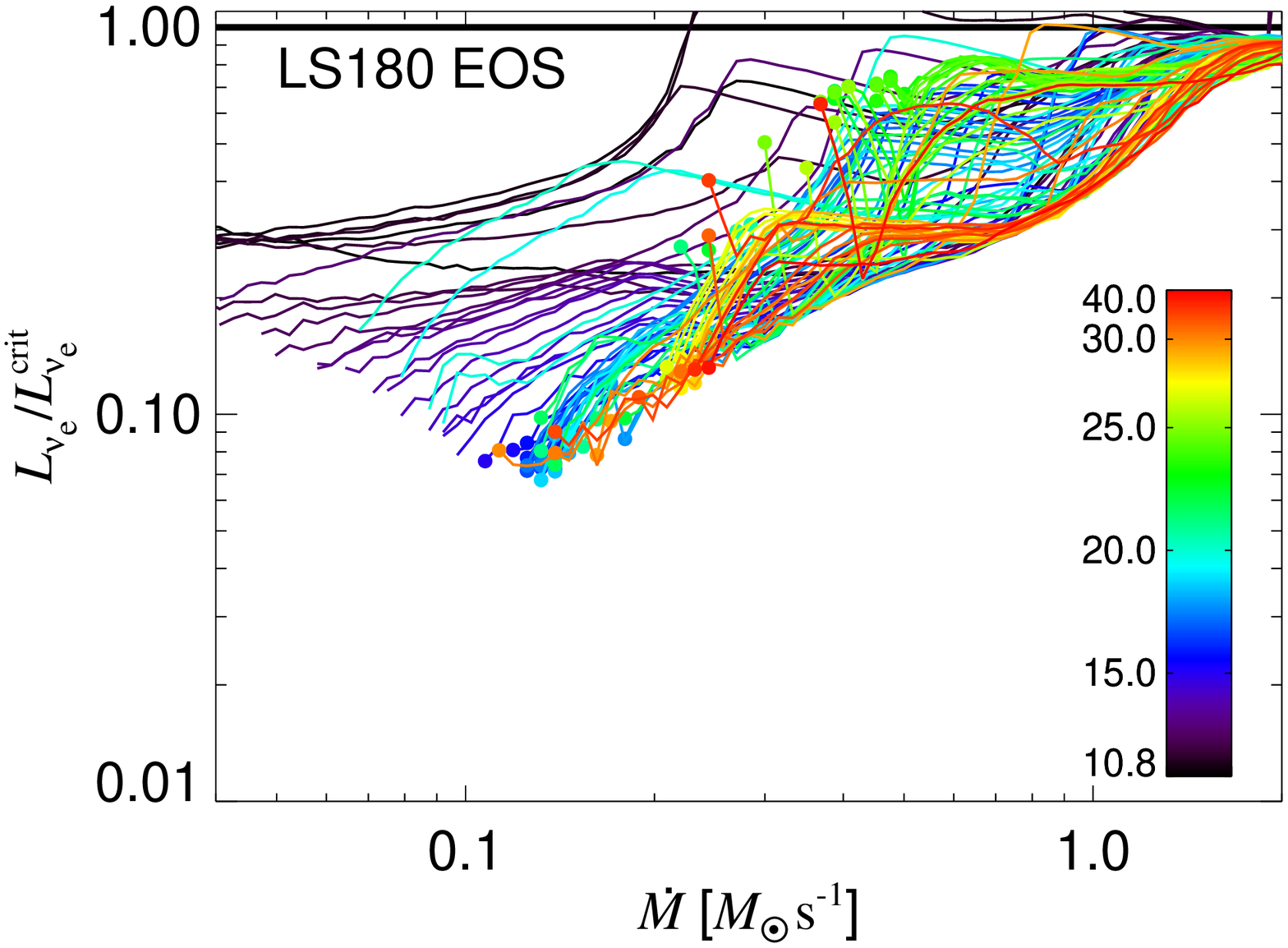}
\caption{The ratios of the actual neutrino luminosity to the critical values, $\lnue/\lcrit$, as a function of $\mdot$ for the ensemble of solar-metallicity progenitors of \citet{woosley02} and HShen (top) and LS180 (bottom) EOS are shown with thin colored lines, where the color denotes the initial progenitor mass. Lines ending with a filled circle correspond to progenitors that formed a black hole in \gr, which is determined as the moment when the lapse function drops below 0.4. The two lines with $\lnue/\lcrit >1$ are the s11.2WHW02 and s11.4WHW02 progenitors where steady-state is likely not achieved, as discussed in Section~\ref{sec:gr1d} and Figure~\ref{fig:rshock}. The two dashed black lines labeled with (a) and (b) denote two possible parameterizations of the neutrino mechanism (Eq.~[\ref{eq:artificial}]) with $(p,q) = (0.18,1.5)$ and $(0.09,1.8)$, respectively, which result in different remnant mass distribution and a different fraction of successful explosions (Fig.~\ref{fig:ns_mass}).}
\label{fig:lratio}
\end{figure}

In the top panel of Figure~\ref{fig:lratio}, we plot $\lnue/\lcrit$ for the full ensemble of solar-metallicity progenitors of \citet{woosley02} evolved with HShen EOS\footnote{In Section~\ref{sec:gr1d}, we discussed a potential problem with the neutrino treatment in \gr, which results in $\enue > \enuebar$ even though $\rnue > \rnuebar$, which we fixed by calculating the neutrino energies from luminosities and neutrinosphere radii assuming black-body emission and Fermi-Dirac spectrum. With neutrino energies instead directly from \gr, we find that the behavior of $\lnue/\lcrit$ changes negligibly and that the remnant mass distribution remains unchanged except for a different outcome for several progenitors. The overall agreement with \citet{ugliano12} is slightly worse, so we continue to use neutrino energies calculated assuming black-body emission.}. All the progenitors except the two low-mass cases discussed in Section~\ref{sec:gr1d} and Figure~\ref{fig:rshock} follow the general trend that $\lnue/\lcrit$ decreases with decreasing $\mdot$. The relatively steep decrease is caused by the fast contraction of the PNS in \gr, which leads to high $\lnue$ at early times and low $\lnue$ later. It is reassuring to see that differences in progenitor structure, which can amount to several order of magnitude range of density profiles (Fig.~\ref{fig:progenitors}), reduce to a much narrower range when the progenitor evolution is normalized by the critical neutrino luminosity. The top panel of Figure~\ref{fig:lratio} also shows with thick dashed lines two possible parameterizations of the neutrino mechanism according to Equation~(\ref{eq:artificial}). We choose parameterization (a) ($p = 0.18$, $q=1.5$) to produce non-negligible fraction of failed explosions with black holes as an end product. This happens either because the progenitor does not reach the explosion threshold or we are not able to evolve the progenitor to lower $\mdot$ with \gr. Since our typical simulation times in \gr\ are longer than 4\,s, we assume that such progenitors form black holes anyway. We choose parameterization (b) ($p=0.09$, $q=1.8$) to make all progenitors explode relatively early. 

The slope and normalization of the progenitor trajectories in Figure~\ref{fig:lratio} changes with assumptions on the physics. For example, using a softer EOS such as LS180, which is shown in the bottom panel of Figure~\ref{fig:lratio}, produces trajectories with shallower slopes. Naturally, this affects the values of $(p,q)$ necessary to achieve approximately similar outcomes for different choices of EOS and other physics. For the subsequent discussion we focus exclusively on the results using the HShen EOS. However, when we marginalize over all possible parameterizations, we find that the pattern for a progenitor population is very similar irrespective of our choice of EOS as we discuss in Section~\ref{sec:eos}.

Similarly, additional parameters could be introduced in Equation~(\ref{eq:artificial}) to achieve more freedom in the range of outcomes. We experimented with adding a constant to Equation~(\ref{eq:artificial}). Looking at Figure~\ref{fig:lratio}, we see that a constant floor to $\lnue/\lcrit$ will not change our answers significantly provided that it is low enough so that the power-law slope at high $\mdot$ remains unaffected. In other words, in most cases the fate of a given progenitor is decided at $\mdot \gtrsim 0.3\,\msun$\,s$^{-1}$ and changes to $\lnue/\lcrit$ at lower $\mdot$ are unimportant. Furthermore, we do not see the gain in adding more phenomenological parameters to Equation~(\ref{eq:artificial}) without a clear physical motivation; the current two-parameter version provides a sufficiently rich ground for investigation, as we show below.

Our eventual goal would be to directly match a physical mechanism important for explosion with specific $(p,q)$. There are several obstacles that generically prevent us from doing so. The values of $(p,q)$ are tied to a specific baseline model, which in our case is \gr\ combined with our steady-state code. Different baseline models would necessarily require different $(p,q)$ to produce the same outcome. Abandoning an effort to make an absolute statement, we still might be able to make a relative statement on how $(p,q)$ change in the presence of a physical effect. We considered the numerical values of $\lcrit$ calculated in 1D, 2D, and 3D by \citet{hanke12}, \citet{couch13b}, and \citet{dolence13} to find that in 2D/3D $p$ decreases to $\sim 75\%$ of its 1D value. Constraints on changes of $q$ are much more uncertain due to differences between individual groups and small number of datapoints, but the numerical results appear consistent with no change in $q$ as a function of dimension. This illustrates that the quantitative importance of various physical effects on $\lcrit$ is not properly characterized even for the well-investigated multi-dimensional effects. The difference in $p$ between our parameterizations (a) and (b) is about $50\%$, which is larger than what would be implied by the multi-dimensional effects. However, the point of the two parameterizations is to illustrate a range of outcomes.

\section{Results}
\label{sec:results}

As a consequence of the different explosion times and progenitor structures, the resulting supernovae will have different properties. The later the explosion, the longer the PNS accretes and its mass increases. The neutrino luminosity and mass of the photo-dissociated region between the PNS and the shock at the moment of explosion will determine the explosion energy. Furthermore, for some parameterizations, there can be progenitors, which do not cross the explosion threshold before a black hole forms. For such cases, a classical supernova explosion would not occur and the possible observable transient would be much weaker and fainter \citep{nadezhin80,kochanek08,lovegrove13,piro13}. All of these issues can be addressed observationally, and we explore the predictions of the neutrino mechanism for these aspects in the following Sections. 

Here, we focus on the two parameterizations (a) and (b) presented in Section~\ref{sec:method} to illustrate how our method works and the range of outcomes. In Section~\ref{sec:rem_mass}, we discuss the mass function of the compact objects remaining after the explosion. In Section~\ref{sec:energy}, we estimate the supernova explosion energies, and in Section~\ref{sec:nickel} we provide the corresponding nickel yields. Finally in Section~\ref{sec:summary}, we summarize our findings as a function of initial mass and metallicity. In Section~\ref{sec:robust}, we the investigate the robustness of our findings for the full range of parameterizations.

\subsection{Remnant mass distribution}
\label{sec:rem_mass}

\begin{figure*}
\centering
\includegraphics[width=0.7\textwidth]{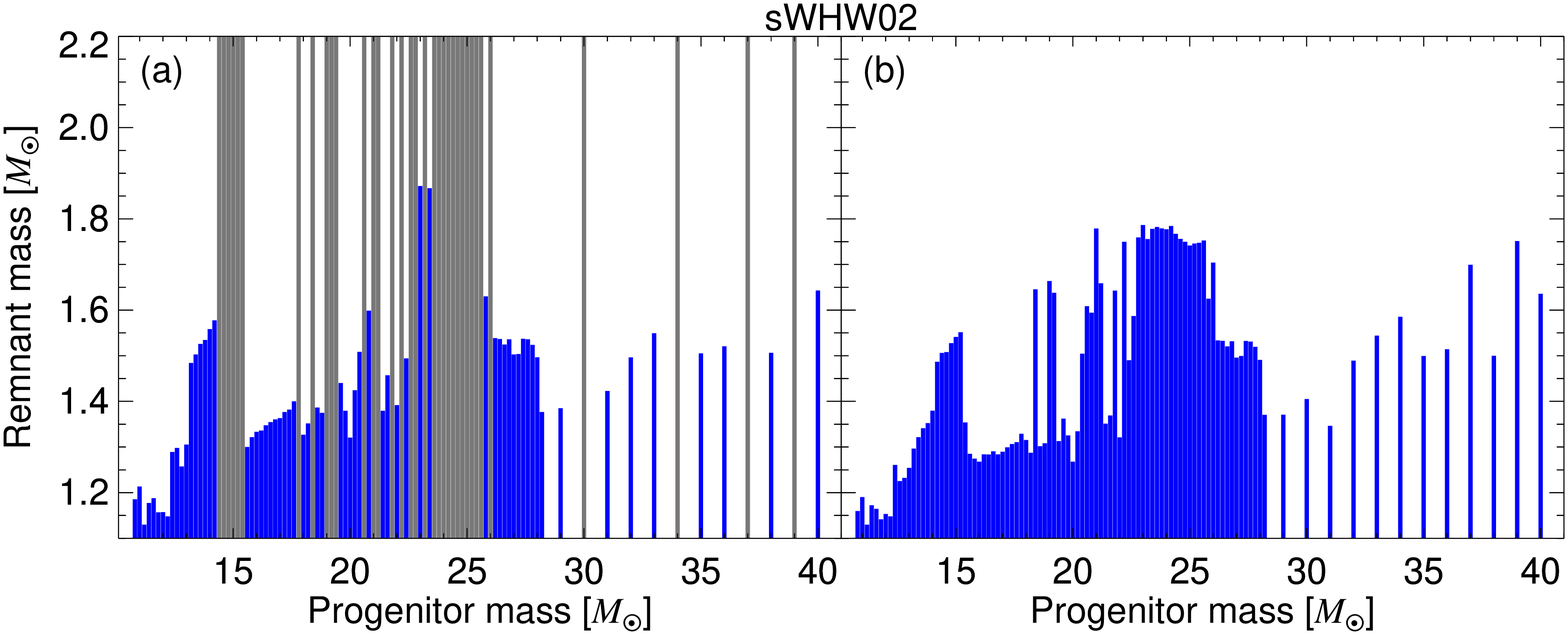}
\includegraphics[width=0.7\textwidth]{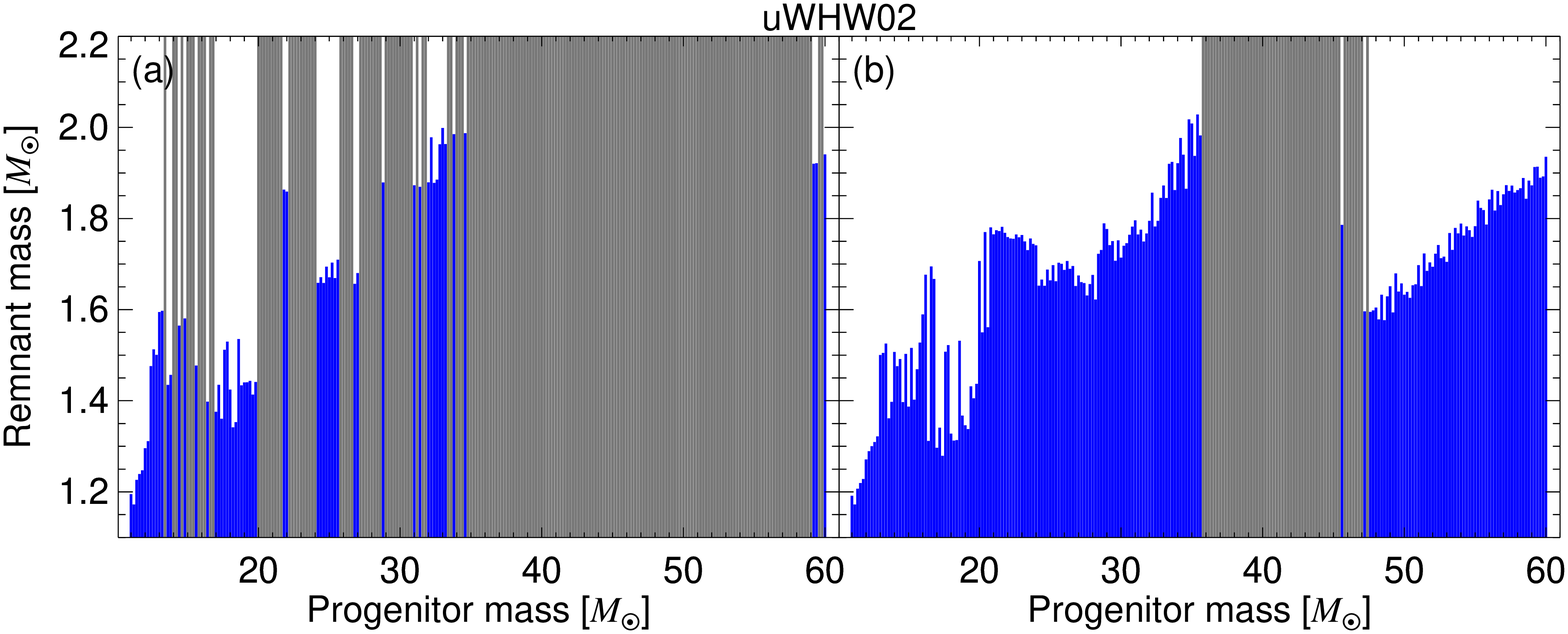}
\includegraphics[width=0.7\textwidth]{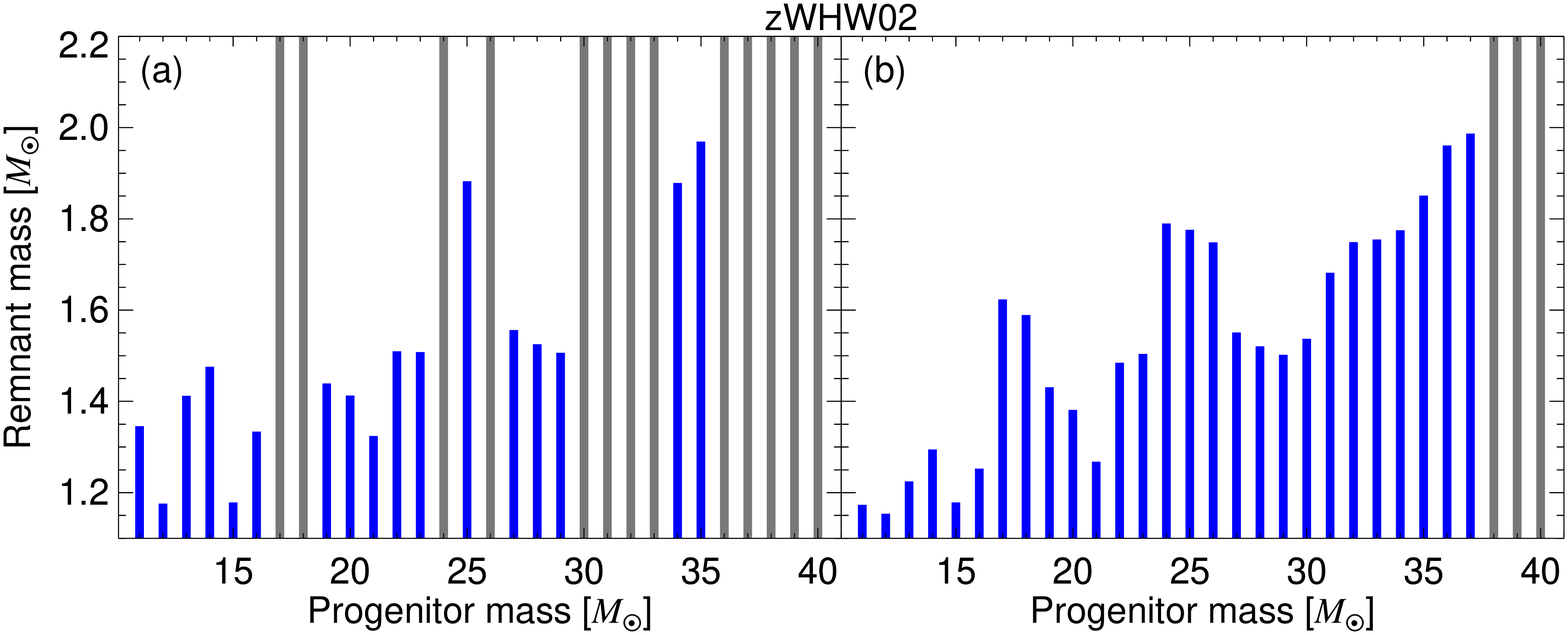}
\includegraphics[width=0.7\textwidth]{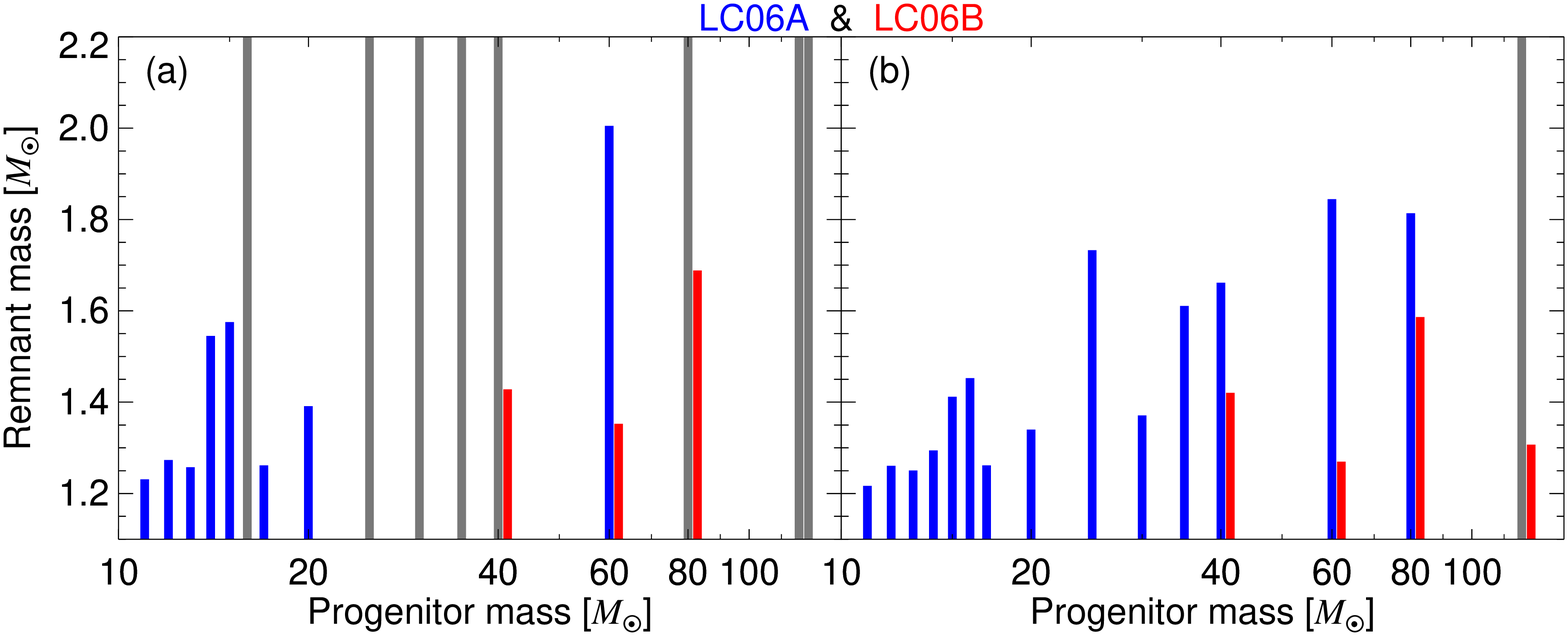}
\caption{Remnant mass functions for \citet{woosley02} progenitors with $Z_\sun$ (top row), $10^{-4}\,Z_\sun$ (second row), and $0$ (third row), and for \citet{limongi06} progenitors (bottom) all using the HShen EOS. The left and right columns are for two different parameterizations of the neutrino mechanism, as shown in Figure~\ref{fig:lratio}. Each panel shows successful explosions by a blue bar with the gravitational mass of the resulting NS and failed explosions assumed to produce black holes by gray bars spanning the vertical range of the plots. In the bottom panel, successful explosions are denoted by blue bars for LC06A and red bars for LC06B. Failed explosions are marked with gray bars. The results for the LC06B models are slightly offset in the progenitor mass for better visibility.}
\label{fig:ns_mass}
\end{figure*}

The top row of Figure~\ref{fig:ns_mass} shows the remnant mass function produced by the two parameterizations of the neutrino mechanism acting on the sWHW02 progenitors (Fig.~\ref{fig:lratio}). For successful explosions (blue bars), the gravitational mass of the remaining neutron stars $M_{\rm grav}$ is calculated from the baryonic mass $M_{\rm bary}$ inside of $\rnue$ at the moment of explosion by solving \citep{timmes96}
\beq
M_{\rm grav} = M_{\rm bary} - 0.075\,\msun \left(\frac{M_{\rm grav}}{\msun}\right)^2.
\label{eq:bary2grav}
\eeq
The commonly assumed value of $0.084\,\msun$ for the numerical factor from \citet{lattimer89} differs slightly from Equation~(\ref{eq:bary2grav}). Although \gr\ consistently calculates gravitational masses, the simulation is not followed until the PNS completely cools off by neutrino emission and we thus prefer to use Equation~(\ref{eq:bary2grav}). We assume that only a negligible amount of baryonic mass is lost from the PNS during the neutrino-driven wind after a successful explosion ensues. We do not take into account the possibility of simultaneous accretion and explosion and we do not include fallback. However, in Section~\ref{sec:energy}, we estimate the explosion energy and we constrain which progenitors likely experience significant fallback during explosions.

The top left panel of Figure~\ref{fig:ns_mass} shows the remnant mass function for parameterization (a), which does not lead to explosion in some of the stars. The failed explosions are not confined to a well-defined range of initial progenitor masses as is usually assumed \citep[e.g.][]{fryer01,woosley02,heger03}, but are instead interwoven between the successful ones, as was found also by \citet{oconnor11} and \citet{ugliano12}. In particular, we see a cluster of failed explosions between about $14$ and $16\,\msun$, fine structure of interwoven explosions and failures between $17$ and $22\,\msun$, a cluster of failed explosions between $23$ and $26\,\msun$, and mostly explosions above about $26\,\msun$.

To illustrate the general pattern of remnant masses, we show in the right panel of Figure~\ref{fig:ns_mass} the results for parameterization (b), which produces explosions for all sWHW02 stars. We see that the general pattern of remnant masses is preserved for different parameterizations and that remnant masses for parameterization (b) are lower than for parameterization (a). We explore remnant mass distributions for many different parameterizations in Section~\ref{sec:correlations} and we show that our results agree with \citet{ugliano12} in Section~\ref{sec:ugliano}.

\subsection{Explosion energy}
\label{sec:energy}

There are a few positive and negative contributions to the supernova explosion energy, which are summarized in \citet{scheck06} and \citet{ugliano12}. Here, we consider three components that are relatively easy to calculate within our model: recombination, the neutrino-driven wind, and the progenitor binding energy. We address these individual contributions in the following Sections, paying special attention to the neutrino-driven wind.

\subsubsection{Recombination energy}

\begin{figure*}
\plotone{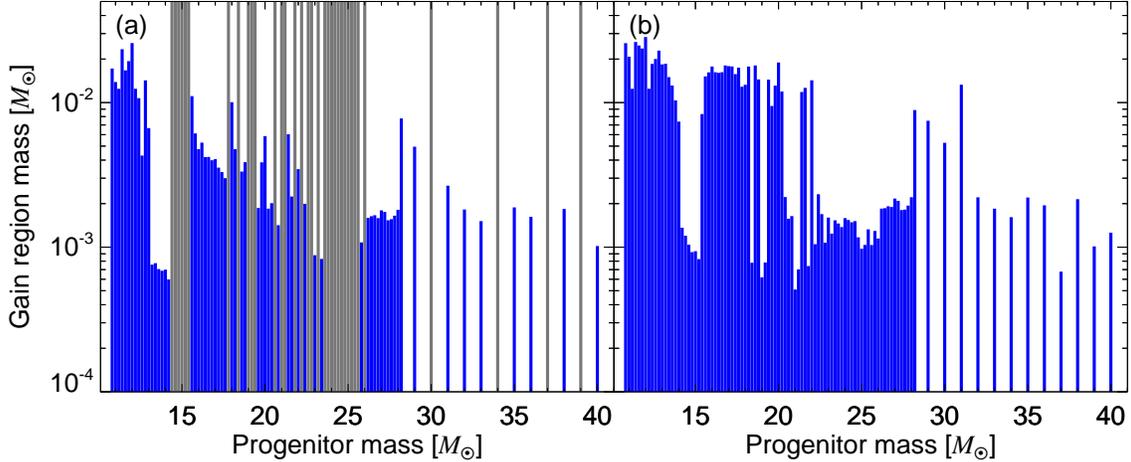}
\caption{Mass in the gain region at the moment of explosion as a function of progenitor mass for sWHW02 progenitors. Parameterizations (a) and (b) from Figure~\ref{fig:lratio} are shown in the left and right panels, respectively.} 
\label{fig:gain_mass}
\end{figure*}

As the supernova ejects material dissociated by the shock, recombination of free nucleons to nuclei gives $\sim 5$\,MeV per baryon or $10^{50}$\,ergs for each $0.01\,\msun$ recombined. This takes into account that some material below the shock is already in the form of $\alpha$ particles \citep{scheck06}. The amount of mass that recombines is not entirely certain, because we are not simulating self-consistent explosions. Usually, it is assumed that material in the gain region with net neutrino heating with mass $M_{\rm gain}$ is ejected \citep{scheck06}. In Figure~\ref{fig:gain_mass}, we plot the gain mass for sWHW02 progenitors and HShen EOS calculated using our steady-state code\footnote{For the same $\lnue$, the gain region mass from \gr\ and the steady-state code agree very well. We choose to use the gain region mass evaluated at $\lcrit$, which is higher than when evaluated at $\lnue$. However, the shape of the distribution is the same except for a multiplicative constant.}. The gain region masses are generally too small to be dynamically important: even if they are higher in 2D and 3D simulations than in 1D by a factor of few, the energy gain associated with recombination would be at best $\sim 10^{50}$\,ergs. The result would not be significantly different if we considered mass between the neutrinosphere and the shock. Hereafter, we thus ignore this contribution to the explosion energy from recombination.

\subsubsection{Neutrino-driven wind}
\label{sec:nu_wind}

The mechanical energy of the neutrino-driven wind $\ewind$ is in many cases the most important contribution to the explosion energy \citep{scheck06,ugliano12}. We assume that the mechanical energy can be obtained by performing a time integral of the wind mechanical power $\pwind$, $\ewind = \int \pwind \intd t$. Following \citet{burrows93}, we assume that $\pwind$ depends on the instantaneous PNS mass and radius, and the neutrino luminosity
\beq
\pwind = A \left(\frac{\lnue}{5 \times 10^{52}\,{\rm ergs\, s}^{-1}} \right)^\kappa \left(\frac{M}{1.3\,\msun} \right)^{-\lambda} \left(\frac{\rnue}{30\,{\rm km}} \right)^{-\mu},
\label{eq:pwind}
\eeq
where $A$, $\kappa$, $\lambda$, and $\mu$ are coefficients characterizing the physics of the neutrino-driven wind. $\ewind$ can be in principle obtained by integrating Equation~(\ref{eq:pwind}) over the wind duration with appropriately changing values of $\lnue$, $\rnue$, and $M$. There are several practical problems when implementing this scheme. For the sake of brevity, we describe these issues and our approach in Appendix~\ref{app:nu_wind} and here we discuss only the results.

\begin{figure*}
\plotone{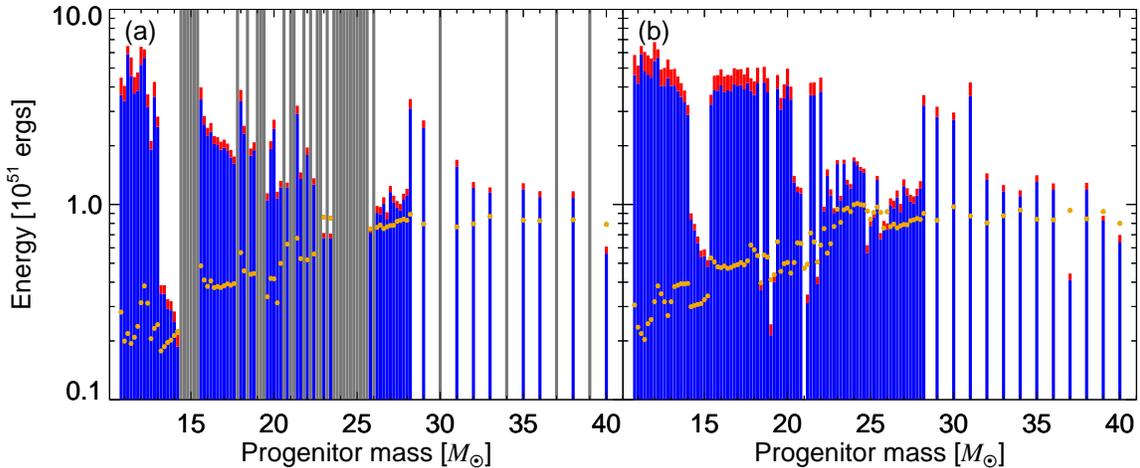}
\caption{Contributions to the explosion energy: time-integrated power of the neutrino-driven wind (blue bars) with recombination energy (red bars on top of the blue bars) for parameterizations (a) (left) and (b) (right). The filled orange circles show the absolute value of the progenitor binding energy above mass coordinate corresponding to the explosion. Explosions with energies less than the binding energy must experience significant fallback. The recombination energy is multiplied by a factor of $5$ for better visibility.}
\label{fig:e_exp}
\end{figure*}

In Figure~\ref{fig:e_exp}, we show $\ewind$ for the two parameterizations of the neutrino mechanism. We see that $\ewind$ spans more than an order of magnitude and that there is again ``fine structure'' in the sense that progenitors close in initial mass can have vastly different explosion energies. When comparing parameterizations (a) and (b) it is interesting to note that failed explosions do not occur exclusively for progenitors with low $\ewind$, for example $15.4$, $17.8$, and $19.4\,\msun$ progenitors do not explode in (a), but have rather normal $\ewind$ in (b). Parameterization (b) generally produces higher $\ewind$ than (a), because the explosions occur earlier, which leads to higher $\lnue$ at explosion. The recombination energy and $\ewind$ appear to be well-correlated \citep{scheck06}, although the recombination energy is always subdominant.

We show in Figure~\ref{fig:e_exp} the progenitor binding energy $E_{\rm bind}$ corresponding to the mass coordinate of the shock at the moment of explosion. For parameterization (a), successful explosions produce $\ewind$ higher than the binding energy of the progenitor except for s14.4, s24.0, s24.4, and s40.0. In our model, these progenitors would likely experience significant fallback. In parameterization (b), all progenitors successfully cross the explosion threshold, but in many cases the progenitor binding energy is significantly larger than $\ewind$, the major component of supernova explosion energy within our model. Although producing an explosion, these progenitors would likely experience significant fallback. Even earlier explosions with higher $\lnue$ would be required to have $\ewind > E_{\rm bind}$. Interestingly, our model predicts $\ewind < E_{\rm bind}$ for s37.0WHW02, which is the only progenitor with significant fallback in \citet{ugliano12}.

We do not take into account the energy production from nuclear burning of the material swept up by the shock during the explosion. As we show in Section~\ref{sec:nickel}, the shock can expose several tenths of $\msun$ of material to temperatures high enough to burn to \nic. However, the material is already composed of heavy elements like silicon and likely produces less than 1\,MeV/baryon if burnt to \nic, making the produced energy subdominant to $\ewind$.

\subsection{Nickel yields}
\label{sec:nickel}

Iron-group elements including \nic\ are produced in nuclear burning of the progenitor material swept up by the supernova shock. As the shock propagates, the temperatures behind the shock become lower, and below a certain value, \nic\ is no longer produced. The region downstream of the shock is radiation dominated \citep[e.g.][]{weaver80,woosley88,thielemann90}, which ties together the explosion energy $\esn$, shock radius $r_{\rm Ni}$, and post-shock temperature $T_{\rm Ni}$ as
\beq
E_{\rm SN} = \frac{4}{3}\pi r_{\rm Ni}^3 aT_{\rm Ni}^4,
\label{eq:nucleo}
\eeq
where $a$ is the radiation density constant. For $E_{\rm SN} \approx 10^{51}$\,ergs and $T_{\rm Ni} \approx 5\times 10^9$\,K, corresponding to the limit of \nic\ production, we obtain $r_{\rm Ni} \approx 3700$\,km. 

\begin{figure}
\plotone{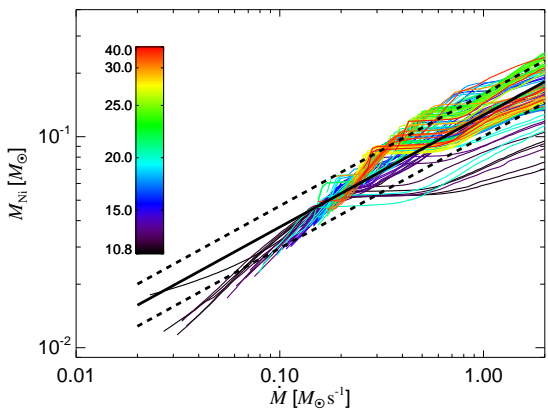}
\caption{The mass of ejected \nic\ as a function of the mass accretion rate through the shock at the moment of explosion for sWHW02 progenitors (denoted by different colors given in the colorbar in the plot) and assuming constant $\esn = 10^{51}$\,ergs. The empirical prediction of Equation~(\ref{eq:mni_fit}) is shown with a thick line along with standard deviation about the fit of $0.1$\,dex indicated with dashed lines.}
\label{fig:nickel_mdot}
\end{figure}

For each progenitor we estimate the mass of ejected \nic, $\mni$, as a function of $E_{\rm SN}$ and the mass coordinate of the cut that separates the PNS and the ejecta. For each mass cut coordinate and progenitor, we calculate the free-fall time to a distance of $100$\,km (the results are not sensitive to this value, as long as it is much smaller than $r_{\rm Ni}$) assuming a quarter gravitational acceleration \citep{yahil82,bethe90} in accordance with the $\mdot$ estimates for the \citet{woosley02} progenitors. We checked that the resulting free-fall times roughly agree with calculations from \gr. At this free-fall time, we evaluate the radii of the infalling overlying mass shells and determine the mass coordinate of $r_{\rm Ni}$. The mass of the material undergoing shock nucleosynthesis is the difference between the mass coordinate at $r_{\rm Ni}$ and the mass cut at the moment when the mass cut reaches $100$\,km multiplied by the \nic\ production efficiency, which depends on $\ye$ and propagation of the shock. We assume that the efficiency is $50\%$ to approximately match the observed $\mni$. We assume that only material with $\ye \ge 0.48$ yields \nic, although the results are essentially unchanged when we keep $\ye \ge 0.49$. 

In Figure~\ref{fig:nickel_mdot}, we show the estimated \nic\ yields for solar-metallicity \citet{woosley02} progenitors as a function of the mass coordinate of the cut separating the ejecta from the PNS and using a constant $\esn = 10^{51}$\,ergs. The mass cut coordinate is expressed in terms of $\mdot$ at the shock at the moment of explosion. This Figure shows that for a constant explosion energy, the \nic\ yield will be smaller for explosions occurring later, when the shock can nucleosynthetically process less material, as expected. We repeated the calculation for $\esn$ smaller and larger by a factor of $10$ and found that 
\beq
\mni \propto \esn^{0.41} \mdot^{0.54},
\label{eq:mni_fit}
\eeq
where $\mdot$ is evaluated at the moment of the explosion and the standard deviation about the fit is about $0.1$\,dex. Equation~(\ref{eq:mni_fit}) implicitly includes the density profiles of progenitors.

\begin{figure*}
\plotone{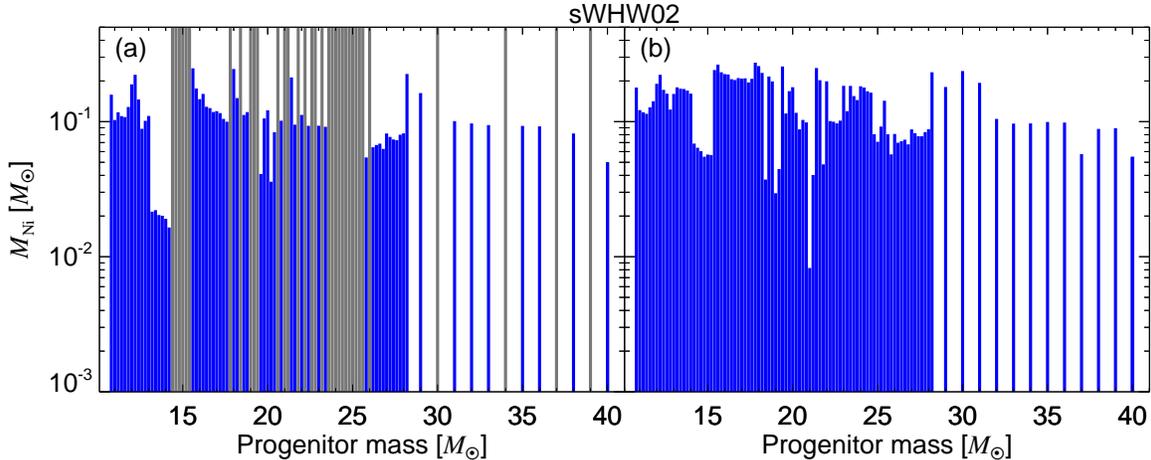}
\caption{Synthesized \nic\ masses as a function of initial progenitor mass for sWHW02 progenitors and parameterizations (a) (left) and (b) (right). Successful explosions are indicated with blue bars, failed core-collapse with gray bars.}
\label{fig:nickel_prog}
\end{figure*}

To obtain $\mni$ estimates for individual progenitors, we assume that the neutrino-driven wind supplies most of the supernova energy, $\esn = \ewind$. We do not include the recombination energy or the negative binding energy of the progenitor. We discuss results for constant $\esn$ later in Section~\ref{sec:obs}, where we also compare our results to the observed low-luminosity supernovae. In Figure~\ref{fig:nickel_prog}, we show the results for the sWHW02 progenitors and the two parameterizations of the neutrino mechanism. Both parameterizations produce $\mni$ between $0.02$ and $0.2\,\msun$, but explosions in parameterization (b) are both earlier and more energetic, which results in higher $\mni$. The nickel yields show similar pattern to what we found in remnant masses and explosion energies. Specifically, there is no monotonic relation between $\mni$ and the initial progenitor mass. Overall, the lowest nickel yields are obtained close to the success/failure boundary for each progenitor. 

\subsection{The landscape of massive single stars deaths}
\label{sec:summary}

\begin{figure*}
 \plottwo{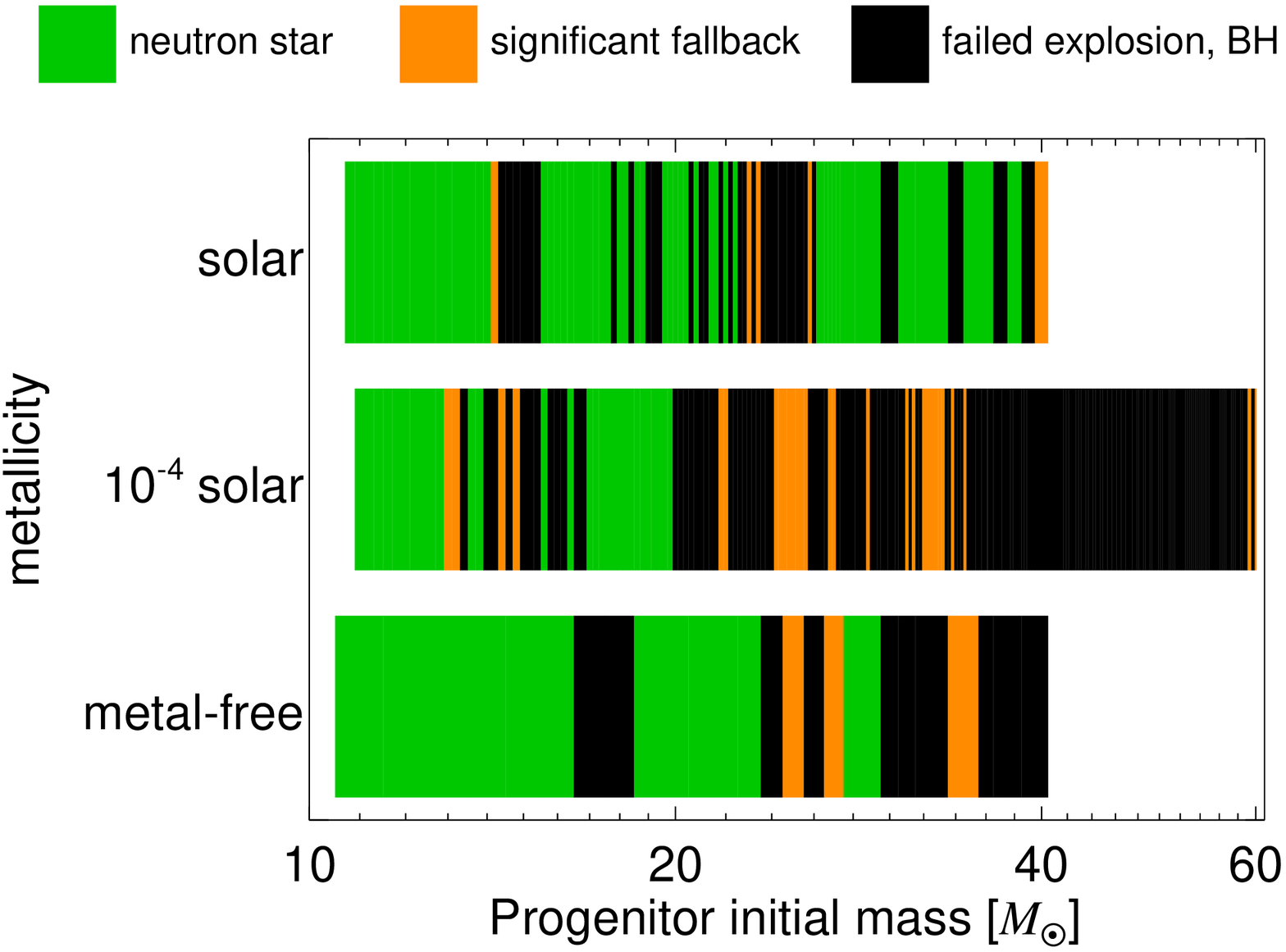}{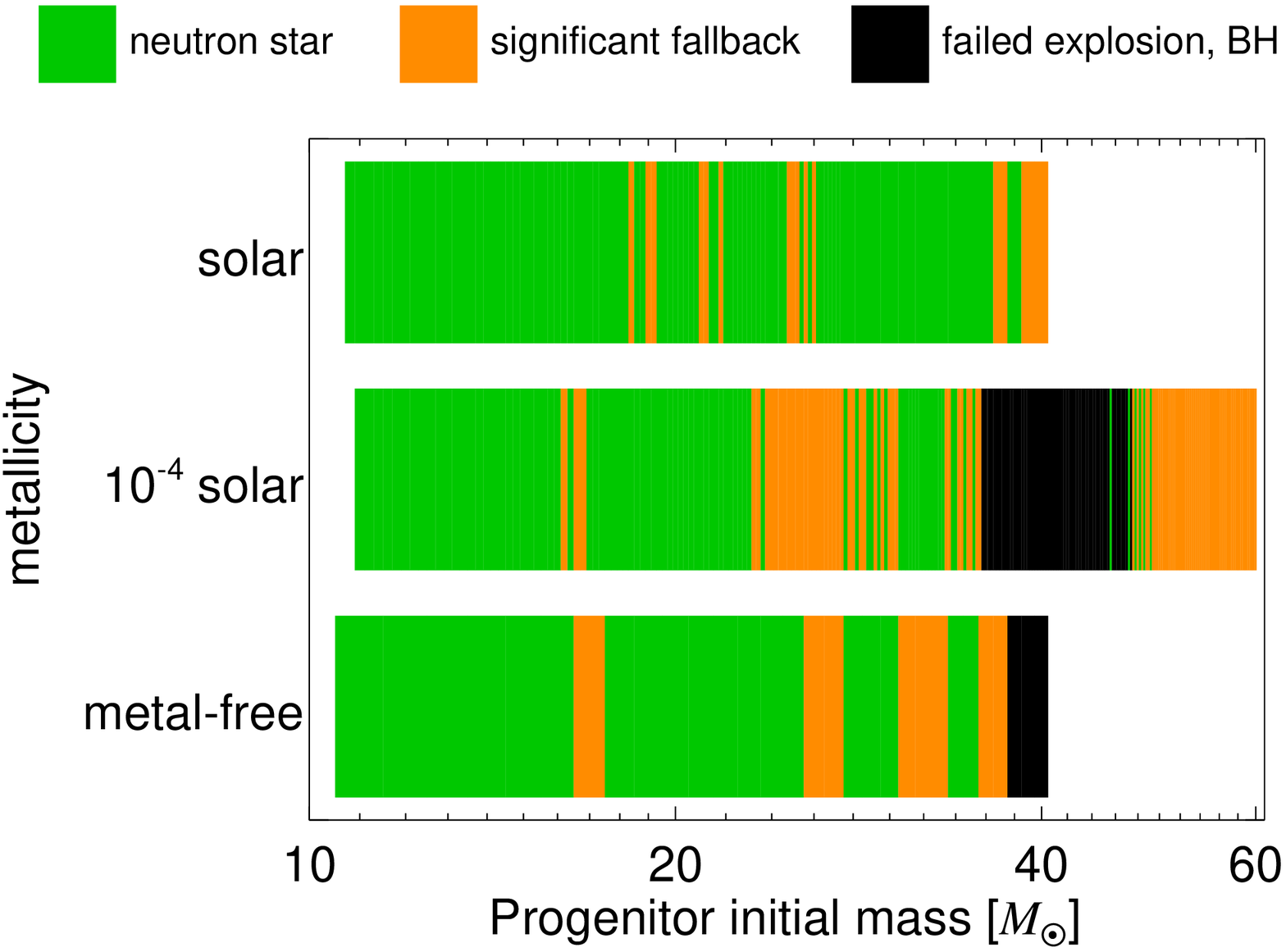}
\caption{Outcomes of core collapse as a function of initial progenitor mass and progenitor metallicity for WHW02 progenitors for parameterizations (a) (left panel) and (b) (right panel). We show successful explosions leaving behind neutron stars (green), successful explosions with significant fallback leaving behind either massive neutron star or a black hole (orange), and failed explosions leaving behind a black hole (black). Our results are very different from the picture presented by \citet{heger03}.}
\label{fig:summary}
\end{figure*}

In Figure~\ref{fig:summary}, we summarize the landscape of the neutrino mechanism as a function of initial progenitor mass and progenitor metallicity. We show results for both parameterizations (a) and (b), which illustrate the range of the possible results (see Sec.~\ref{sec:constraint}). Our model has three possible outcomes, as discussed in this Section. The first is a successful explosion leaving behind a neutron star with mass determined by the mass coordinate at the shock at the moment of explosion (green). The second is a successful explosion with explosion energy approximated by the energy of the neutrino-driven wind that is smaller than the binding energy of the overlying material (orange). In this case, significant fallback likely results and the remnant is either a massive neutron star or a black hole. Finally, progenitors that never satisfy the explosion condition continue accreting and form a black hole (black). All three metallicities show all three outcomes intertwined in a complex pattern that is not related to the initial progenitor mass. The structure of the progenitor stars changes sufficiently as a function of metallicity, which results in no discernible overall pattern as a function of the metallicity.


\section{The predictions of the neutrino mechanism are robust}
\label{sec:robust}

In this Section, we explore the robustness of the predictions and the overall diversity of the neutrino mechanism presented in Section~\ref{sec:results} by studying many different parameterizations and correlations between different observables (Section~\ref{sec:correlations}), equations of state (Section~\ref{sec:eos}), and progenitor populations (Section~\ref{sec:metallicity}).

\subsection{Dependence on neutrino mechanism parameterization}
\label{sec:correlations}

In this Section, we exploit the greatest advantage of our approach: the ability to make a systematic study of observables with respect to the parameterizations of the neutrino mechanism. By varying the parameters in Equation~(\ref{eq:artificial}), we obtain a continuum of outcomes ranging from all stars failing to all stars exploding. We emphasize that we do not arbitrarily decide which progenitors explode and which do not. Instead, our parameterization of the neutrino mechanism is very restricted in the sense that the outcome of progenitors with similar behavior of $\lnue/\lcrit$ must be similar given the assumptions on the EOS and evolution.

We explore the neutrino mechanism by varying $0 \le p \le 1$ and $0.5 \le q \le 5.5$ on a uniform $100\times 100$ grid (Eq.~[\ref{eq:artificial}]). For each $(p,q)$, we repeat the procedure from Sections~\ref{sec:method} and \ref{sec:results} and determine which progenitors explode, the masses of the remnants, the explosion energies, and nickel masses.  We find that our results are insensitive to making the $(p,q)$ grid larger or denser.

Since there are significant differences between progenitors with close initial masses, we make use of global properties such as means and widths of the observable quantities for a population of progenitors to capture the full range of structural variances. To construct a population from progenitor sets, we draw the initial masses of the progenitors $\mathscr{M}$ from a \citet{salpeter55} IMF so that their relative number $N$ is
\beq
\frac{\intd N}{\intd \mathscr{M}} \propto \mathscr{M}^{-2.35}.
\label{eq:salpeter}
\eeq
To characterize the properties of successful and failed explosions independently and as a function of $(p,q)$, we use define $\intd N_{\rm SN}(\mathscr{M},p,q)$, which is equal to $\intd N(\mathscr{M})$ for progenitors with successful explosions and zero otherwise. The global quantities are defined by integrating over $\mathscr{M}$, for example, the fraction of successful supernova explosions $\fsn(p,q)$ is
\beq
\fsn(p,q) = \frac{\int \frac{\intd N_{\rm SN}}{\intd \mathscr{M}} \intd \mathscr{M}}{\int \frac{\intd N}{\intd \mathscr{M}} \intd \mathscr{M}}.
\eeq
The integration is performed by using the mid-point rule centered on the given progenitor $\mathscr{M}$ and with $\intd \mathscr{M}$ equal to half the distance in $\mathscr{M}$ to the nearest progenitor models. The range of initial masses of progenitors we use (Table~\ref{tab:progenitors}) likely does not cover the full range of core collapse. For example, if core collapse occurs for progenitors with $\mathscr{M} \ge 8\,\msun$, Equation~(\ref{eq:salpeter}) implies that $\sim 38\%$ of core collapse events might not be represented by the progenitor samples. Although in some cases it is possible to manually add the results of core collapse of such low-mass progenitors \citep[e.g.][]{zhang08,pejcha_ns}, we prefer to not do so to keep the effects of the neutrino mechanism as clean as possible. However, we strongly encourage production of progenitor models with $8 \lesssim \mathscr{M} \lesssim 11\,\msun$ to make the picture of core-collapse supernovae more complete.

\begin{figure}
\centering
\includegraphics[width=0.35\textwidth]{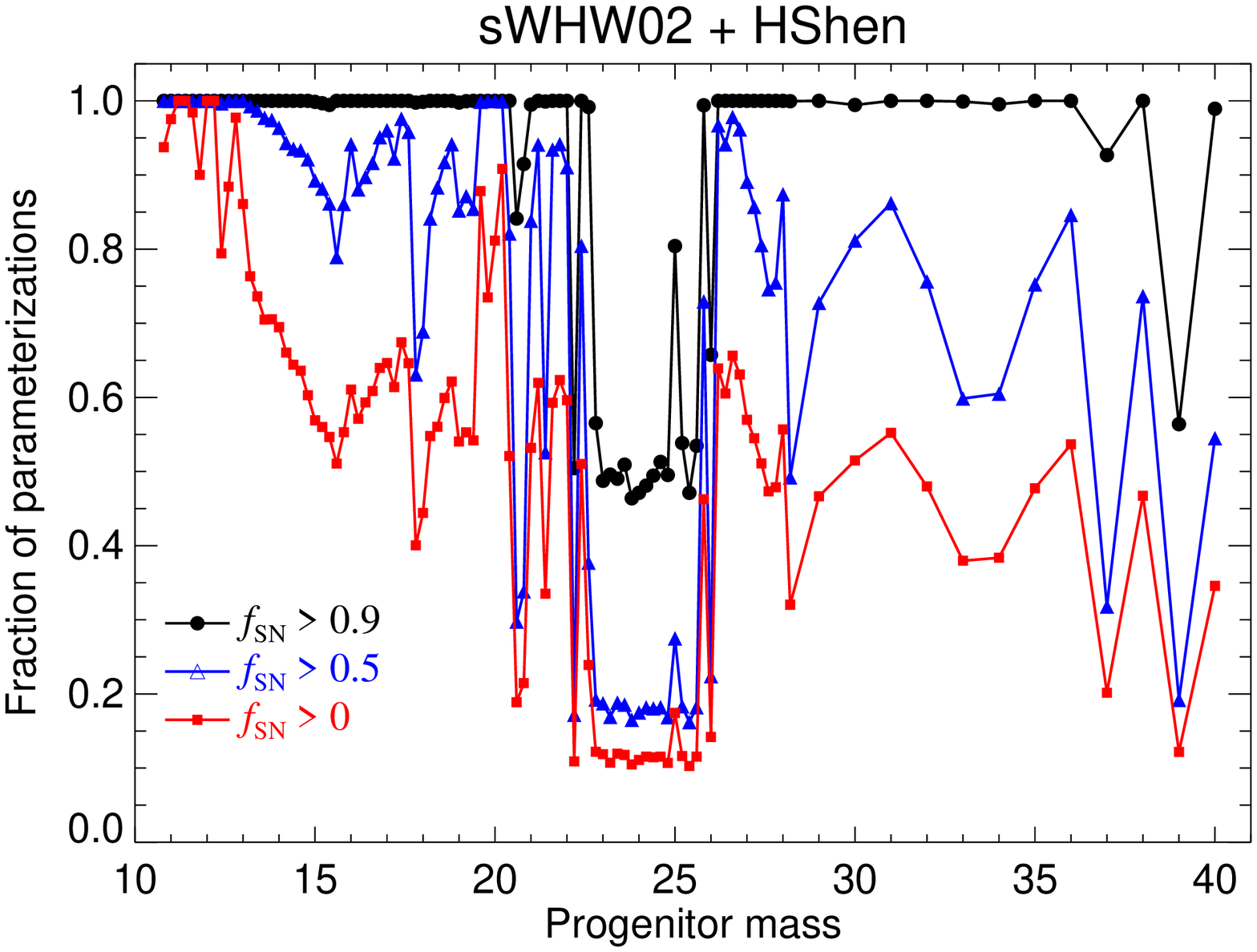}\\
\includegraphics[width=0.35\textwidth]{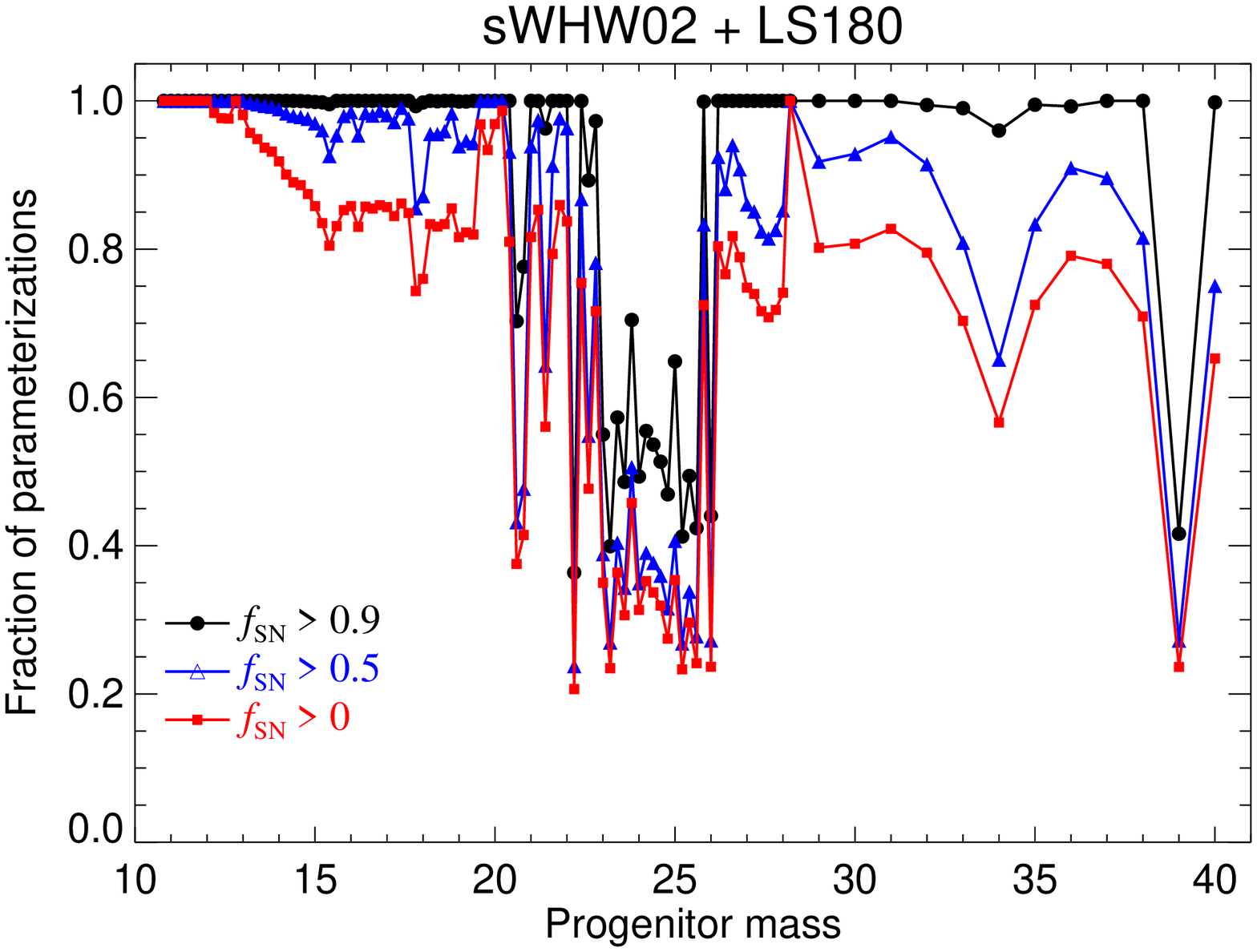}\\
\includegraphics[width=0.35\textwidth]{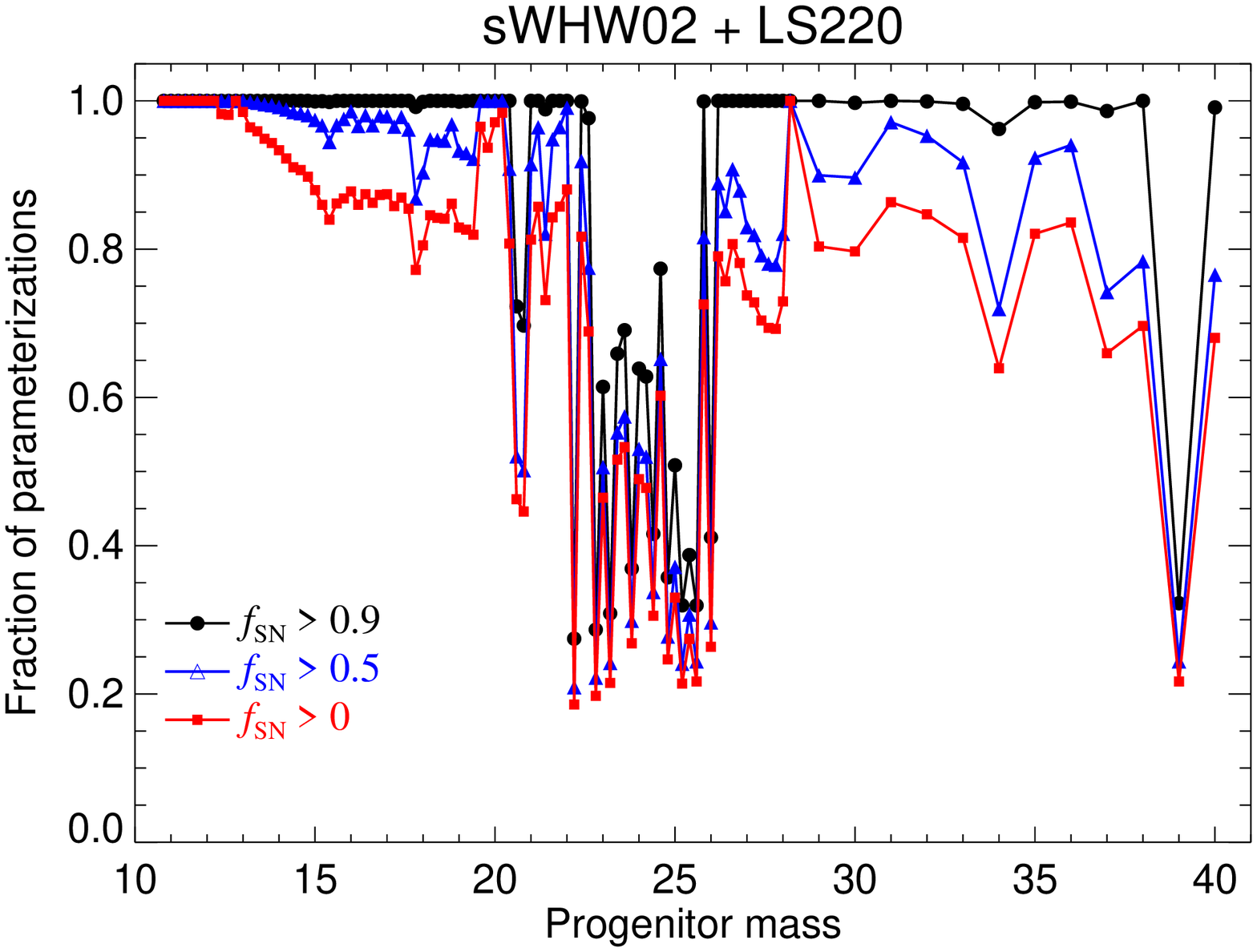}\\
\includegraphics[width=0.35\textwidth]{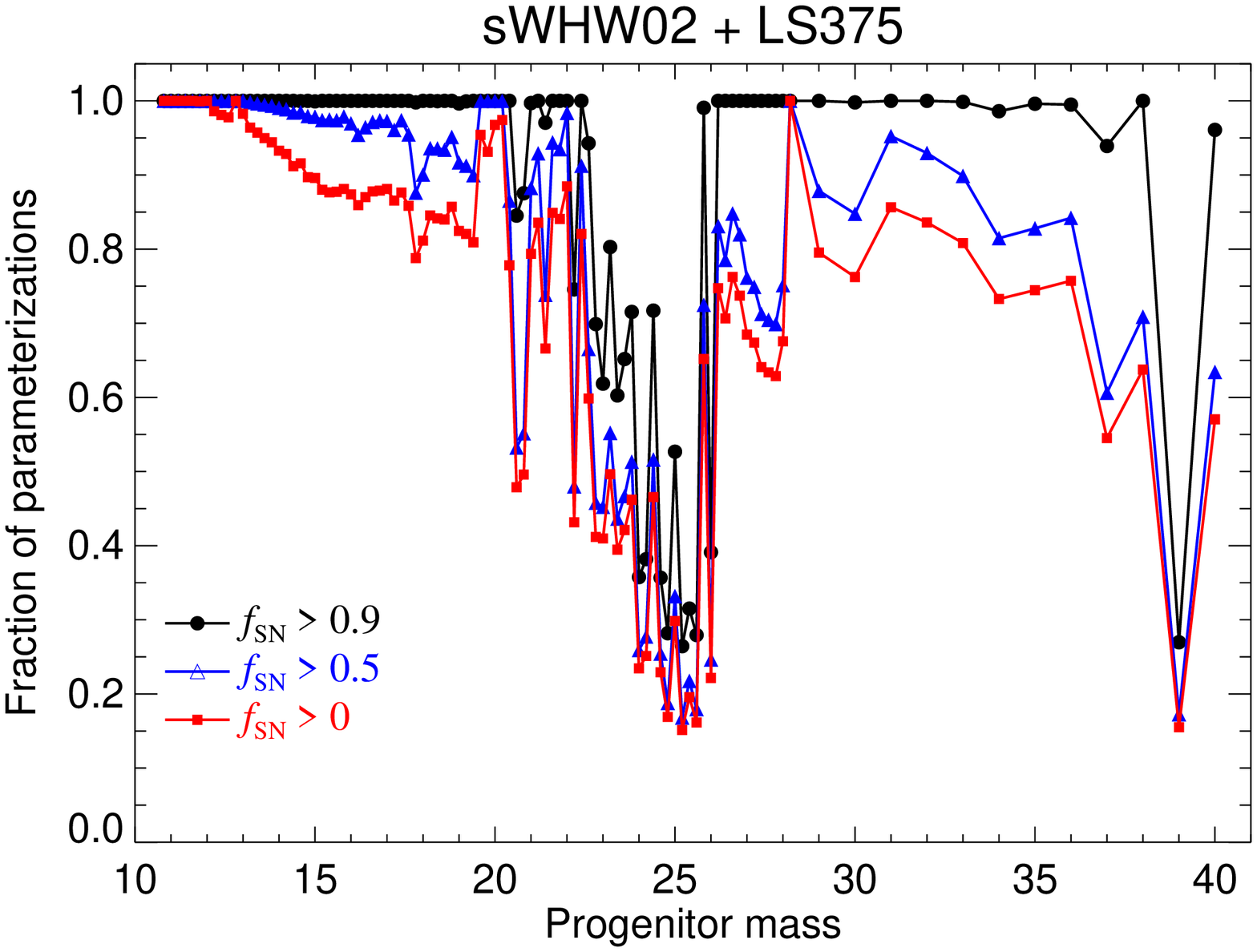}
\caption{Fraction of parameterizations that yield a successful explosion for a given sWHW02 progenitor from the HSHen (top), LS180 (upper middle), LS220 (lower middle), and LS375 (bottom) EOSs obtained by integrating over $(p,q)$ (Eq.~[\ref{eq:fa}]). Each line corresponds to a subset of parameterizations that yield overall $\fsn$ higher than $0.9$ (black circles), $0.5$ (blue triangles), and $0.0$ (red squares). Each fraction is normalized to a total number of parameterizations in a given subset. }
\label{fig:prog_frac}
\end{figure}

In Figure~\ref{fig:prog_frac}, we first investigate the robustness of the neutrino mechanism by measuring the fraction of parameterizations $(p,q)$ that yield successful explosions for individual progenitors from the sWHW02 set by integrating $\intd N_{\rm SN}(\mathscr{M},p,q)$ over $(p,q)$
\beq
f_a(\mathscr{M}) = \frac{\sum\limits_{p,q: \fsn>a} \frac{\intd N_{\rm SN}}{\intd \mathscr{M}}}{\sum\limits_{p,q: \fsn >a}\frac{\intd N}{\intd \mathscr{M}}},
\label{eq:fa}
\eeq
where we split the parameterizations into three groups based on the overall fraction of successful supernova explosions $\fsn$ calculated over all progenitors weighted by Equation~(\ref{eq:salpeter}). Specifically, we show $f_{0.9}$, $f_{0.5}$, and $f_{0.0}$ that yield $\fsn \ge 0.9$, $0.5$, and $0.0$, respectively. From the top panel of Figure~\ref{fig:prog_frac}, which uses the HShen EOS following the discussion in Section~\ref{sec:method}, we see that some progenitors have a consistently lower fraction of successful explosions no matter the overall fraction of successful explosions for all progenitors. Primarily, this includes progenitors around $16$, $18$, and $21\,\msun$ and a well defined mass range between $22$ and $26\,\msun$. Some of the higher-mass progenitors also exhibit lower explosion fractions, although the progenitor grid is too coarse to identify specific trends. This shows that our method yields consistent results and the decision between explosion and failure is not arbitrary, but driven by the progenitor structure.

\begin{figure}
\plotone{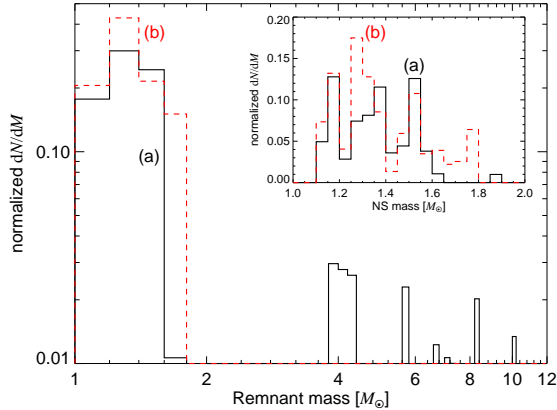}
\caption{Distribution of remnant masses from a population of sWHW02 progenitors. Solid black and dashed red lines are the parameterizations of the neutrino mechanism (a) and (b), respectively, from Figure~\ref{fig:lratio} and \ref{fig:ns_mass}. Failed explosions are assumed to produce black holes with mass equal to the mass of the helium core of the progenitor star \citep{kochanek14a,kochanek14b}. The inset shows the distribution of NS masses with a finer binning.}
\label{fig:rem_dist}
\end{figure}

Although individual progenitors consistently explode or fail in different parameterizations, there can be substantial differences in properties like remnant masses between the many parameterizations that produce successful explosions for a given progenitor. To understand this issue further, we use Equation~(\ref{eq:salpeter}) to produce remnant mass distributions for our parameterizations. In Figure~\ref{fig:rem_dist}, we illustrate the resulting mass distribution for sWHW02 progenitors and parameterizations (a) ($\fsn=0.741$) and (b) ($\fsn=1$). To get a remnant mass distribution of black holes originating from failed explosions, we follow \citet{kochanek14a,kochanek14b}, based on the works of \citet{nadezhin80} and \citet{lovegrove13}, and we use the pre-explosion helium core mass as the black hole mass. Using the pre-supernova progenitor mass as the black hole mass would yield black holes with $M \gtrsim 10\,\msun$ as in \citet{ugliano12}, which is not observed \citep[e.g.][]{burrows87,ozel10}. Similarly to \citet{ugliano12} and \citet{kochanek14a}, we obtain a distinct gap between neutron star and black hole masses as illustrated by parameterization (a). This is caused by the fact that progenitors, which would ordinarily experience fallback and would yield a smooth distribution connecting neutron stars and black hole, simply do not explode and collapse to a black hole. The gap between neutron star and black hole masses is found in black-hole low-mass X-ray binaries, where the probability distribution peaks at $\sim 5\,\msun$ \citep[e.g.][]{bailyn98,ozel10,farr11,kreidberg12}. We predict the lowest-mass black holes should have mass of $\sim 4\,\msun$. Note that in Section~\ref{sec:nu_wind} we suggested the possibility of fallback in several progenitors based on comparison of $\ewind$ with the progenitor binding energy. The number of such progenitors and the corresponding range of masses is so small that they likely would not change Figure~\ref{fig:rem_dist} very much.

The predicted neutron star mass distribution peaks just between $1.2$ and $1.5\,\msun$, which is broadly consistent with observations \citep[e.g.][]{finn94,thorsett99,schwab10,valentim11,ozel12,pejcha_ns,kiziltan13}. In reality, the relative probability in this peak would be even higher, because we do not include the lowest-mass progenitors ($8\lesssim \mathscr{M} \lesssim 10.8\,\msun$), which likely produce low-mass neutron stars. For parameterization (b), the relative frequency of neutron stars with $1.6 \lesssim M \lesssim 1.8\,\msun$ increases, because progenitors with $23 \lesssim \mathscr{M} \lesssim 26\,\msun$ are assumed to explode. Fallback would increase the relative probability in this mass range. However, we do not find neutron stars with $M \gtrsim 1.9\,\msun$ even though the maximum gravitational mass for HShen EOS is $2.24\,\msun$ \citep{oconnor11}, which suggests that the progenitor structure does not allow the supernova explosion mechanism to produce very massive neutron stars in the absence of significant fallback.
 
\begin{figure*}
\plotone{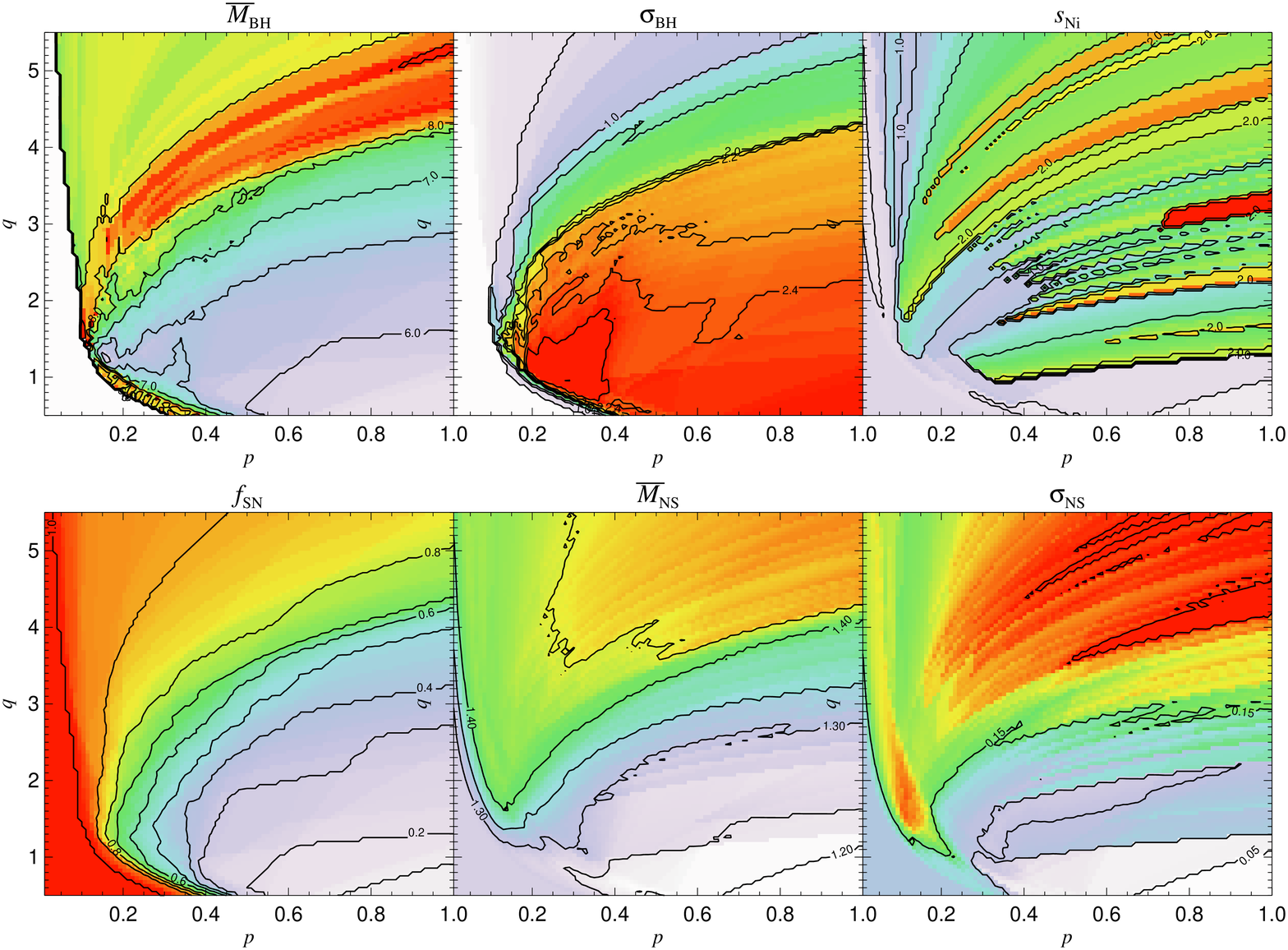}
\caption{Global parameters of an IMF-weighted set of sWHW02 progenitors in the space of parameters of the neutrino mechanism $(p,q)$ (Eq.~[\ref{eq:artificial}]). We show the fraction of successful explosions $\fsn$, means ($\mns$ and $\mbh$) and widths ($\sns$ and $\sbh$) of neutron star and black hole mass distributions, and the width of nickel mass yield $s_{\rm Ni} = \log (\max \mni / \min \mni)$. The resulting values are color-coded from white (lowest value) to red (highest value) and we show labeled contours.}
\label{fig:dep_2d}
\end{figure*}

To generalize these results to the full set of parameterizations, we show in Figure~\ref{fig:dep_2d} the dependence of $\fsn$, means ($\mns$ and $\mbh$) and widths ($\sns$ and $\sbh$) of the neutron star and black hole distributions and the width of the produced nickel masses $s_{\rm Ni} = \log (\max \mni / \min \mni)$ on the parameterization $(p,q)$. We see that each parameter behaves in a somewhat different way and we use this to constrain $(p,q)$ in Section~\ref{sec:constraint}.

\begin{figure}
\centering
\includegraphics[width=0.4\textwidth]{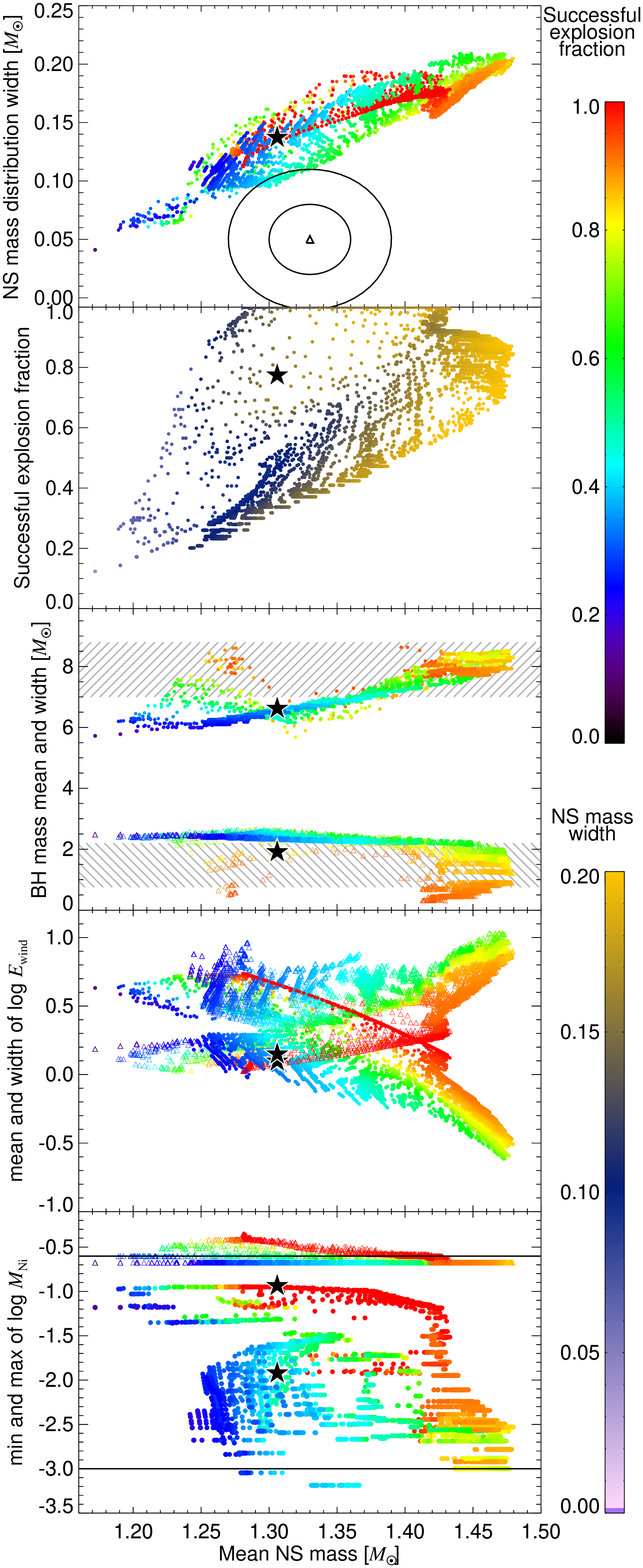}
\caption{Relation between the mean neutron star mass $\mns$ and the width of neutron star mass distribution $\sns$, the fraction of successful explosions $\fsn$, the mean $\mbh$ (solid circles) and width $\sbh$ (open triangles) of the black hole mass distribution, the mean (solid circles) and width (open triangles) of the energy of the neutrino-driven wind $\ewind$ (in units of $10^{51}$\,ergs), and minimum (solid circles) and maximum (open triangles) synthesized nickel mass (in units of $\msun$) for many parameterizations of the neutrino mechanism. Each point is a result of a single calculation with distinct $(p,q)$ color-coded either by successful explosion fraction or the NS mass width, as indicated by the colorbars. The black stars show the results of \citet{ugliano12} convolved with a progenitor population in the same way as for our results. The triangle with two concentric circles shows the mean and width of the double neutron star population and their approximate $68\%$ and $95\%$ confidence intervals \citep{ozel12}. The horizontal hatched bars in the black hole mass distribution are approximate $68\%$ confidence intervals on the observed BH mass distribution from \citet{ozel10}. The horizontal lines in the bottom panel show minimum and maximum nickel masses in the observed samples of \citet{hamuy03} and \citet{spiro14}.}
\label{fig:dependencies}
\end{figure}

At this point, we are not interested in the specific values $(p,q)$ corresponding to specific values of parameters, but we want to uncover overall scalings and relations. We show in the top panel of Figure~\ref{fig:dependencies} the neutron star mass distribution width $\sns$ as a function of the mean neutron star mass $\mns$. We see a distinct correlation between the mean and width of the neutron star mass distribution, which is the sole result of the parameterized neutrino mechanism applied to an ensemble of progenitors. The individual points are color-coded according to the fraction of successful explosions and the colors indicate that the correlation is present for cases with $\fsn\approx 1$ (red) as well as for the remaining points with lower $\fsn$. The correlation for $\fsn\approx 1$ cases comes from parameterizations with small $p$, which make almost everything explode. Small $q$ makes the explosions relatively early at high $\mdot$ (small mean and width of NS masses), while for big $q$ they occur somewhat later when the system had time to accrete more mass (large mean and width of NS masses). For $\fsn < 1$, the correlation is due to changes in relative contributions of different progenitor mass ranges to the final mass distribution so that higher $\fsn$ usually results in including progenitors producing more massive NS and hence larger mean and width of the NS mass distribution. The red branch of calculations with $\fsn \approx 1$ has slightly higher NS mass width, but this depends to some extent on the treatment of neutrino energies as we mentioned in Section~\ref{sec:gr1d}.

In the second panel of Figure~\ref{fig:dependencies}, we show the successful explosion fraction as a function of the mean NS mass. This Figure is primarily useful as an illustration of how our models can exclude parts of parameter space. For example, the mean observed neutron star mass of $\sim 1.33\,\msun$ \citep[e.g.][]{pejcha_ns,ozel12} immediately implies $\fsn \gtrsim 0.35$. This comes from the fact that getting higher-mass NS requires more progenitors exploding and thus higher $\fsn$.

The third panel of Figure~\ref{fig:dependencies} shows the mean $\mbh$ and width $\sbh$ of the black hole mass distribution as a function of the mean NS mass. To make the calculation easier, we opt to characterize the black hole masses by mean and width instead of the more appropriate cutoff mass and decay scale \citep{ozel10,clausen14}. The cutoff mass would be similar to the lowest obtained BH mass, which is $\sim 4\,\msun$ in our calculations. The width of our BH mass distribution is essentially constant at $\sim 2\,\msun$ if the mean NS mass is $\lesssim 1.40\,\msun$. The mean BH mass, however, shows a clear correlation with the mean NS mass as well as with the successful explosion fraction, indicating a shift towards higher-mass black holes.

The fourth panel of Figure~\ref{fig:dependencies} shows the mean and width of the distribution of $\ewind$, which we assume as a proxy of the explosion energy. In Section~\ref{sec:nu_wind}, we discussed the issues with the absolute scale and width of $\ewind$ distribution. Despite the caveats, the qualitative correlations between quantities should be preserved in a set of calculations with a consistent set of assumptions.  We see there is anti-correlation between the mean of $\ewind$, $\mns$, and the width of $\ewind$. This is caused in part by later explosions producing higher NS masses also having lower explosion energies, and partially by the fact that successful explosions in progenitors close to the success/failure border will produce high NS masses and low explosion energies as is visible by comparing parameterizations (a) and (b) in Figure~\ref{fig:e_exp}.

The bottom panel of Figure~\ref{fig:dependencies} shows the minimum and maximum achieved amount of $\mni$ in a given parameterization. We see that the maximum $\mni$ is essentially constant in all parameterizations, however, the minimum shows a significant spread with little overall correlation except that lower $\mni$ minima are typically obtained in parameterizations with low $\fsn$. This is due to the fact that low $\mni$ can only be produced in some progenitors that explode close to success/failure border, where the tenuous progenitor layers yield little $\mni$.


The main result of this Section is that {\em despite our very wide range of parameterizations of the neutrino mechanism, the neutron star and black hole observational properties exhibit surprisingly small variance}. For example, within our method it is impossible to produce a population of neutron stars with mean mass higher than about $1.6\,\msun$. While this illustrates the robustness of the neutrino mechanism and the crucial dependence on the progenitor structure, it also makes it hard to observationally constrain the mechanism properties based on observations of remnant masses. We found that the width of the nickel yield distribution, and especially the low envelope of $\mni$, is surprisingly sensitive diagnostic and varies greatly among parameterizations, which offers interesting prospects for diagnosing the explosion mechanism and progenitor structure.

\subsection{Dependence on the equation of state}
\label{sec:eos}

The equation of state significantly affects the cooling properties of the PNS and hence the neutrino mechanism. To investigate the sensitivity of our results to the EOS, we calculated the evolution of the sWHW02 progenitors for LS180, LS220, and LS375 EOS in addition to the HShen EOS used in the rest of our work. For some progenitors, the LS EOS leads to a potentially unphysically sudden upturn in $\lnue/\lcrit$ at low $\mdot$ that is connected with the increase in neutrino luminosity and energy before black hole formation,  which can be visible in the bottom panel of Figure~\ref{fig:lratio}. This does not occur with HShen EOS. We remove this potential artifact by assuming that the progenitor forms a BH at the last minimum of $\lnue/\lcrit$, but find the results do not change very much when we leave the feature untouched.

We find that by varying $(p,q)$ we can reproduce a similar pattern of success and failure to that seen in Figure~\ref{fig:ns_mass}, although the exact values of $(p,q)$ are naturally different for each EOS. We find only small changes in the remnant masses for progenitors that robustly explode. To further illustrate the similarity of the results, we expand Figure~\ref{fig:prog_frac} to include results for all the EOSs. We see that the general pattern is essentially unchanged, namely the progenitors that likely fail to explode cluster between $21$ and $26\,\msun$. The fraction of parameterizations that yield failure in low- to moderate-mass progenitors ($13$ to $19\,\msun$) is relatively lower for the LS EOS compared to HShen. We also do not find any systematic trend with the incompressibility factor. For softer EOS LS180 and LS220 we find typically higher $\ewind$ than for the fiducial calculation with HShen due to generally higher neutrino luminosities. The width of $\ewind$ is also smaller for softer EOS. We find little difference between LS375 and HSHen.

\subsection{Dependence on progenitor metallicity}
\label{sec:metallicity}

Another advantage of our formalism is that we can obtain relative results between different progenitor sets provided that the EOS and the specific values of $(p,q)$ remain the same. We illustrate this in Figure~\ref{fig:ns_mass}, where we show the resulting remnant masses for all progenitor sets from Table~\ref{tab:progenitors} using the HShen EOS. We see that the uWHW02 progenitors produce a relatively higher fraction of failed explosions even for parameterization (b), which makes all sWHW02 progenitors explode. In particular, progenitors with $36 \le \mathscr{M} \le 47\,\msun$ are difficult to explode. The metal-free progenitors (zWHW02) also exhibit the interwoven pattern of successful and failed explosions. The number of \citet{limongi06} progenitors is significantly smaller, but for LC06A we can see a pattern partially similar to sWHW02 in the sense that NS masses increase between progenitor masses of $11$ and $15\,\msun$.

The explosion energies of uWHW02 and zWHW02 are not substantially different from sWHW02, although it is worth pointing out that progenitors with $\mathscr{M} \gtrsim 23\,\msun$ in the uWHW02 series have binding energies significantly larger than $\ewind$. In zWHW02 progenitors, the binding energy is comparable to $\ewind$ for $\mathscr{M} \gtrsim 25\,\msun$. Nickel yields are also similar, but uWHW02 progenitors with $47 \lesssim \mathscr{M} \lesssim 51\,\msun$ have $\mni \approx 0.5\,\msun$ in parameterization (b). These progenitors do not explode in parameterization (a).

\begin{figure}
 \plotone{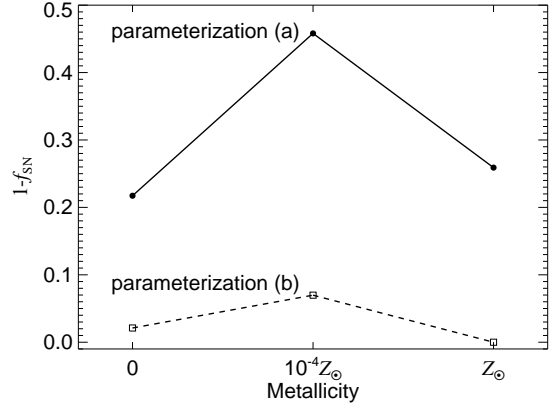}
\caption{Fraction of progenitors forming black hole, $1-\fsn$, as a function of metallicity for WHW02 progenitors and parameterizations (a) (solid circles) and (b) (open squares).}
\label{fig:fbh}
\end{figure}

In Figure~\ref{fig:fbh} we show the fraction of black holes formed, $1-\fsn$, as a function of metallicity for WHW02 progenitors. We see that black hole fraction is about twice as high at $10^{-4}Z_\sun$ than at zero or solar metallicity for parameterization (a). For parameterization (b), the differences are even more pronounced. This could go in the way of explaining why gamma-ray bursts seem to prefer low-metallicity regions \citep[e.g.][]{fruchter06,stanek06,savaglio09}. Our black hole fractions were calculated assuming a \citet{salpeter55} IMF; if the IMF favors more massive stars at low metallicity, the black hole fraction would be even higher.

\section{Comparison with previous works and observations}
\label{sec:disc}

\subsection{Comparison to \citet{ugliano12}}
\label{sec:ugliano}

\citet{ugliano12} performed a set of time-dependent 1D calculations of sWHW02 progenitors parameterized by the evolution of the inner parts of the PNS. By matching the explosion properties of s19.8WHW08 to SN1987A they produced remnant masses, explosion energies, and nickel yields for the same set of progenitors as in this work. Comparison with their work based on different codes and assumptions provides an external check on our results and on the robustness of the neutrino mechanism.

Our parameterization (a) was chosen to produce a similar pattern of successful and failed explosions as in \citet{ugliano12}. We emphasize that the tuning of $(p,q)$ was not extreme; our results from Section~\ref{sec:correlations} show that many parameterizations yield similar results. Both our results and the work of \citet{ugliano12} produce successful and failed explosion interwoven between each other. We agree on a cluster of failed explosions between about $14$ and $16\,\msun$, fine structure of interweaved explosions and failures between $17$ and $22\,\msun$, a cluster of failed explosions between $23$ and $26\,\msun$, and mostly explosions above about $26\,\msun$. In general it is possible to get better agreement with \citet{ugliano12} in a narrow range of progenitor masses at the expense of some other progenitor mass range. For example, making $q$ in Equation~(\ref{eq:artificial}) smaller would produce generally higher remnant masses for progenitors around $15\,\msun$, but at the same time we would obtain explosions for progenitors between about $23$ and $26\,\msun$.

\begin{figure}
 \plotone{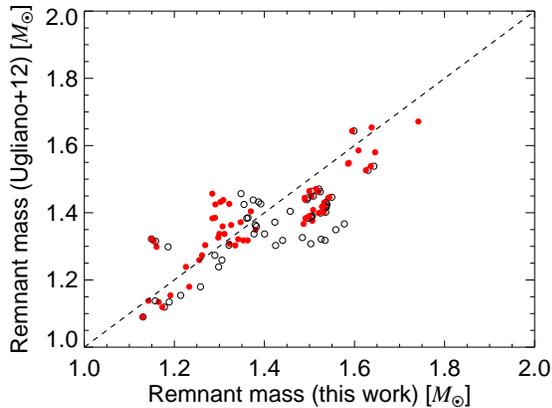}
\caption{Comparison of gravitational neutron star masses without fallback from \citet{ugliano12} and for our neutrino mechanism parameterizations (a) and (b) shown with open black and solid red circles, respectively.}
\label{fig:ugliano_comp_m}
\end{figure}

In Figure~\ref{fig:ugliano_comp_m}, we directly compare our NS masses with those of \citet{ugliano12} before fallback. We see excellent agreement for both our parameterizations, in particular, the relative difference (and its scatter) is $0.06$ ($0.09$) and $0.02$ ($0.08$) for parameterizations (a) and (b), respectively. This demonstrates that despite the differences in the overall approach, codes used and input physics, we obtain very similar results as \citet{ugliano12}.

To illustrate the good agreement, we also show in Figure~\ref{fig:dependencies} with a black star the means and widths of the NS and BH mass distributions and $\fsn$ of \citet{ugliano12} calculated in the same way as for our results. There is a good agreement between our results and \citet{ugliano12} in all global parameters except for the small offset in the width of the BH mass distribution\footnote{For progenitors that did not explode in \citet{ugliano12} we calculated the BH mass in the same way as for our results, ie.\ equal to the He core mass.}.

\begin{figure}
\plotone{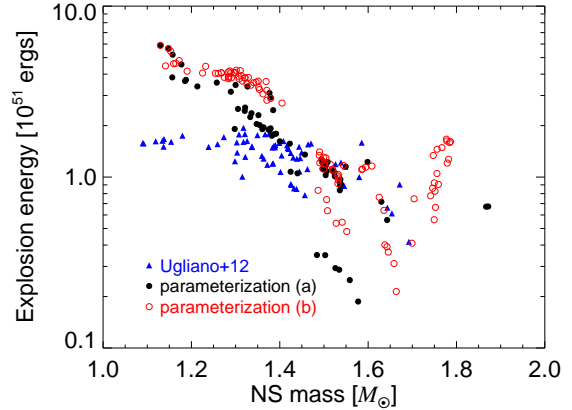}
\caption{Explosion energy as a function of the remnant mass for parameterizations (a) and (b) (solid black and open red circles, respectively), and \citet[][solid blue triangle]{ugliano12}. For our results we assume that supernova explosion energy is equal to the energy of the neutrino-driven wind.}
\label{fig:cor_m_e}
\end{figure}

As discussed in Section~\ref{sec:energy}, a reasonable guess for the supernova explosion energy is the energy of the neutrino-driven wind. There are, however, many other contributions such as recombination and nuclear burning. As a result, we obtain far worse relative agreement with \citet{ugliano12} for individual progenitors than for NS masses. To evaluate whether our results are reasonable, we focus on relative trends of explosion energy and other variables. In Figure~\ref{fig:cor_m_e}, we show the explosion energy as a function of the remnant mass. For $M \lesssim 1.3\,\msun$, we predict explosion energies a factor of $\sim 2$ larger than \citet{ugliano12}, but the agreement is somewhat better for higher-mass NS, although we produce some substantially weaker explosions and a ``turn-over'' to higher $\ewind$ at high NS masses. There is a weak anti-correlation between $\esn$ and NS mass in \citet{ugliano12} and both of our parameterizations (a) and (b). This comes from the fact that typically higher NS masses are achieved in more delayed explosions where the PNS luminosity has already faded. Our explosion energies have greater spread than those of \citet{ugliano12}. 

\begin{figure*}
 \plotone{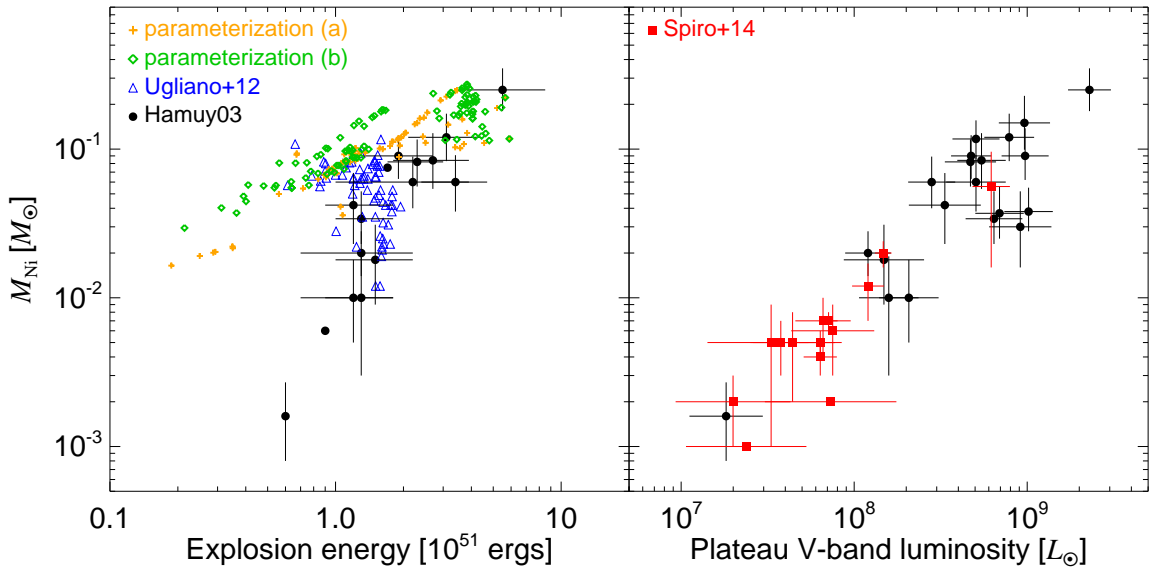}
\caption{Nickel mass as a function of the supernova kinetic explosion energy (left panel) and absolute $V$-band luminosity at the middle of the plateau (right panel). We shows theoretical results for sWHW02 progenitors from this work (orange plus signs and green diamonds for parameterizations (a) and (b), respectively) and \citet[][blue triangles]{ugliano12}, and observations from \citet[][black circles]{hamuy03} and \citet[][red squares]{spiro14}. }
\label{fig:iip}
\end{figure*}

The last quantity to compare is the mass of ejected \nic. Our estimate of $\mni$ is based on an estimate of the progenitor mass swept up by the propagating shock (Eq.~[\ref{eq:nucleo}]), while \citet{ugliano12} solve the time-dependent evolution with a small reaction network. As a result, we again focus on reproducing general trends rather than nickel masses for individual progenitors. In Figure~\ref{fig:iip}, we show the $\mni$ as a function of the explosion energy. We see that results of both our parameterizations produce a distinct correlation between the explosion energy and $\mni$. The slope of this relation is not entirely constrained in our model due to uncertainties in determining $\ewind$. If the distribution of $\ewind$ was narrower, the correlation would be preserved, but with slightly narrower $\mni$, because the dependence of $\mni$ on $\esn$ is relatively weak (Eq.~[\ref{eq:mni_fit}]). What is surprising is that \citet{ugliano12} do not produce any significant correlation between $\mni$ and $\esn$. We do not know the reason for this lack of correlation.

\subsection{Comparison to compactness}
\label{sec:compactness}

\citet{oconnor11} proposed that a good metric of propensity of a supernova progenitor to explode is the compactness parameter
\beq
\xi_M = \left.\frac{M/\msun}{R(M_{\rm bary} ={M})/1000\,{\rm km}}\right|_{t_{\rm bounce}}
\label{eq:compactness}
\eeq
evaluated at the moment of core bounce. \citet{oconnor11} suggested that $\xi_{2.5}$ is an indicator of black hole formation and \citet{oconnor13} argued that $\xi_{1.75}$ characterizes the supernova neutrino luminosity evolution. \citet{nakamura14b} studied the outcomes of core collapse for many progenitors with respect to $\xi_{1.5}$. The compactness of various supernova progenitors was recently studied by \citet{sukhbold14}. Despite these works, little attention has been devoted to evaluate the internal consistency of $\xi_M$ and comparing it with other possible definitions of compactness such as $\xi_{\ye=0.497}$, $\xi_{\ye=0.499}$, and $\xi_{^{28}{\rm Si}}$. We explore these issues in Appendix~\ref{app:compactness} finding that $\xi_{M}$ is remarkably internally consistent, but mostly inconsistent with other possible definitions. None of the definitions is favored {\em a priori} and here we thus focus on comparing our results with these possible definitions of compactness. 

\begin{deluxetable}{cccccc}
\tabletypesize{\scriptsize} 
\tablecaption{Rank correlation coefficients}
\tablewidth{0pt}
\tablehead{
\colhead{} & \colhead{$\xi_{1.75}$} & \colhead{$\xi_{2.5}$} & \colhead{$\xi_{\ye=0.497}$} & \colhead{$\xi_{\ye=0.499}$} & \colhead{$\xi_{^{28}{\rm Si}}$}
}
\startdata
\cutinhead{HShen}
$\fsn\ge 0.9$ & $-0.79$ & $-0.75$ & $0.19$ & $0.26$ & $-0.58$\\
$\fsn\ge 0.5$ & $-0.86$ & $-0.90$ & $0.20$ & $0.34$ & $-0.58$\\
$\fsn\ge 0.0$ & $-0.87$ & $-0.92$ & $0.21$ & $0.34$ & $-0.58$\\
\cutinhead{LS180}
$\fsn\ge 0.9$ & $-0.78$& $-0.70$ & $  0.25$ & $  0.20$ & $ -0.49$\\
$\fsn\ge 0.5$ & $-0.90$ &$-0.94$ & $  0.28$ & $  0.18$ & $ -0.64$\\
$\fsn\ge 0.0$ & $-0.88$ &$-0.95$ & $  0.25$ & $  0.18$ & $ -0.62$\\
\cutinhead{LS220}
$\fsn\ge 0.9$ & $-0.78$ & $ -0.75$ & $  0.18$ & $  0.21$ & $ -0.53$\\
$\fsn\ge 0.5$ & $-0.90$ & $ -0.96$ & $  0.26$ & $  0.19$ & $ -0.65$\\
$\fsn\ge 0.0$ & $-0.90$ & $ -0.96$ & $  0.26$ & $  0.18$ & $ -0.63$\\
\cutinhead{LS375}
$\fsn\ge 0.9$ & $-0.81$ & $ -0.75$ & $  0.24$ & $  0.18$ & $ -0.53$\\
$\fsn\ge 0.5$ & $-0.90$ & $ -0.96$ & $  0.26$ & $  0.18$ & $ -0.64$\\
$\fsn\ge 0.0$ & $-0.90$ & $ -0.96$ & $  0.27$ & $  0.18$ & $ -0.63$\\
\enddata
\label{tab:spearman}
\tablecomments{The Table shows Spearman's rank correlation coefficients $\varrho$ for the five definitions of compactness shown in Figure~\ref{fig:compactness} (table columns) and the fraction of neutrino mechanism parameterizations yielding successful explosions for each progenitor shown in Figure~\ref{fig:prog_frac} split in three categories according to an overall $\fsn$ (table rows) and for all EOSs. Since $\xi$ should be higher for progenitors less propitious for explosion, negative values of $\varrho$ indicate better agreement between our results and the particular compactness definition.}
\end{deluxetable}

To find the best definition of compactness, we focus on the sWHW02 progenitors and compare the data from Figure~\ref{fig:compactness} with the data from Figure~\ref{fig:prog_frac}, which shows the fraction of parameterizations of the neutrino mechanism that yields successful explosions for each progenitor for three subsets of parameterizations defined by the overall minimum $\fsn$. The fraction of parameterizations is convenient for comparison with compactness, because it evaluates the propensity of a progenitor to explode on a continuous scale. The resulting values of the Spearman rank correlations coefficient $\varrho$ are given in Table~\ref{tab:spearman} and indicate that $\xi_{1.75}$ and $\xi_{2.5}$ present the best agreement with our results. Results of Appendix~\ref{app:compactness} imply, that any $\xi_{M}$ with $M > 1.4\,\msun$ will give a comparably good match. Definitions of $\xi$ based on $\ye$ give a poor match to our results. The results for LS EOS are essentially the same. \citet{kochanek14b} found that to describe the BH mass function, $\xi_{2.0}$ and $\xi_{2.5}$ work significantly better than $\xi_{3.0}$, the iron core mass or the mass enclosed by the oxygen burning shell. 

We emphasize that the comparison in Table~\ref{tab:spearman} is based on the full ensemble of all parameterizations and that the agreement between any single neutrino mechanism parameterization and the compactness might be considerably worse. For example, failed explosion of s14.4WHW02 in parameterization (a) would imply that progenitors with $\xi_{2.5} > 0.12$ or $\xi_{1.75} > 0.52$ would fail as well. By comparing Figures~\ref{fig:ns_mass} and \ref{fig:compactness} we see that this is clearly not the case: the failure condition $\xi_{2.5} > 0.12$ means that all solar-metallicity progenitors with $\mathscr{M} > 15\,\msun$ (except s20.2WHW02) should fail, but this is not the case in parameterization (a). The overall agreement is better with  $\xi_{1.75} = 0.52$ as the border between success and failure, but there are important differences as well. For example, sWHW02 progenitors with $26 \le \mathscr{M} \le 28\,\msun$ explode in parameterization (a), but should fail based on the compactness condition. The compactness condition would also predict failures above $32\,\msun$, but we obtain a mixture of successes and failures. Now the question is whether we can find a threshold in $\xi_M$ as well as $M$ that will give the best agreement with the results of parameterization (a). We varied both $M$ and $\xi_M$ on a dense grid and found that $\xi_M$ can predict the outcome correctly for a maximum of $88\%$ of the progenitors (see Fig.~\ref{fig:comp2d} in Appendix~\ref{app:compactness}). The best agreement was obtained for $1.7 \ge M \ge 2.2\,\msun$. The requisite values of $M$, $\xi_M$ and the overall agreement will be different for each parameterization, as can be seen by applying this logic to parameterization (b), where all sWHW02 progenitors explode. For the results of \citet{ugliano12}, the highest fraction of correctly predicted outcome is about $83\%$, and the best compactness is $\xi_{1.42}$, which significantly lower than in our parameterization (a). To summarize, for plausible parameterizations such as (a) (see Sec.~\ref{sec:constraint}), the feedbacks of the neutrino mechanism involving the evolution of the mass accretion rate, and PNS mass, radius and luminosity result in a more complicated behavior than what would be predicted based on compactness only. From this point of view, the ``critical compactness'' used in studying black hole formation \citep[e.g.][]{oconnor11,kochanek14b} is only an approximate estimate of the outcome of the neutrino mechanism acting on a set of progenitors.

Recently, \citet{nakamura14b} performed 2D core collapse calculations on sWHW02 progenitors and found good correlation of resulting neutron star masses, explosion energies and nickel yields with the compactness. All of their calculations explode. Our results are somewhat similar, since we find that compactness roughly correlates with NS masses (comparing Figures~\ref{fig:ns_mass} and \ref{fig:compactness}), which in turn correlates with explosion energy (Fig.~\ref{fig:cor_m_e}) and nickel mass (Fig.~\ref{fig:iip}). However, \citet{nakamura14b} finds very high NS masses going up to about $2.5\,\msun$, and only a limited range of  explosion energies (about $0.2$ to $0.7\times 10^{51}$\,ergs) and nickel masses (about $0.008$ to $0.05\,\msun$). If the dependency of observed properties on compactness found by \citet{nakamura14b} is confirmed, the range of observed of properties such as explosion energy and nickel mass would be an immediate constraint on the range of compactnesses of progenitor stars. \citet{suwa14} performed 1D and 2D simulations of \citet{woosley07} progenitors and did not find any clear correlation between the compactness and the outcome of their simulations.

\subsection{Comparison to observations}
\label{sec:obs}

The final judgment on the viability and properties of the neutrino mechanism must come from comparison of models with the observed properties of supernovae. In Figure~\ref{fig:rem_dist}, we presented the remnant mass distributions obtained from the two representative parameterizations of the neutrino mechanism. We see that there are observationally testable differences: for parameterization (b) explosions occur earlier and for all progenitors, which results in lower NS masses for the bulk of the population and higher probability of NS with $M \sim 1.8\,\msun$ than for parameterization (a). Direct comparison of our models with the observed population of double neutron stars similar to our previous work in \citet{pejcha_ns} is beyond the scope of the present paper. First, we do not include progenitors with initial masses below the minimum mass of our progenitor models (Table~\ref{tab:progenitors}). These progenitors are relatively abundant and are commonly thought to produce relatively low-mass neutron stars. The result would be relatively large probability for NS with $M\sim 1.15\,\msun$ and potential increase of the distribution width. Second, even a modest amount of fallback can significantly change the relative probability of models. For example, a mere $0.1\,\msun$ of fallback observed in the low-mass progenitors by \citet{ugliano12} would significantly worsen the relative probability of parameterization (a) and improve the probability of parameterization (b). Such mild fallback in some progenitors might require more significant fallback in others, potentially affecting the black hole mass distribution. However, we cannot calculate fallback in our models. Instead of a detailed comparison, we show in the top panel of Figure~\ref{fig:dependencies} the observed mean and width of double neutron stars along with their confidence intervals from \citet{ozel12}. Although this comparison is not fully appropriate, as we discuss in \citet{pejcha_ns}, our predictions overlap with the $95\%$ confidence interval of the observations. We also emphasize that we theoretically predict a correlation between the mean and width of NS mass distribution without the inclusions of any fallback, making it an inherent feature of the neutrino mechanism convolved with a set of progenitors.

Although the masses of double neutron stars can be precisely measured, it is arguable whether their prior binary evolution does not hamper inferences about the supernova models, for example by making the progenitors converge on a similar structure and thus reducing the dispersion in resulting double neutron star masses. We point out that the neutron star mass distribution predicted by our models (Fig.~\ref{fig:ns_mass}) as well as by \citet{ugliano12} exhibits fine structure in the sense that even a population with a small range of initial progenitor masses will yield NS masses of considerable width (for example, progenitors with $12 \le \mathscr{M} \le 15\,\msun$). This is the result of variations in progenitor structure, which originate in the latest stages of stellar evolution and even similar CO core masses might lead to different outcomes as measured by the compactness \citet{sukhbold14}. The core collapse outcome and final NS mass of a progenitor are thus determined by the evolution immediately before collapse and should depend little on the past binary evolution if the sampled range of CO core masses is wide enough.

The second panel of Figure~\ref{fig:dependencies} shows the fraction of successful supernova explosions $\fsn$ of a progenitor population, but this quantity unfortunately does not show any strong connection with the mean neutron star mass or width of the distribution. However, we find a new connection between the mean NS mass and a lower limit on $\fsn$. For example, $\mns \approx 1.33\,\msun$ implies $\fsn \gtrsim 0.35$. This is compatible with the current measurements of $\fsn \gtrsim 0.5$ based on direct supernova detection, estimates of supernova production in the Galaxy, star formation rate or diffuse neutrino background \citep[e.g.][]{kochanek08,kochanek14b,smartt09,horiuchi11,adams13,nakazato13,clausen14,mathews14}.

In the third panel of Figure~\ref{fig:dependencies}, we show with hatched bars the $68\%$ confidence intervals of the mean and Gaussian width of the observed population of galactic black holes. Our results are also broadly consistent with observationally determined mean and width of the BH mass distribution \citep{ozel10} assuming that BH masses are set by the progenitor helium core mass \citep{kochanek14a}. The lowest BH mass we obtain is $\sim 4\,\msun$ from a $\mathscr{M} \sim 15\,\msun$ progenitor.

\begin{figure}
 \plotone{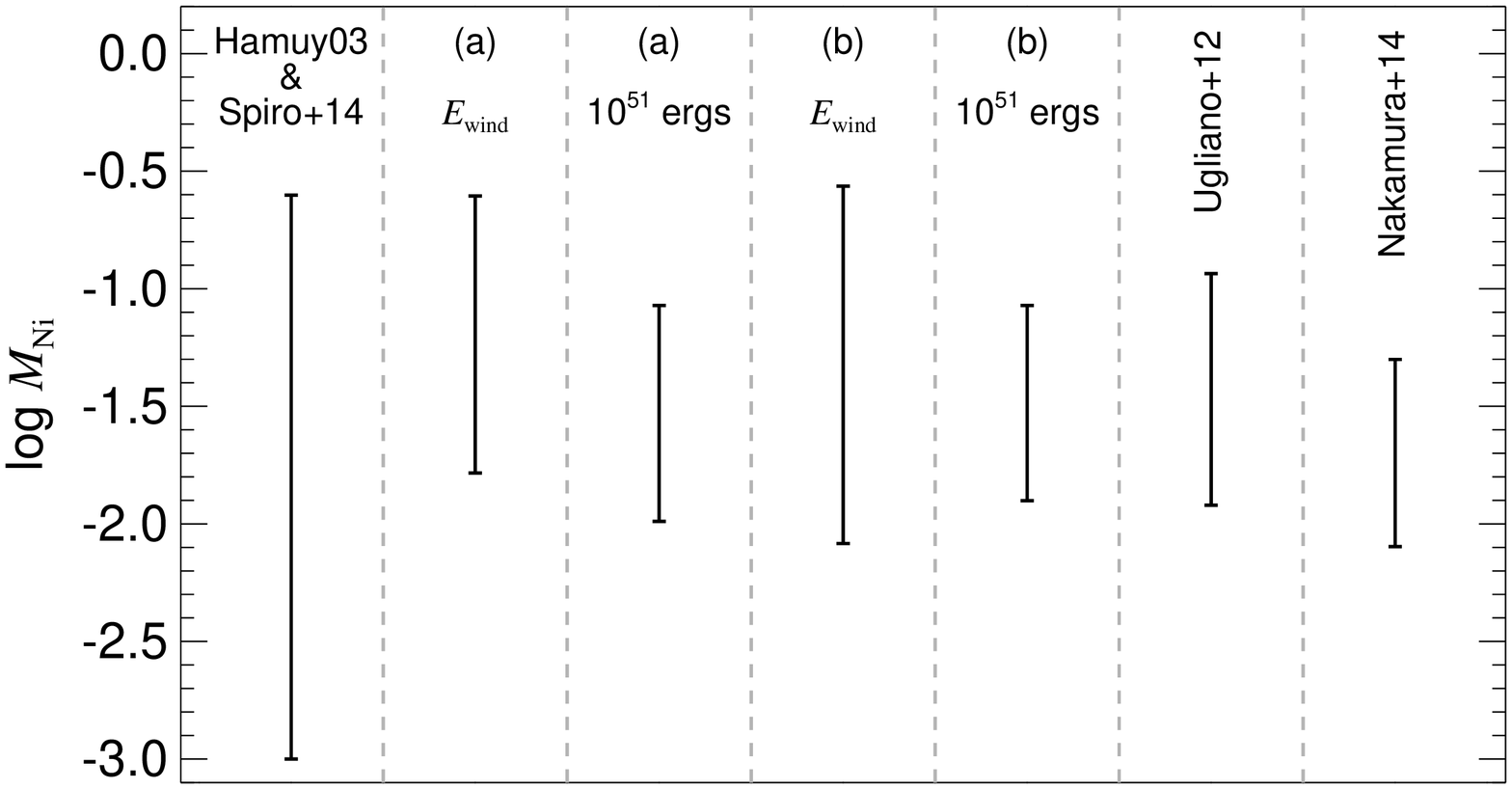}
\caption{Range of nickel yields observed in supernovae in the combined sample of \citet{hamuy03} and \citet{spiro14} and predicted theoretically in this work (parameterizations (a) and (b) with $\esn=\ewind$ or $\esn=10^{51}$\,ergs) and by \citet{ugliano12} and \citet[estimated from their Fig.~5]{nakamura14b}. The vertical bars show minimum and maximum of $\log \mni$ in each sample.}
\label{fig:ni_comp}
\end{figure}

In the left panel of Figure~\ref{fig:iip}, we show theoretical results from our work and \citet{ugliano12} with observations of explosion energies and nickel masses from \citet{hamuy03}. The observations show a clear correlation between the explosion energy and nickel masses, which is reproduced by our parameterizations (a) and (b), but is lacking in the results of \citet{ugliano12}. This correlation should also be present in \citet{nakamura14b}, but they do not give the slope and normalization in relation to observations.

The theoretical results severely underpredict the range of observed $\mni$, which is about $2.5$\,dex. This is summarized in Figure~\ref{fig:ni_comp}, which plots minimum and maximum values of $\mni$ of observations and theoretical results. The results are very similar if we use $5\%$ and $95\%$ quantiles instead. The theoretical results predict a range of $\mni$ smaller by $1$\,dex or more in some cases. All of the theoretical results do not include progenitors with $\mathscr{M} < 10.8\,\msun$ and these low-mass progenitors with tenuous envelopes likely produce little $\mni$ \citep{chugai00}. However, the simulations of \citet{kitaura06} of the $8.8\,\msun$ progenitor of \citet{nomoto84,nomoto87} produce $\log \mni \lesssim -1.8$, where the upper limit comes from the total mass of the ejecta. This limit is about $1$\,dex above what is observed in some supernovae. If $8\le \mathscr{M} \le 10\,\msun$ progenitors are responsible for low $\mni$ events, then they should comprise $\sim 27\%$ of core collapse and even a higher fraction of IIp supernovae. However, the observational estimates of low-luminosity SN frequency range from several per cent to $\sim 20\%$ \citep{chugai00,richardson02,richardson14,pastorello04} suggesting a somewhat smaller range of masses.

Within our models, the progenitors producing low $\mni$ do not come from a single well-defined range of initial masses, but are instead interweaved between high $\mni$ progenitors as seen in Figure~\ref{fig:nickel_prog}. In parameterization (a), the lowest $\mni$ are produced by moderate-mass progenitors with $13 \lesssim \mathscr{M} \lesssim 14\,\msun$, which comprise about $10\%$ of all successful explosions. This is in some agreement with \citet{spiro14}, who found that at least some low-luminosity type II supernovae come from intermediate-mass stars with $10 \lesssim \mathscr{M} \lesssim 15\,\msun$. In parameterization (b), only a couple progenitors with $\mathscr{M} \approx 20\,\msun$ produce low enough $\mni$ and the overall fraction is correspondingly smaller.

How robust are the observed quantities? Since the electromagnetic luminosity is tiny compared to the kinetic energy of the ejecta, obtaining the explosion energies is not easy. Often, the modeled ejecta masses (and thus possibly also the explosion energies) are overestimated \citep[e.g.][]{utrobin07,utrobin09,dessart11,dessart13}. Conversely, the nickel masses are relatively straightforward to determine from the normalization of the bolometric light curve during the optically-thin exponential decay. Hence, the uncertainties concerning extinction and distance probably dominate the systematic uncertainty in $\mni$ \citep{pp14}, provided that the $\gamma$-ray escape fraction does not vary drastically. Nonetheless, there is considerable diversity among type II supernovae as illustrated in the right panel of Figure~\ref{fig:iip}, where we plot $\mni$ against the plateau $V$-band luminosities of from \citet{hamuy03} and \citet{spiro14}. The observations clearly show a continuous distribution of $\mni$ spanning more than $2$\,dex.

We emphasize that our results on the range of $\mni$ and the $E_{\rm SN}$--$\mni$ and $E_{\rm SN}$--$M_{\rm NS}$ correlations do not require any fallback, which is not included within our model. These correlations are a generic feature and a natural consequence of the neutrino mechanism applied to a set of progenitor models. As a result, low-luminosity supernovae are an excellent probe of the neutrino mechanism and progenitor structure on the borderline between success and failure. For example, their explosion energy and $\mni$ should constrain $\mdot$ through the shock at the moment of explosion (Fig.~\ref{fig:nickel_mdot}).

Finally, as we point out here, the properties of supernova explosions are set by the progenitor structure in the central $\sim 2\,\msun$ and not by the initial mass. As a result, we do not predict any clear correlation with the initial progenitor mass (see Figures~\ref{fig:ns_mass}, \ref{fig:e_exp}, and \ref{fig:nickel_prog}), unlike what was found from observations by \citet{nomoto06} and \citet{poznanski13}.

\section{Constraining the neutrino mechanism with observations}
\label{sec:constraint}

In this Section, we illustrate how parameters of the neutrino mechanism can be directly constrained by observations. We regard this exercise as a proof of principle rather than definite answer due to various issues and caveats with producing observable quantities in our framework. In Figures~\ref{fig:dep_2d} we show the global observable parameters of sWHW02 progenitors on the grid of the parameters of the neutrino mechanism $(p,q)$ (Eq.~[\ref{eq:artificial}]).  We see that the landscapes of individual parameters are substantially different, which in principle allows empirical constraints on $(p,q)$. For example, high values of $\fsn$ favor low $p$ and high $q$, but observations would constrain $\mns$ to a narrow nearly-diagonal ridge. Combining just these two parameters will yield a constraint on $(p,q)$.

\begin{figure}
\plotone{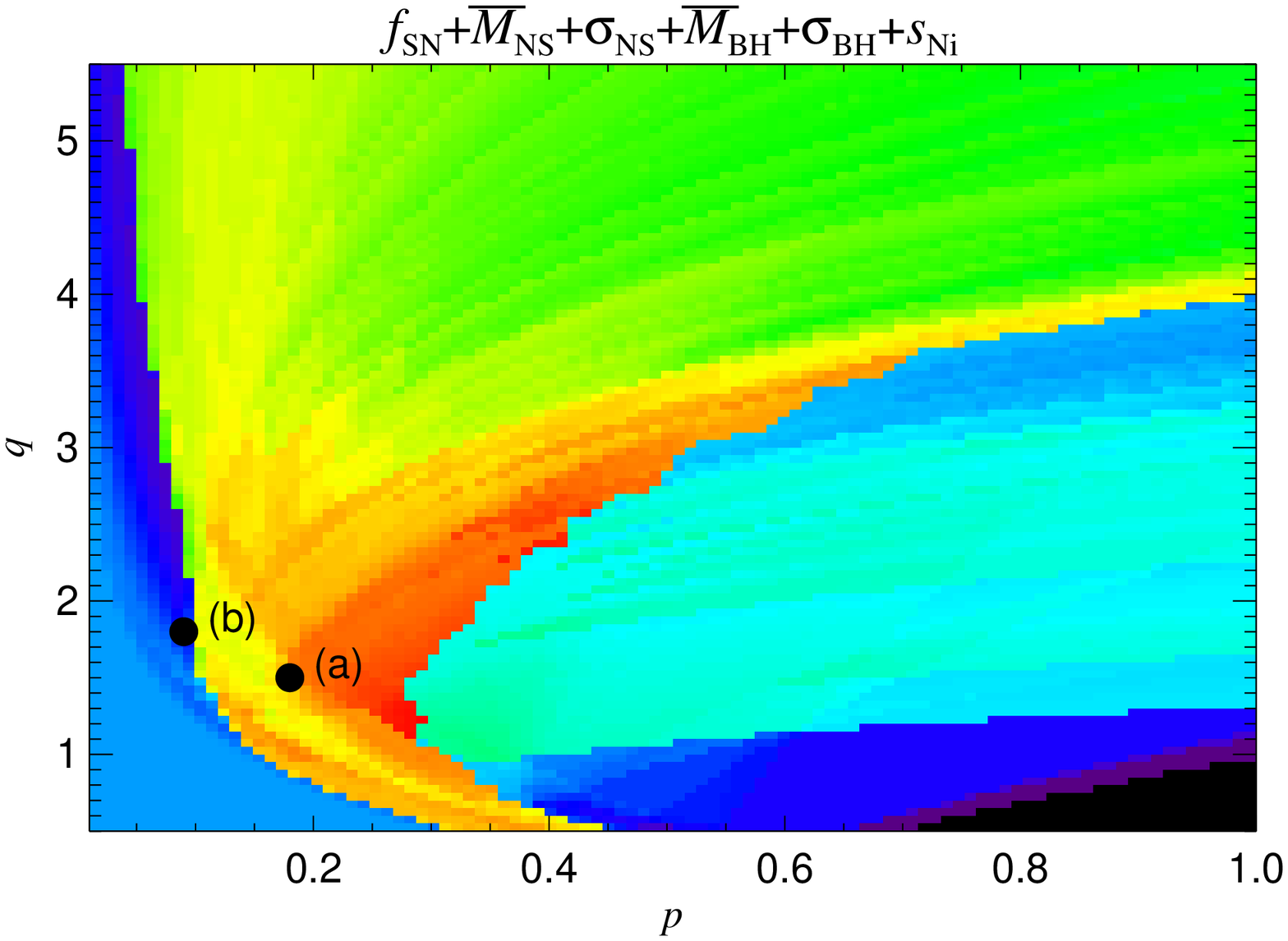}
\caption{Likelihood of the parameters of neutrino mechanism obtained by combining constraints on $\fsn$, $\mns$, $\sns$, $\mbh$, $\sbh$, and nickel mass distribution width $s_{\rm Ni}$ (see text for more details). The combined likelihood is on a logarithmic scale with lowest values denoted by black and blue colors and highest values by orange and red. The two points show the positions of the parameterizations (a) and (b). }
\label{fig:dep_weight_tot}
\end{figure}

To illustrate how a global constraint on the neutrino mechanism could be obtained in observations, we assign to each position in $(p,q)$ space a likelihood based on the value of observable quantities produced by sWHW02 progenitors for that particular parameterization. We assume a Gaussian likelihood with mean equal to the observed value and width equal to the uncertainty. Specifically, we use $\mns = 1.33\pm 0.03\,\msun$, $\sns = 0.05 \pm 0.03\,\msun$, $\mbh = 7.9 \pm 0.9\,\msun$, $\sbh = 1.475 \pm 0.725 \,\msun$, and $s_{\rm Ni} = \log (\max \mni / \min\mni) = 2.4 \pm 0.5$\,dex. We do not put constraints on $\esn$ due to the uncertainties in zero-point and width in our model. We also assign a uniform unity likelihood to $\fsn \ge 0.5$ and very small likelihood to $\fsn < 0.5$. Regarding likelihoods on each quantity as independent, we can combine them to produce a likelihood map for $(p,q)$, which we show in Figure~\ref{fig:dep_weight_tot}. We see that the combined constraint produces high likelihood in a ridge between $0.2 \le p \le 0.5$ and $1.2 \le q \le 3.0$. The right boundary of the ridge is due to the constraint $\fsn \ge 0.5$. We also see that our parameterization (a), chosen to resemble \citet{ugliano12}, is located close to the maximum likelihood, while parameterization (b) has a low likelihood resulting from no production of black holes. It is worth noting that the ratio of $p$ in parameterizations (a) and (b) is $2$, similar to the differences between 1D, 2D, and 3D simulations \citep[e.g.][]{nordhaus10,hanke12,takiwaki12,couch13b,dolence13}. Figure~\ref{fig:dep_weight_tot} shows that already in our rather simplistic inference, there are significant differences in overall likelihoods on this scale, potentially allowing a direct constraint on the supernova mechanism.

Posterior predictions of observables can be obtained by using Figure~\ref{fig:dep_weight_tot} to assign a weight to each parameterization $(p,q)$ and then constructing a weighted histogram using data in Figure~\ref{fig:dep_2d}. We find that this does not yield robust constraints, because the high-likelihood area at $(p,q) \approx (0.3,2)$ is overwhelmed by the moderate-likelihood, but significantly larger, area at $q \gtrsim 3$. This is due to the fact that we have implicitly used a uniform prior on $p$ and $q$. In principle, with better understanding of the neutrino mechanism it is either possible to assign weights or narrow down individual parameters leading to a better constraint on the mechanism.

Finally, we emphasize that there is a long way to go before a supernova model that ``fits'' the observations is obtained. The current model has tensions between some observables. For example the small $\sns$ favors lower right corner of $(p,q)$ space, while $\mbh$ favors the upper left corner. Better constraints on the black hole and neutron star mass function would definitely be helpful in this aspect.

\section{Discussions and Conclusions}
\label{sec:conclusions}

We have constructed a semi-analytic model of core-collapse supernova explosions that is inexpensive to calculate and predicts success or failure of core collapse, remnant masses, explosion energies and nickel yields. We aimed to construct a model based on a minimum set of assumptions: (i) the progenitor structure fully determines the core collapse outcome by specifying the instantaneous boundary conditions of a sonically-connected region below the accretion shock and by driving the long-term evolution by neutrino diffusion through the initial plus accreted mass; (ii) the supernova accretion phase is quasi-static; (iii) the neutrino mechanism works and the necessary critical neutrino luminosity is lowered by systematic effects like turbulence and SASI, but also by potentially yet unknown processes such as rotation, magnetic fields and pre-collapse turbulence in the progenitor. The decrease of critical luminosity is similar for progenitors with similar structure. Specifically, we assume that the fractional reduction of critical neutrino luminosity is the same for all progenitors at a given mass accretion rate. This fractional reduction is monotonic in $\mdot$ and described by two parameters $(p,q)$ (Eq.~[\ref{eq:artificial}]); (iv) explosion occurs when the actual PNS luminosity achieves the reduced critical luminosity.

We apply this core collapse decision logic to more than $400$ progenitors from two groups with three different metallicities  and four equations of state (Table~\ref{tab:progenitors}, Figures~\ref{fig:progenitors} and \ref{fig:limongi}). We make predictions for observable quantities. For successful explosions, the remnant mass is equal to the PNS at the moment of explosion corrected for binding energy release (Fig.~\ref{fig:ns_mass}). For failures, the remnant is a black hole with a mass equal to the mass of the helium core of the progenitor star (Fig.~\ref{fig:rem_dist}). The explosion energy is estimated as the time integrated power of the neutrino-driven wind (Fig.~\ref{fig:e_exp}) based on the PNS characteristics of each progenitor. Nickel mass is proportional to the mass swept up by the shock to the radius of silicon burning for given explosion energy (Fig.~\ref{fig:nickel_prog}).

We find that successful explosions and failures are intertwined in a complex pattern unrelated to the initial progenitor mass (Fig.~\ref{fig:summary}). For a parameterization of the neutrino mechanism devised to match the similar previous study of \citet{ugliano12}, we obtain very good match for the outcome of core collapse and remnant masses of individual progenitors (Fig.~\ref{fig:ugliano_comp_m}). The agreement of explosion energies and nickel yields is poor on an individual bases, but we reproduce the global trends such as the anticorrelation between remnant mass and explosion energy (Fig.~\ref{fig:cor_m_e}).

By varying the parameterization of the neutrino mechanism $(p,q)$ we obtain a range of outcomes of the observable quantities (Fig.~\ref{fig:lratio}). We found that for a population of progenitors, a wide range of parameterizations robustly yields remnant masses broadly compatible with observations (Fig.~\ref{fig:dependencies}). We show that the mean and width of the neutron star mass distribution are correlated and that the NS mean mass is correlated with the mean black hole mass. Measuring mean neutron star mass immediately implies a constraint on the fraction of successful explosions, for example, $\mns = 1.33\,\msun$ suggests $\fsn \gtrsim 0.35$. We qualitatively reproduce the correlation between explosion energies and nickel masses, but fail to achieve the lowest observed nickel masses, similar to other theoretical studies (Fig.\ref{fig:iip}). We find that these low nickel masses are produced primarily by progenitors with low propensity for explosions close to the success/failure boundary. Low luminosity supernovae are thus an excellent probe of the neutrino mechanism and progenitor structure near the border of black hole production. 

The parameterizations of the neutrino mechanism to obtain successful explosions differ for the considered equations of state, but the relative outcomes of individual progenitors as well as the bulk population properties remain very similar (Fig.~\ref{fig:prog_frac}). We find that progenitors with $Z=10^{-4}\,\msun$ are difficult to explode compared to solar and zero metallicity progenitors, if these stars explode by the neutrino mechanism (Figs.~\ref{fig:ns_mass} and \ref{fig:fbh}). This is potentially important for understanding the affinity of gamma-ray burst to low-metallicity environments.

As a proof of principle, we evaluated the relative likelihood of the different parameterizations of the neutrino mechanism by comparing the global properties such as neutron star and black hole masses and nickel yields to observations. This exercise yields a meaningful area of high probability (Fig.~\ref{fig:dep_weight_tot}), which includes our parameterization (a). Some of the observables like mean the black hole mass and the width of the neutron star mass distribution appear to be somewhat mutually inconsistent in the sense that each observable favors different completely different part of the parameterization space $(p,q)$. More work is definitely needed to provide more robust ways of determining the observable properties and to ensure correct comparison to observations.

We study the progenitor ``compactness'' $\xi_M$ (Eq.~[\ref{eq:compactness}]) as a measure of the propensity of a star to produce a successful supernova explosion. We find that $\xi_M$ is well-defined if parameterized by a mass coordinate fixed for all progenitors. A large range of mass coordinates gives similar results. However, compactness can be evaluated at composition interfaces or at specified abundance thresholds. We find that several such definitions give wildly different results from $\xi_M$ based on mass (Figures~\ref{fig:compactness} and \ref{fig:compactness_limongi}). We compare our determination of the progenitor propensity to explode (Fig.~\ref{fig:prog_frac}) based on averaging all of our parameterizations with the considered definitions of compactness and find that the best agreement is with compactness based on mass (Tab.~\ref{tab:spearman}) as found also by \citet{kochanek14b} using the observed black hole mass function. The compactness of individual progenitors, however, might poorly predict outcomes of any single neutrino mechanism parameterization, specifically, the compactness can correctly predict success or failure for at most $88\%$ of progenitors in parameterization (a), even if both $M$ and $\xi_M$ are varied. From this point of view, the ``critical compactness'' used in studying black hole formation \citep[e.g.][]{oconnor11,kochanek14b} is only an approximate estimate of an outcome of the action of the neutrino mechanism on a set of progenitors, and would fail to predict the successful explosion of many progenitors (see Sec.~\ref{sec:compactness}).

We showed that our model reproduces the basic observed properties of the supernova population, but there are many effects that we in principle cannot include. For example, the assumption of quasi-static evolution might not be justified for some progenitors and this can have effect on how the explosion develops \citep[e.g.[]{janka01}. We saw indications of that in Figure~\ref{fig:rshock}, where the progenitors s11.2WHW02 and s11.4WHW02 accreted a density jump much earlier than other progenitors. Nonetheless, our assumption of early explosions in these two progenitors does not give results too discrepant from time-dependent calculations of \citet{ugliano12}. In other progenitors, the quasi-static approximation might be violated more insidiously, because the density drops are accreted later. Nonetheless, by comparing shock radii from time-dependent calculations with \gr\ and our steady-state code in Figure~\ref{fig:rshock}, we find little disagreement for $\mdot < 1\,\msun$\,s$^{-1}$ other than for the two anomalous progenitors. From Figure~\ref{fig:lratio} we see that most explosions in parameterization (a) do indeed  occur for sufficiently small $\mdot$. It is worth emphasizing that the density drops in supernova progenitors might be more pronounced both in spherical average and in individual directions as a result of unstable convective burning just before the collapse \citep{arnett11}. The response of the neutrino mechanism to accretion of significant radial and angular density variations is only beginning to be explored \citep{couchott13,couchott14}.

Our quasi-static supernova model does not include fallback, which is inherently a time-dependent phenomenon. In terms of the observables we considered, fallback will necessarily increase the mass of the remnant and lower the mass of the ejecta. In some cases, a large amount of fallback will convert the proto-neutron star to a black hole. Fallback will also reduce the mass of ejected nucleosynthetic products including \nic. However, there are several reasons why we think fallback is not the principal ingredient in forming the properties of compact objects and supernovae. First, the usual simulations based on a momentum or energy bomb at a specific progenitor mass coordinate \citep[e.g.][]{woosley95,thielemann96,timmes96} do not take into account the details of pushing the ejecta out on a timescale $\gtrsim 1$\,s by the neutrino-driven wind forming the double-shock structure \citep[e.g.][]{arcones07,arcones11}. In fact, the recent 1D simulations by \citet{ugliano12} of neutrino-driven explosions initiated consistently in a set of progenitors show fallback $\lesssim 0.1\,\msun$ for most progenitors. Only for one progenitor was the fallback strong enough to make a black hole after a successful supernova explosion. Second, we are able to reproduce observed values and their correlations without resorting to fallback. Third, observations of the remnant mass functions prefer no fallback \citep{pejcha_ns,kochanek14a,kochanek14b}. However, abundance analysis and proper motions of binary black holes suggest a successful supernova explosion in the primary star \citep[e.g.][]{israelian99,orosz01,willems05,gonzalez06,wong12}. Thus, the time-dependent physics of transition from accretion to explosion needs to be calculated to understand what sets supernova explosion energies and nucleosynthetic yields and how to reproduce the observed parameters.

Fallback in relatively high-mass progenitors has been suggested as an explanation of low $\mni$ in some low-luminosity supernovae \citep{turatto98,zampieri98}. We do not favor this explanation, because it implies a fine-tuning problem: since most of the nickel from the burned silicon layer must fall back to the remnant it is not clear what guarantees that at least some nickel is ejected. In other words, why are there no supernovae with no ejected nickel? Second, low-luminosity supernovae also lie on the same explosion energy-nickel mass correlation (Fig.~\ref{fig:iip}) suggesting their low $\mni$ is an extreme version of processes that determine $\mni$ in more ordinary supernovae. Third, one of the main points of this paper is that it is not possible to arbitrarily assign explosion energies and ejecta mass cut offs to individual progenitors, but instead the structure of each progenitor implies what these quantities should be in relation to other progenitors. 

We assumed that the combination of physical effects that lowers the critical neutrino luminosity compared our quasi-static 1D calculation is a function of only $\mdot$. In principle, the reduction of $\lcrit$ depends also on the mass and radius of the PNS and likely also on the elapsed time to allow for the growth of instabilities. In principle, all these variables can be included and parameterized in a form similar to Equation~(\ref{eq:artificial}) and the parameters can be easily marginalized over like we did in Figures~\ref{fig:dependencies}, \ref{fig:dep_2d}, and \ref{fig:dep_weight_tot}. However, we think that the resulting changes will be only minimal, because the variations in these additional parameters are relatively small and correlated with $\mdot$. Furthermore, we explored a very wide range of parameterizations and found that the results are quite robust. This will likely not change significantly with additional degrees of freedom.

A multi-dimensional effect that can completely change the answers is the possibility of simultaneous accretion and explosion, where the accretion can give an additional boost to the explosion. If important, this mechanism could achieve high explosion energies and high NS masses. However, the problem with the neutrino mechanism is that observations require relatively early explosions with relatively small NS masses.

We operated under the assumption that the neutrino mechanism, possibly modulated by other physical processes, is {\em the} mechanism of supernova explosions and constructed the artificial explosions to explore its predictions. Although the knowledge of the ``true'' values of $(p,q)$ in the language of Equation~(\ref{eq:artificial}) need to come from self-consistent multi-dimensional simulations rich in physics, our results strongly suggest that there are indeed many choices of $(p,q)$ giving results compatible with observations (except extreme cases like hypernovae and gamma-ray bursts) and that the neutrino mechanism is robust. 

A method similar to the model developed here is indispensable in comparing the theory with observations, because the theory ought to reproduce the full range of observed parameters. In other words, not all supernovae explode with $E_{\rm SN} =10^{51}$\,ergs and $\mni = 0.07\,\msun$, there is a spread of more than two orders-of-magnitude in these quantities. Drawing any conclusions from a sophisticated simulation of a single or a few progenitors with a single set of physics is impossible, because it is uncertain how to generalize to a population of progenitors and whether some small change in the input physics would not have dramatic consequences. Similarly, extensive parameter studies of a single progenitor offer limited information, because in many cases the parameterization can be tuned to achieve a wide range of results. For example, \citet{yamamoto13} found a correlation between explosion energy and nickel mass in parameterized explosions of a single $15\,\msun$ progenitor. However, there is no guarantee that this correlation transfers to a population of progenitors, where each has different structure, moment of explosion and different explosion energy. Furthermore, progenitor structure measured by the compactness can change by large factors in response to relatively mild changes in the input physics as shown by \citet{west13}. Therefore, we think that the only way to verify the neutrino mechanism is with ensembles of parameterized simulations of progenitor populations as was done here and by \citet{ugliano12}, \citet{nakamura14b}, and \citet{suwa14}. 

\acknowledgements
We are very grateful to Evan O'Connor for discussions and help with \gr. We thank Chris Kochanek for discussions, comments and detailed reading of the manuscript. We thank Adam Burrows, Josh Dolence, Sean Couch for discussions. We thank Thomas Janka for providing us with machine-readable results of \citet{ugliano12}, and Bernhard M\"{u}ller for comments on the manuscript. We thank the anonymous referee whose comments helped to improve the manuscript. Support for Program number HST-HF-51327.01-A was provided by NASA through a Hubble Fellowship grant from the Space Telescope Science Institute, which is operated by the Association of Universities for Research in Astronomy, Incorporated, under NASA contract NAS5-26555.


\appendix

\section{Estimates of energy of neutrino-driven wind}
\label{app:nu_wind}

There are several practical problems when calculating the time-integrated neutrino-driven wind power according to Equation~(\ref{eq:pwind}). The first problem is the choice of numerical values of the coefficients in Equation~(\ref{eq:pwind}). \citet{burrows93} rescale the results of \citet{duncan86} to obtain $A=4.2\times 10^{51}$\,\ergs, $\kappa = 3.5$, $\lambda=2.0$, and $\mu=4/3$ implying that $\pwind$ is very sensitive to the neutrino luminosity. \citet{thompson01} performed more detailed calculations of the neutrino-driven wind and found $\kappa \approx 3.2$ to $3.4$. However, the value of $A$ implied by Table~1 of \citet{thompson01} is more than order of magnitude smaller than $A$ of \citet{burrows93}, implying completely negligible $\pwind$. A similar discrepancy is present with the low-magnetization slow-rotation results of \citet{metzger07}. The normalization of \citet{burrows93} is probably more appropriate, because it roughly predicts $\pwind \sim 10^{51}$\,\ergs\ for $\lnue \sim 2\times 10^{52}$\,\ergs\ in the time-dependent simulations of \citep[e.g. Fig.~C3 of][]{scheck06}, where the luminosities and wind powers are sustained over a timescale of a few $100$\,ms.

\begin{figure}
 \plotone{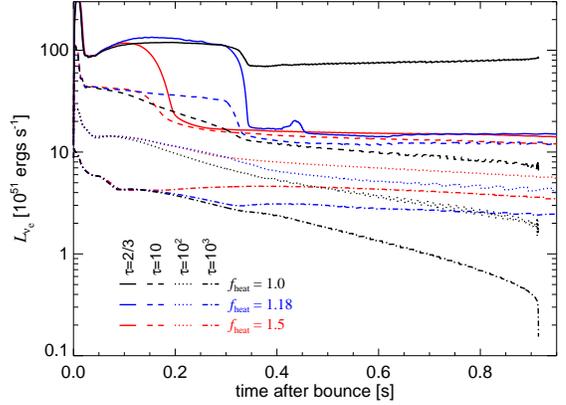}
\caption{Neutrino luminosity $\lnue$ evolution for the s23.0WHW02 progenitor evolved with LS220 EOS and $\fheat =1.0$ (black), $1.18$ (blue), and $1.5$ (red). The luminosity is evaluated at the neutrinosphere (solid), $\tau=10$ (dashed), $10^2$ (dotted), and $10^3$ (dash-dotted).}
\label{fig:single_prog_lnue}
\end{figure}

Second, due to the high sensitivity of $\pwind$ to the neutrino luminosity, care must be taken to supply the time evolution of $\lnue$. In particular, we are interested in estimating $\lnue$ of exploding models from data on the default non-exploding models. In Figure~\ref{fig:single_prog_lnue} we illustrate some of the challenges with the s23.0WHW02 model evolved using the LS220 EOS. Results for other progenitors are very similar. Our default calculation (black lines) forms a BH $\sim 0.9$\,s after bounce. We also show two calculations with enhanced neutrino heating parameterized by $\fheat$ \citep{oconnor10,oconnor11} that produce explosions at two different times\footnote{We failed to achieve explosions occurring later than $0.35$\,s after bounce despite adjusting $\fheat$ to a relative accuracy of about $10^{-4}$. This moment of time corresponds to a significant drop in $\mdot$.}, $0.2$ and $0.35$\,s after bounce, separated by about $150$\,ms and resulting in baryonic NS masses of $1.87$ and $2.15\,\msun$. The non-exploding model experiences a drop in the neutrinospheric $\lnue$ (solid black line) at about $0.35$\,s after bounce, when a drop in $\mdot$ is encountered. This suggests that the contribution of accretion luminosity proportional to the instantaneous $\mdot$ is significant at the neutrinosphere. This influence is much weaker deeper in the PNS as illustrated by $\lnue$ evaluated at $\tau=10$ (dashed black line), where $\tau$ is the optical depth to electron neutrino absorption. The accretion luminosity disappears during the transition from accretion to explosion and the luminosity correspondingly drops (red and blue solid lines). Interestingly, simulations with different values of $\fheat$ converge\footnote{We found that in many cases $\lnue$ at the neutrinosphere experiences suspicious unphysical jumps up as is visible in a relatively mild case in the $\fheat = 1.18$ calculation at about $0.45$\,s after bounce. These features are not present at higher $\tau$.} to an essentially constant value of $\lnue$. We find that this value is nearly constant for all progenitors, but depends on the EOS. The drop in $\lnue$ is smaller at $\tau=10$ and at $\tau=10^2$ and $10^3$ the luminosity increases after the explosion. 

\begin{figure}
 \plotone{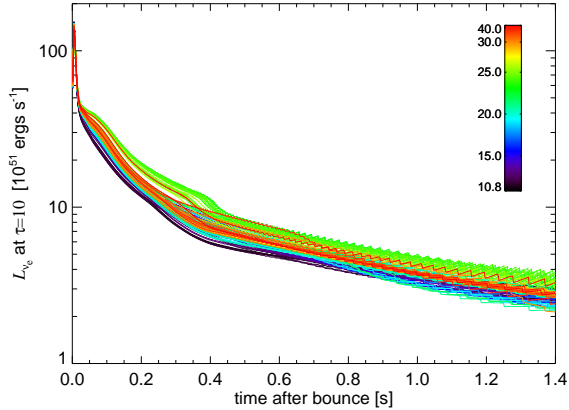}
\caption{Neutrino luminosities at $\tau=10$ in sWHW02 progenitors evolved with the HSHen EOS.}
\label{fig:lumtau}
\end{figure}

For the purposes of predicting the neutrino luminosity evolution after explosion from the non-exploding model, we note in Figure~\ref{fig:single_prog_lnue} that accretion luminosity is unimportant at $\tau=10$ in the non-exploding model and that $\lnue(\tau=10)$ is very similar for exploding and non-exploding models. We thus take the following approach to determine the contribution of the neutrino-driven wind to the explosion energy. We assume that the accretion luminosity during the transition from accretion to explosion does not contribute significantly to the explosion energy. Equation~(\ref{eq:pwind}) is not valid before the quasi-static wind is established and applying its high value of $\kappa$ to the potentially high value of accretion luminosity means that the transient could overwhelm the long-term contribution from the wind in an undefined way. This is not observed: time-dependent simulations instead show that wind power over many $100$\,ms is the dominant part of $\ewind$ \citep[Fig.~C3 of][]{scheck06}. We use the time evolution of $\lnue(\tau=10)$ from the default non-exploding models from the moment of explosion as input for Equation~(\ref{eq:pwind}) and we show it for sWHW02 progenitors in Figure~\ref{fig:lumtau}. The evolution is very similar for many different progenitors and the gradual decay in all cases ensures the desired result that later explosions will generally be less energetic. We assume that $M$ is constant and $\rnue$ after explosion evolves according to
\beq
\rnue(t) = \rnue^{\rm f} + (\rnue^{\rm i}-\rnue^{\rm f}) \exp\left[-\frac{(t-\texp)}{t_{\rm wind}}\right],
\label{eq:rnu_time}
\eeq
where $\rnue^{\rm i} = \rnue(\texp)$, $\rnue^{\rm f} = 10$\,km, and $t_{\rm wind} = 1$\,s. Due to low values of $\lambda$ and $\mu$ compared to $\kappa$, $\ewind$ is insensitive to the exact values or evolution of $M$ and $\rnue$. In Equation~(\ref{eq:pwind}), we choose to keep the power-law indices $\kappa$, $\lambda$, and $\mu$ to the values of \citet{burrows93} and we choose $A$ so that the typical explosion energies are around $10^{51}$\,ergs. This adjustment is necessary, because neutrino luminosities at $\tau=10$ are lower than at the neutrinosphere.

\begin{figure}
 \plotone{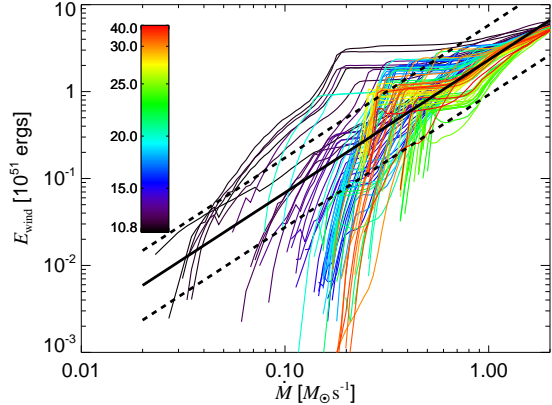}
\caption{Energy of the neutrino-driven wind for sWHW02 progenitors evaluated as a function of $\mdot$. $\ewind$ was calculated by time-integration of Equation~(\ref{eq:pwind}) assuming that an explosion initiated at each value of $\mdot$. The best fit to the data of $\esn \propto \mdot^{1.5}$ is shown with a thick black line along with the standard deviation around the fit of $0.4$\,dex indicated by dashed lines.}
\label{fig:expe_mdot}
\end{figure}

In Figure~\ref{fig:expe_mdot}, we show our estimates of $\ewind$ for sWHW02 progenitors if the explosion occurs at a particular time parameterized by the instantaneous $\mdot$. As expected, $\ewind$ is smaller at later times when the luminosity driving the wind is smaller. We also see that for most progenitors $\ewind$ stays relatively flat for a range of $\mdot$ and then drops precipitously. This is due to a sudden drop in $\mdot$ as a shell interface is accreted, which occurs over short-enough time that $\lnue(\tau=10)$ is relatively unaffected. If explosion occurs after accretion of the shell interface, it is likely to be very weak. Overall, energy of the neutrino-driven wind scales as $\ewind \propto \mdot^{1.5}$ with a considerable scatter of about $0.4$\,dex.

Finally, we emphasize that the procedure to obtain $\ewind$ presented  here is by no means unique or fully correct. We experimented with many different prescriptions for the post-explosion evolution of $\lnue$ and we found that the differences were predominantly in the normalization $A$ in Equation~(\ref{eq:pwind}) required to approximately match the typical supernova energies, and in the width of the distribution. In other words, our formalism can rather reliably differentiate between high and low $\ewind$, but the absolute values and ratios are much more uncertain. For example, in all our attempts, sWHW02 progenitors with initial masses between $13$ and $15\,\msun$ had lower energies than progenitors with $<13\,\msun$. In most cases, prescriptions for $\ewind$ were able to approximately reproduce the expected trend of later explosions having smaller energies, and the correlation between $\ewind$ and the recombination energy. When relevant, we also consider cases with $\esn \equiv 10^{51}$\,ergs in subsequent discussions to show how the results depend on the explosion energy.

\section{Internal consistency of the compactness parameter}
\label{app:compactness}

Although the compactness parameter defined in Equation~(\ref{eq:compactness}) has been used to study supernova progenitors and their explosions, little attention has been devoted to evaluating whether $\xi_M$ is well-defined internally in the sense that ordering of progenitors by $\xi_M$ is preserved for different values of $M$.

\begin{figure}
\plotone{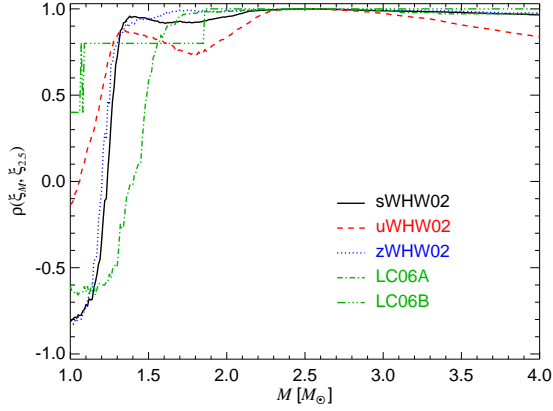}
\caption{Internal consistency of compactness expressed through the rank correlation coefficient between $\xi_{2.5}$ and $\xi_M$, $1 \le M \le 4\,\msun$, for sWHW02 (solid black line), uWHW02 (dashed red), zWHW02 (dotted blue), LC06A (dash-dotted green), and LC06B (dash-dot-dot green).}
\label{fig:compact_2.5}
\end{figure}

\begin{figure}
\plotone{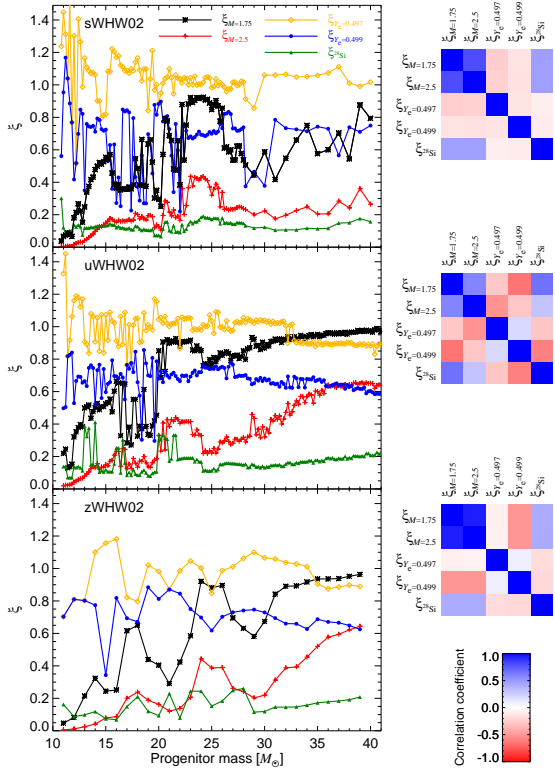}
\caption{Compactness of \citet{woosley02} progenitors evaluated at $M=1.75\,\msun$ (black crosses), $M=2.5\,\msun$ (red plus), $\ye=0.497$ (orange diamonds), $\ye=0.499$ (blue circles), and at the mass shell where the abundance of $^{28}$Si gets below $10^{-3}$ (green triangles). Each panel is for a different metallicity as given in Table~\ref{tab:progenitors}, and the side maps show Spearman rank correlation coefficients between individual sets of compactness displayed in a matrix, where blue and red indicate correlation and anti-correlation, respectively, as explained on the colorbar. }
\label{fig:compactness}
\end{figure}

\begin{figure}
\plotone{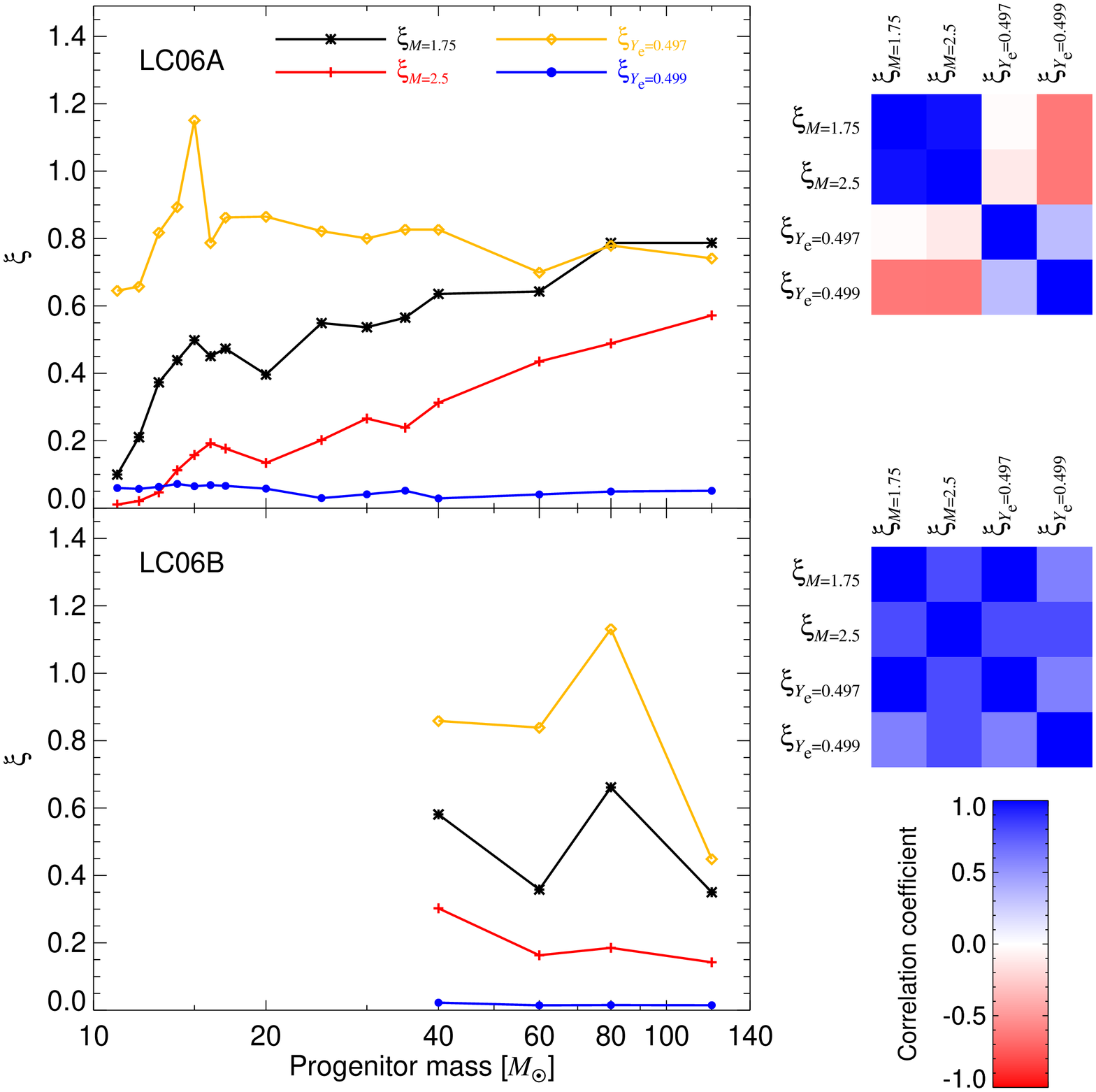}
\caption{Same as in Figure~\ref{fig:compactness}, but for \citet{limongi06} progenitors.}
\label{fig:compactness_limongi}
\end{figure}

\begin{figure}
\plotone{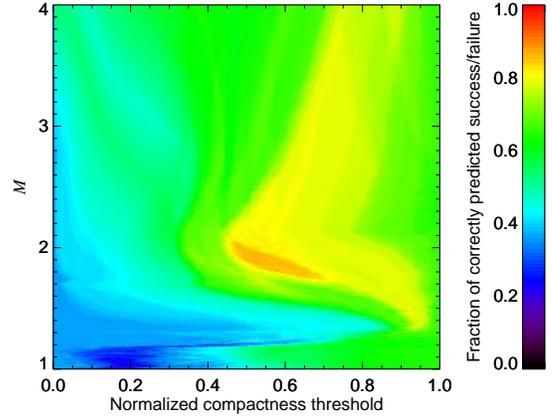}
\caption{Fraction of correctly predicted outcome (successful explosion or failure) in parameterization (a) as a function of critical threshold in $\xi_M$ and $M$. For each $M$, the critical $\xi_M$ is varied between the smallest and largest value in sWHW02 progenitors and normalized to this range. The highest achieved fraction of correct predictions is $88\%$ for $1.7 \le M \le 2.2\,\msun$.}
\label{fig:comp2d}
\end{figure}

Within our model, we calculate $\xi_M$ by obtaining the mass coordinate and radius of the shock immediately after the core bounce, and then adjusting the radii of the progenitor mass shells relative to the mass element at the shock and assuming one quarter of the free fall acceleration as in Section~\ref{sec:nickel}. For each of the sWHW02 progenitors, we calculate $\xi_{M_i}$ at $300$ values of $M_i$ equidistantly spaced between $1.0$ and $4.0\,\msun$. To study the internal consistency, we calculate the Spearman's rank correlation coefficient $\varrho_{ij} = \varrho(\xi_{M_i},\xi_{M_j})$ for each pair of progenitor ensembles of $(\xi_{M_i},\xi_{M_j})$ with $M_i \ne M_j$. The rank correlation coefficient is particularly useful for this goal, because it compares a rank of quantities rather than their absolute number and can thus take into account non-linear correlations. We repeated our calculations with Kendall's $\tau$ coefficient and found little change in our conclusions. In Figure~\ref{fig:compact_2.5}, we show a slice through $\varrho_{ij}$ with respect to the most common $\xi_{2.5}$. We see that for all WHW02 progenitors, $\varrho > 0.9$ for $M > 1.4\,\msun$, which corresponds to the mass scale of the iron core. This implies very good internal consistency of compactness defined with respect to the mass coordinate. Specifically, it does not matter much whether one uses $\xi_{2.5}$ \citep{oconnor11}, $\xi_{1.75}$ \citep{oconnor13}, or $\xi_{1.5}$ \citep{nakamura14b} -- the results should be very similar. The range of good internal consistency is narrower for LC06 progenitors, although their small number precludes making a sufficiently robust conclusion.

Here we have argued in agreement with \citet{ugliano12} that the explosion should primarily develop at the composition interface and the associated density drop. Furthermore, the remnant masses of different progenitors can be quite different. This implies that using a single mass $M$ in the definition of compactness for all progenitors might be too simplistic. To reflect this issue, we generalize Equation~(\ref{eq:compactness}) and calculate $\xi_{\ye=0.497}$ and $\xi_{\ye=0.499}$ measured at the mass coordinate corresponding to these specific $\ye$ chosen to reflect composition interfaces, and $\xi_{^{28}{\rm Si}}$ measured at the mass coordinate where the abundance of $^{28}$Si drops below $10^{-3}$. In Figures~\ref{fig:compactness} and \ref{fig:compactness_limongi}, we show these alternative definitions of compactness along with $\xi_{1.75}$ and $\xi_{2.5}$ considered by \citet{oconnor11,oconnor13}. In addition, we also display the rank correlation matrix of the plotted compactnesses. We see that the behavior of these alternative definitions is quite different from $\xi_{1.75}$ and $\xi_{2.5}$. In fact, $\xi_{1.75}$ and $\xi_{2.5}$  are well mutually correlated as also discussed above, but the correlation with $\xi_{^{28}{\rm Si}}$ is much weaker and there is in fact an anti-correlation with $\xi_{\ye=0.497}$ and $\xi_{0.499}$. This is particularly striking for $\xi_{1.75}$ and $\xi_{\ye=0.499}$, which have comparable numerical values implying similar position within the star, but the behavior of $\xi$ is completely different. We find this for all metallicities of WHW02 and for LC06A progenitors. This illustrates that other reasonable choices for the position where compactness is evaluated lead to very different results and it is not clear which parameterization should actually be preferred. \citet{kochanek14b} found that $\xi_{2.0}$ and $\xi_{2.5}$ provide much better match with the observed black hole mass function than $\xi_{3.0}$ or the iron core mass or the mass enclosed by the oxygen burning shell.

In Figure~\ref{fig:comp2d}, we show a map of how well can a condition on the compactness predict the outcomes in our parameterization (a). For each $M$, we varied the critical compactness threshold between the smallest and largest $\xi_M$ in sWHW02 progenitors and recorded the fraction of correctly predicted outcomes, which are either successful explosion (compactness smaller than critical) or failure (compactness larger than critical). The highest achieved agreement fraction is $88\%$.


\begin{thebibliography}{}
\bibitem[Adams et al.(2013)]{adams13} Adams, S.~M., Kochanek, C.~S., Beacom, J.~F., Vagins, M.~R., \& Stanek, K.~Z.\ 2013, \apj, 778, 164 
\bibitem[Arcones et al.(2007)]{arcones07} Arcones, A., Janka, H.-T., \& Scheck, L.\ 2007, \aap, 467, 1227 
\bibitem[Arcones \& Janka(2011)]{arcones11} Arcones, A., \& Janka, H.-T.\ 2011, \aap, 526, A160 
\bibitem[Arnett \& Meakin(2011)]{arnett11} Arnett, W.~D., \& Meakin, C.\ 2011, \apj, 733, 78 
\bibitem[Bailyn et al.(1998)]{bailyn98} Bailyn, C.~D., Jain, R.~K., Coppi, P., \& Orosz, J.~A.\ 1998, \apj, 499, 367 
\bibitem[Baron et al.(1985a)]{baron85a} Baron, E., Cooperstein, J., \& Kahana, S.\ 1985a, Physical Review Letters, 55, 126 
\bibitem[Baron et al.(1985b)]{baron85b} Baron, E., Cooperstein, J., \& Kahana, S.\ 1985b, Nuclear Physics A, 440, 744 
\bibitem[Baron et al.(1989)]{baron89} Baron, E., Myra, E.~S., Cooperstein, J., \& van den Horn, L.~J.\ 1989, \apj, 339, 978 
\bibitem[Baron \& Cooperstein(1990)]{baron90} Baron, E., \& Cooperstein, J.\ 1990, \apj, 353, 597 
\bibitem[Bethe \& Wilson(1985)]{bethe85} Bethe, H.~A., \& Wilson, J.~R.\ 1985, \apj, 295, 14 
\bibitem[Bethe(1990)]{bethe90} Bethe, H.~A.\ 1990, Reviews of Modern Physics, 62, 801
\bibitem[Blondin et al.(2003)]{blondin03} Blondin, J.~M., Mezzacappa, A., \& DeMarino, C.\ 2003, \apj, 584, 971 
\bibitem[Blondin \& Mezzacappa(2006)]{blondin06} Blondin, J.~M., \& Mezzacappa, A.\ 2006, \apj, 642, 401 
\bibitem[Bruenn(1988)]{bruenn88} Bruenn, S.~W.\ 1988, \apss, 143, 15 
\bibitem[Bruenn et al.(2001)]{bruenn01} Bruenn, S.~W., De Nisco, K.~R., \& Mezzacappa, A.\ 2001, \apj, 560, 326 
\bibitem[Bruenn et al.(2013)]{bruenn13} Bruenn, S.~W., Mezzacappa, A., Hix, W.~R., et al.\ 2013, \apjl, 767, L6 
\bibitem[Buras et al.(2006)]{buras06} Buras, R., Rampp, M., Janka, H.-T., \& Kifonidis, K.\ 2006, \aap, 447, 1049 
\bibitem[Burrows(1987)]{burrows87} Burrows, A.\ 1987, Physics Today, 40, 28
\bibitem[Burrows \& Lattimer(1983)]{burrows83} Burrows, A., \& Lattimer, J.~M.\ 1983, \apj, 270, 735 
\bibitem[Burrows \& Goshy(1993)]{burrows93} Burrows, A., \& Goshy, J.\ 1993, \apjl, 416, L75 
\bibitem[Burrows et al.(1995)]{burrows95} Burrows, A., Hayes, J., \& Fryxell, B.~A.\ 1995, \apj, 450, 830 
\bibitem[Burrows et al.(2007a)]{burrows07a} Burrows, A., Dessart, L., \& Livne, E.\ 2007a, Supernova 1987A: 20 Years After: Supernovae and Gamma-Ray Bursters, ed. S. Immler \& R. McCray, 370-380
\bibitem[Burrows et al.(2007b)]{burrows07b} Burrows, A., Livne, E., Dessart, L., Ott, C.~D., \& Murphy, J.\ 2007b, \apj, 655, 416 
\bibitem[Burrows et al.(2012)]{burrows12} Burrows, A., Dolence, J.~C., \& Murphy, J.~W.\ 2012, \apj, 759, 5 
\bibitem[Chugai \& Utrobin(2000)]{chugai00} Chugai, N.~N., \& Utrobin, V.~P.\ 2000, \aap, 354, 557 
\bibitem[Clausen et al.(2014)]{clausen14} Clausen, D., Piro, A.~L., \& Ott, C.~D.\ 2014, arXiv:1406.4869 
\bibitem[Colgate \& White(1966)]{colgate66} Colgate, S.~A., \& White, R.~H.\ 1966, \apj, 143, 626 
\bibitem[Couch(2013a)]{couch13a} Couch, S.~M.\ 2013a, \apj, 765, 29 
\bibitem[Couch(2013b)]{couch13b} Couch, S.~M.\ 2013b, \apj, 775, 35
\bibitem[Couch \& O'Connor(2014)]{couch14} Couch, S.~M., \& O'Connor, E.~P.\ 2014, \apj, 785, 123 
\bibitem[Couch \& Ott(2013)]{couchott13} Couch, S.~M., \& Ott, C.~D.\ 2013, \apjl, 778, L7 
\bibitem[Couch \& Ott(2014)]{couchott14} Couch, S.~M., \& Ott, C.~D.\ 2014, arXiv:1408.1399 
\bibitem[Dall'Ora et al.(2014)]{dallora14} Dall'Ora, M., Botticella, M.~T., Pumo, M.~L., et al.\ 2014, \apj, 787, 139 
\bibitem[Dessart et al.(2006)]{dessart06} Dessart, L., Burrows, A., Livne, E., \& Ott, C.~D.\ 2006, \apj, 645, 534 
\bibitem[Dessart \& Hillier(2011)]{dessart11} Dessart, L., \& Hillier, D.~J.\ 2011, \mnras, 410, 1739 
\bibitem[Dessart et al.(2012)]{dessart12} Dessart, L., O'Connor, E., \& Ott, C.~D.\ 2012, \apj, 754, 76 
\bibitem[Dessart et al.(2013)]{dessart13} Dessart, L., Hillier, D.~J., Waldman, R., \& Livne, E.\ 2013, \mnras, 433, 1745 
\bibitem[Dolence et al.(2013)]{dolence13} Dolence, J.~C., Burrows, A., Murphy, J.~W., \& Nordhaus, J.\ 2013, \apj, 765, 110 
\bibitem[Dolence et al.(2014)]{dolence14} Dolence, J.~C., Burrows, A., \& Zhang, W.\ 2014, arXiv:1403.6115 
\bibitem[Duncan et al.(1986)]{duncan86} Duncan, R.~C., Shapiro, S.~L., \& Wasserman, I.\ 1986, \apj, 309, 141
\bibitem[Farr et al.(2011)]{farr11} Farr, W.~M., Sravan, N., Cantrell, A., et al.\ 2011, \apj, 741, 103 
\bibitem[Fern{\'a}ndez(2010)]{fernandez10} Fern{\'a}ndez, R.\ 2010, \apj, 725, 1563 
\bibitem[Fern{\'a}ndez(2012)]{fernandez12} Fern{\'a}ndez, R.\ 2012, \apj, 749, 142 
\bibitem[Fern{\'a}ndez et al.(2014)]{fernandez14} Fern{\'a}ndez, R., M{\"u}ller, B., Foglizzo, T., \& Janka, H.-T.\ 2014, \mnras, 440, 2763 
\bibitem[Finn(1994)]{finn94} Finn, L.~S.\ 1994, Physical Review Letters, 73, 1878 
\bibitem[Foglizzo(2002)]{foglizzo02} Foglizzo, T.\ 2002, \aap, 392, 353 
\bibitem[Foglizzo et al.(2006)]{foglizzo06} Foglizzo, T., Scheck, L., \& Janka, H.-T.\ 2006, \apj, 652, 1436 
\bibitem[Foglizzo et al.(2007)]{foglizzo07} Foglizzo, T., Galletti, P., Scheck, L., \& Janka, H.-T.\ 2007, \apj, 654, 1006 
\bibitem[Fruchter et al.(2006)]{fruchter06} Fruchter, A.~S., Levan, A.~J., Strolger, L., et al.\ 2006, \nat, 441, 463 
\bibitem[Fryer \& Heger(2000)]{fryer00} Fryer, C.~L., \& Heger, A.\ 2000, \apj, 541, 1033 
\bibitem[Fryer \& Kalogera(2001)]{fryer01} Fryer, C.~L., \& Kalogera, V.\ 2001, \apj, 554, 548 
\bibitem[Fryer \& Warren(2002)]{fryer02} Fryer, C.~L., \& Warren, M.~S.\ 2002, \apjl, 574, L65 
\bibitem[Fryer \& Warren(2004)]{fryer04} Fryer, C.~L., \& Warren, M.~S.\ 2004, \apj, 601, 391 
\bibitem[Goldreich \& Weber(1980)]{goldreich80} Goldreich, P., \& Weber, S.~V.\ 1980, \apj, 238, 991 
\bibitem[Gonz{\'a}lez Hern{\'a}ndez et al.(2006)]{gonzalez06} Gonz{\'a}lez Hern{\'a}ndez, J.~I., Rebolo, R., Israelian, G., et al.\ 2006, \apjl, 644, L49 
\bibitem[Hamuy(2003)]{hamuy03} Hamuy, M.\ 2003, \apj, 582, 905 
\bibitem[Hanke et al.(2012)]{hanke12} Hanke, F., Marek, A., M{\"u}ller, B., \& Janka, H.-T.\ 2012, \apj, 755, 138 
\bibitem[Hanke et al.(2013)]{hanke13} Hanke, F., M{\"u}ller, B., Wongwathanarat, A., Marek, A., \& Janka, H.-T.\ 2013, \apj, 770, 66  
\bibitem[Heger et al.(2003)]{heger03} Heger, A., Fryer, C.~L., Woosley, S.~E., Langer, N., \& Hartmann, D.~H.\ 2003, \apj, 591, 288 
\bibitem[Herant et al.(1992)]{herant92} Herant, M., Benz, W., \& Colgate, S.\ 1992, \apj, 395, 642 
\bibitem[Herant et al.(1994)]{herant94} Herant, M., Benz, W., Hix, W.~R., Fryer, C.~L., \& Colgate, S.~A.\ 1994, \apj, 435, 339 
\bibitem[Horiuchi et al.(2011)]{horiuchi11} Horiuchi, S., Beacom, J.~F., Kochanek, C.~S., et al.\ 2011, \apj, 738, 154 
\bibitem[H{\"u}depohl et al.(2010)]{hudepohl10} H{\"u}depohl, L., M{\"u}ller, B., Janka, H.-T., Marek, A., \& Raffelt, G.~G.\ 2010, Physical Review Letters, 104, 251101 
\bibitem[Ibeling \& Heger(2013)]{ibeling13} Ibeling, D., \& Heger, A.\ 2013, \apjl, 765, L43 
\bibitem[Israelian et al.(1999)]{israelian99} Israelian, G., Rebolo, R., Basri, G., Casares, J., \& Mart{\'{\i}}n, E.~L.\ 1999, \nat, 401, 142 
\bibitem[Iwakami et al.(2008)]{iwakami08} Iwakami, W., Kotake, K., Ohnishi, N., Yamada, S., \& Sawada, K.\ 2008, \apj, 678, 1207 
\bibitem[Iwakami et al.(2014)]{iwakami14}  Iwakami, W., Nagakura, H., \& Yamada, S.\ 2014, \apj, 786, 118 
\bibitem[Janka(2001)]{janka01} Janka, H.-T.\ 2001, \aap, 368, 527 
\bibitem[Janka \& Mueller(1996)]{janka96} Janka, H.-T., \& Mueller, E.\ 1996, \aap, 306, 167 
\bibitem[Janka et al.(2008)]{janka08} Janka, H.-T., M{\"u}ller, B., Kitaura, F.~S., \& Buras, R.\ 2008, \aap, 485, 199 
\bibitem[Keil et al.(1996)]{keil96} Keil, W., Janka, H.-T., \& Mueller, E.\ 1996, \apjl, 473, L111 
\bibitem[Kitaura et al.(2006)]{kitaura06} Kitaura, F.~S., Janka, H.-T., \& Hillebrandt, W.\ 2006, \aap, 450, 345 
\bibitem[Kiziltan et al.(2013)]{kiziltan13} Kiziltan, B., Kottas, A., De Yoreo, M., \& Thorsett, S.~E.\ 2013, \apj, 778, 66 
\bibitem[Kochanek et al.(2008)]{kochanek08} Kochanek, C.~S., Beacom, J.~F., Kistler, M.~D., et al.\ 2008, \apj, 684, 1336 
\bibitem[Kochanek(2014)]{kochanek14a} Kochanek, C.~S.\ 2014, \apj, 785, 28
\bibitem[Kochanek(2015)]{kochanek14b} Kochanek, C.~S.\ 2015, \mnras, 446, 1213
\bibitem[Kreidberg et al.(2012)]{kreidberg12} Kreidberg, L., Bailyn, C.~D., Farr, W.~M., \& Kalogera, V.\ 2012, \apj, 757, 36 
\bibitem[Langanke et al.(2003)]{langanke03} Langanke, K., Mart{\'{\i}}nez-Pinedo, G., Sampaio, J.~M., et al.\ 2003, Physical Review Letters, 90, 241102 
\bibitem[Langer(1989)]{langer89} Langer, N.\ 1989, \aap, 220, 135 
\bibitem[Lattimer \& Yahil(1989)]{lattimer89} Lattimer, J.~M., \& Yahil, A.\ 1989, \apj, 340, 426 
\bibitem[Lattimer \& Swesty(1991)]{lattimer91} Lattimer, J.~M., \& Swesty, F.~D.\ 1991, Nuclear Physics A, 535, 331 
\bibitem[Liebend{\"o}rfer et al.(2001)]{liebendorfer01} Liebend{\"o}rfer, M., Mezzacappa, A., Thielemann, F.-K., et al.\ 2001, \prd, 63, 103004 
\bibitem[Limongi \& Chieffi(2006)]{limongi06} Limongi, M., \& Chieffi, A.\ 2006, \apj, 647, 483
\bibitem[Lovegrove \& Woosley(2013)]{lovegrove13} Lovegrove, E., \& Woosley, S.~E.\ 2013, \apj, 769, 109 
\bibitem[Marek \& Janka(2009)]{marek09} Marek, A., \& Janka, H.-T.\ 2009, \apj, 694, 664 
\bibitem[Mathews et al.(2014)]{mathews14} Mathews, G.~J., Hidaka, J., Kajino, T., \& Suzuki, J.\ 2014, \apj, 790, 115 
\bibitem[Metzger et al.(2007)]{metzger07} Metzger, B.~D., Thompson, T.~A., \& Quataert, E.\ 2007, \apj, 659, 561 
\bibitem[Mezzacappa et al.(1998)]{mezzacappa98} Mezzacappa, A., Calder, A.~C., Bruenn, S.~W., et al.\ 1998, \apj, 493, 848 
\bibitem[Mezzacappa et al.(2001)]{mezzacappa01} Mezzacappa, A., Liebend{\"o}rfer, M., Messer, O.~E., et al.\ 2001, Physical Review Letters, 86, 1935 
\bibitem[M{\"u}ller et al.(2012a)]{muller12a} M{\"u}ller, B., Janka, H.-T., \& Marek, A.\ 2012a, \apj, 756, 84 
\bibitem[M{\"u}ller et al.(2012b)]{muller12b} M{\"u}ller, B., Janka, H.-T., \& Heger, A.\ 2012b, \apj, 761, 72 
\bibitem[M{\"u}ller et al.(2013)]{muller13} M{\"u}ller, B., Janka, H.-T., \& Marek, A.\ 2013, \apj, 766, 43 
\bibitem[Murphy \& Burrows(2008)]{murphy08} Murphy, J.~W., \& Burrows, A.\ 2008, \apj, 688, 1159 
\bibitem[Murphy \& Meakin(2011)]{murphy11} Murphy, J.~W., \& Meakin, C.\ 2011, \apj, 742, 74 
\bibitem[Murphy et al.(2013)]{murphy13} Murphy, J.~W., Dolence, J.~C., \& Burrows, A.\ 2013, \apj, 771, 52 
\bibitem[Myra et al.(1987)]{myra87} Myra, E.~S., Bludman, S.~A., Hoffman, Y., et al.\ 1987, \apj, 318, 744 
\bibitem[Nadezhin(1980)]{nadezhin80} Nadezhin, D.~K.\ 1980, \apss, 69, 115 
\bibitem[Nakamura et al.(2014a)]{nakamura14a} Nakamura, K., Kuroda, T., Takiwaki, T., \& Kotake, K.\ 2014, \apj, 793, 45 
\bibitem[Nakamura et al.(2014b)]{nakamura14b} Nakamura, K., Takiwaki, T., Kuroda, T., \& Kotake, K.\ 2014b, arXiv:1406.2415 
\bibitem[Nakazato(2013)]{nakazato13} Nakazato, K.\ 2013, \prd, 88, 083012 
\bibitem[Nomoto(1984)]{nomoto84} Nomoto, K.\ 1984, \apj, 277, 791 
\bibitem[Nomoto(1987)]{nomoto87} Nomoto, K.\ 1987, \apj, 322, 206 
\bibitem[Nomoto \& Hashimoto(1988)]{nomoto88} Nomoto, K., \& Hashimoto, M.\ 1988, \physrep, 163, 13 
\bibitem[Nomoto et al.(2006)]{nomoto06} Nomoto, K., Tominaga, N., Umeda, H., Kobayashi, C., \& Maeda, K.\ 2006, Nuclear Physics A, 777, 424 
\bibitem[Nordhaus et al.(2010)]{nordhaus10} Nordhaus, J., Burrows, A., Almgren, A., \& Bell, J.\ 2010, \apj, 720, 694 
\bibitem[Obergaulinger et al.(2014)]{obergaulinger14}  Obergaulinger, M., Janka, H.-T., \& Aloy, M.~A.\ 2014, \mnras, 445, 3169 
\bibitem[O'Connor \& Ott(2010)]{oconnor10} O'Connor, E., \& Ott, C.~D.\ 2010, Classical and Quantum Gravity, 27, 114103
\bibitem[O'Connor \& Ott(2011)]{oconnor11} O'Connor, E., \& Ott, C.~D.\ 2011, \apj, 730, 70 
\bibitem[O'Connor \& Ott(2013)]{oconnor13} O'Connor, E., \& Ott, C.~D.\ 2013, \apj, 762, 126 
\bibitem[Ohnishi et al.(2006)]{ohnishi06} Ohnishi, N., Kotake, K., \& Yamada, S.\ 2006, \apj, 641, 1018 
\bibitem[Orosz et al.(2001)]{orosz01} Orosz, J.~A., Kuulkers, E., van der Klis, M., et al.\ 2001, \apj, 555, 489 
\bibitem[Ott et al.(2013)]{ott13} Ott, C.~D., Abdikamalov, E., M{\"o}sta, P., et al.\ 2013, \apj, 768, 115 
\bibitem[{\"O}zel et al.(2010)]{ozel10} {\"O}zel, F., Psaltis, D., Narayan, R., \& McClintock, J.~E.\ 2010, \apj, 725, 1918 
\bibitem[{\"O}zel et al.(2012)]{ozel12} {\"O}zel, F., Psaltis, D., Narayan, R., \& Santos Villarreal, A.\ 2012, \apj, 757, 55 
\bibitem[Pastorello et al.(2004)]{pastorello04} Pastorello, A., Zampieri, L., Turatto, M., et al.\ 2004, \mnras, 347, 74 
\bibitem[Pejcha \& Thompson(2012)]{pt12} Pejcha, O., \& Thompson, T.~A.\ 2012, \apj, 746, 106 
\bibitem[Pejcha et al.(2012a)]{pejcha_ns} Pejcha, O., Thompson, T.~A., \& Kochanek, C.~S.\ 2012a, \mnras, 424, 1570 
\bibitem[Pejcha et al.(2012b)]{pejcha_cno} Pejcha, O., Dasgupta, B., \& Thompson, T.~A.\ 2012b, \mnras, 425, 1083 
\bibitem[Pejcha \& Prieto(2014)]{pp14} Pejcha, O., \& Prieto, J.~L.\ 2014, accepted to \apj, arXiv:1409.2500 
\bibitem[Piro(2013)]{piro13} Piro, A.~L.\ 2013, \apjl, 768, L14 
\bibitem[Poznanski(2013)]{poznanski13} Poznanski, D.\ 2013, \mnras, 436, 3224
\bibitem[Rampp \& Janka(2000)]{rampp00} Rampp, M., \& Janka, H.-T.\ 2000, \apjl, 539, L33 
\bibitem[Richardson et al.(2002)]{richardson02} Richardson, D., Branch, D., Casebeer, D., et al.\ 2002, \aj, 123, 745 
\bibitem[Richardson et al.(2014)]{richardson14} Richardson, D., Jenkins, R.~L., III, Wright, J., \& Maddox, L.\ 2014, \aj, 147, 118 
\bibitem[Salpeter(1955)]{salpeter55} Salpeter, E.~E.\ 1955, \apj, 121, 161 
\bibitem[Savaglio et al.(2009)]{savaglio09} Savaglio, S., Glazebrook, K., \& Le Borgne, D.\ 2009, \apj, 691, 182 
\bibitem[Sawai et al.(2013)]{sawai13} Sawai, H., Yamada, S., \& Suzuki, H.\ 2013, \apjl, 770, L19 
\bibitem[Sawai \& Yamada(2014)]{sawai14} Sawai, H., \& Yamada, S.\ 2014, \apjl, 784, L10 
\bibitem[Scheck et al.(2006)]{scheck06} Scheck, L., Kifonidis, K., Janka, H.-T., M\"{u}ller, E.\ 2006, \aap, 457, 963
\bibitem[Scheck et al.(2008)]{scheck08} Scheck, L., Janka, H.-T., Foglizzo, T., \& Kifonidis, K.\ 2008, \aap, 477, 931 
\bibitem[Schwab et al.(2010)]{schwab10} Schwab, J., Podsiadlowski, P., \& Rappaport, S.\ 2010, \apj, 719, 722 
\bibitem[Shen et al.(2011)]{shen11} Shen, H., Toki, H., Oyamatsu, K., \& Sumiyoshi, K.\ 2011, \apjs, 197, 20 
\bibitem[Smartt(2009)]{smartt09} Smartt, S.~J.\ 2009, \araa, 47, 63
\bibitem[Smartt et al.(2009)]{smartt_etal09} Smartt, S.~J., Eldridge, J.~J., Crockett, R.~M., \& Maund, J.~R.\ 2009, \mnras, 395, 1409 
\bibitem[Spiro et al.(2014)]{spiro14} Spiro, S., Pastorello, A., Pumo, M.~L., et al.\ 2014, \mnras, 439, 2873 
\bibitem[Stanek et al.(2006)]{stanek06} Stanek, K.~Z., Gnedin, O.~Y., Beacom, J.~F., et al.\ 2006, \actaa, 56, 333 
\bibitem[Sukhbold \& Woosley(2014)]{sukhbold14} Sukhbold, T., \& Woosley, S.~E.\ 2014, \apj, 783, 10  
\bibitem[Suwa et al.(2010)]{suwa10} Suwa, Y., Kotake, K., Takiwaki, T., et al.\ 2010, \pasj, 62, L49 
\bibitem[Suwa et al.(2013)]{suwa13} Suwa, Y., Takiwaki, T., Kotake, K., et al.\ 2013, \apj, 764, 99 
\bibitem[Suwa et al.(2014)]{suwa14} Suwa, Y., Yamada, S., Takiwaki, T., \& Kotake, K.\ 2014, arXiv:1406.6414 
\bibitem[Takiwaki et al.(2012)]{takiwaki12} Takiwaki, T., Kotake, K., \& Suwa, Y.\ 2012, \apj, 749, 98 
\bibitem[Takiwaki et al.(2014)]{takiwaki14} Takiwaki, T., Kotake, K., \& Suwa, Y.\ 2014, \apj, 786, 83 
\bibitem[Thielemann et al.(1990)]{thielemann90} Thielemann, F.-K., Hashimoto, M.-A., \& Nomoto, K.\ 1990, \apj, 349, 222 
\bibitem[Thielemann et al.(1996)]{thielemann96} Thielemann, F.-K., Nomoto, K., \& Hashimoto, M.-A.\ 1996, \apj, 460, 408 
\bibitem[Thompson et al.(2001)]{thompson01} Thompson, T.~A., Burrows, A., \& Meyer, B.~S.\ 2001, \apj, 562, 887 
\bibitem[Thompson et al.(2003)]{thompson03} Thompson, T.~A., Burrows, A., \& Pinto, P.~A.\ 2003, \apj, 592, 434 
\bibitem[Thompson et al.(2005)]{thompson05} Thompson, T.~A., Quataert, E., \& Burrows, A.\ 2005, \apj, 620, 861 
\bibitem[Thorsett \& Chakrabarty(1999)]{thorsett99} Thorsett, S.~E., \& Chakrabarty, D.\ 1999, \apj, 512, 288 
\bibitem[Timmes et al.(1996)]{timmes96} Timmes, F.~X., Woosley, S.~E., \& Weaver, T.~A.\ 1996, \apj, 457, 834 
\bibitem[Timmes \& Swesty(2000)]{timmes00} Timmes, F.~X., \& Swesty, F.~D.\ 2000, \apjs, 126, 501 
\bibitem[Turatto et al.(1998)]{turatto98} Turatto, M., Mazzali, P.~A., Young, T.~R., et al.\ 1998, \apjl, 498, L129 
\bibitem[Ugliano et al.(2012)]{ugliano12} Ugliano, M., Janka, H.-T., Marek, A., \& Arcones, A.\ 2012, \apj, 757, 69 
\bibitem[Utrobin(2007)]{utrobin07} Utrobin, V.~P.\ 2007, \aap, 461, 233 
\bibitem[Utrobin \& Chugai(2009)]{utrobin09} Utrobin, V.~P., \& Chugai, N.~N.\ 2009, \aap, 506, 829 
\bibitem[Valentim et al.(2011)]{valentim11} Valentim, R., Rangel, E., \& Horvath, J.~E.\ 2011, \mnras, 414, 1427 
\bibitem[Weaver \& Woosley(1980)]{weaver80} Weaver, T.~A., \& Woosley, S.~E.\ 1980, Ninth Texas Symposium on Relativistic Astrophysics, 336, 335
\bibitem[West et al.(2013)]{west13} West, C., Heger, A., \& Austin, S.~M.\ 2013, \apj, 769, 2 
\bibitem[Willems et al.(2005)]{willems05} Willems, B., Henninger, M., Levin, T., et al.\ 2005, \apj, 625, 324  
\bibitem[Wong et al.(2012)]{wong12} Wong, T.-W., Valsecchi, F., Fragos, T., \& Kalogera, V.\ 2012, \apj, 747, 111 
\bibitem[Woosley(1988)]{woosley88} Woosley, S.~E.\ 1988, \apj, 330, 218 
\bibitem[Woosley \& Weaver(1995)]{woosley95} Woosley, S.~E., \& Weaver, T.~A.\ 1995, \apjs, 101, 181 
\bibitem[Woosley et al.(2002)]{woosley02} Woosley, S.~E., Heger, A., \& Weaver, T.~A.\ 2002, Reviews of Modern Physics, 74, 1015 
\bibitem[Woosley \& Heger(2007)]{woosley07} Woosley, S.~E., \& Heger, A.\ 2007, \physrep, 442, 269 
\bibitem[Yahil \& Lattimer(1982)]{yahil82} Yahil, A., \& Lattimer, J.~M.\ 1982, NATO ASIC Proc.~90: Supernovae: A Survey of Current Research, 53 
\bibitem[Yahil(1983)]{yahil83} Yahil, A.\ 1983, \apj, 265, 1047 
\bibitem[Yamamoto et al.(2013)]{yamamoto13} Yamamoto, Y., Fujimoto, S.-i., Nagakura, H., \& Yamada, S.\ 2013, \apj, 771, 27 
\bibitem[Yamasaki \& Yamada(2005)]{yamasaki05} Yamasaki, T., \& Yamada, S.\ 2005, \apj, 623, 1000 
\bibitem[Yamasaki \& Yamada(2006)]{yamasaki06} Yamasaki, T., \& Yamada, S.\ 2006, \apj, 650, 291 
\bibitem[Yamasaki \& Yamada(2007)]{yamasaki07} Yamasaki, T., \& Yamada, S.\ 2007, \apj, 656, 1019 
\bibitem[Zampieri et al.(1998)]{zampieri98} Zampieri, L., Shapiro, S.~L., \& Colpi, M.\ 1998, \apjl, 502, L149 
\bibitem[Zhang et al.(2008)]{zhang08} Zhang, W., Woosley, S.~E., \& Heger, A.\ 2008, \apj, 679, 639 
\bibitem[Zhang et al.(2011)]{zhang11} Zhang, C.~M., Wang, J., Zhao, Y.~H., et al.\ 2011, \aap, 527, A83 


\end{thebibliography}
\end{document}